\documentclass[twocolumn]{aastex6}
\usepackage{times}
\usepackage{amsmath}
\usepackage{textcomp}
\usepackage{amssymb}
\usepackage{natbib}
\usepackage{float}
\usepackage{color}
\usepackage{graphicx}
\usepackage{epstopdf}
\newcommand{\Msun}{\ensuremath{M_{\odot}}}
\newcommand{\lum}{erg\,s$^{-1}$}
\newcommand{\fermi}{{\it Fermi}}
\newcommand{\nustar}{{\it NuSTAR}}
\newcommand{\xmm}{{\it XMM-Newton}}
\newcommand{\swift}{{\it Swift}}
\newcommand{\chandra}{{\it Chandra}}

\newcommand{\ergflux}{\mbox{${\rm \, erg \,\, cm^{-2} \, s^{-1}}$}}
\newcommand{\gm}{$\gamma$}
\newcommand{\gl}{$\gamma$-ray loud}
\newcommand{\gq}{$\gamma$-ray quiet}
\slugcomment{ApJ accepted}

\shorttitle{{\it CGRaBS} Blazars}
\shortauthors{Paliya et al.}

\begin{document}
\title{General Physical Properties of {\it CGRaBS} Blazars}

\author{Vaidehi S. Paliya$^{1}$, L. Marcotulli$^{1}$, M. Ajello$^{1}$, M. Joshi$^{2}$, S. Sahayanathan$^{3}$, A. R. Rao$^{4}$, and D. Hartmann$^{1}$} 
\affil{$^1$Department of Physics and Astronomy, Clemson University, Kinard Lab of Physics, Clemson, SC 29634-0978, USA}
\affil{$^2$Institute for Astrophysical Research, Boston University, 725 Commonwealth Avenue, Boston, MA 02215, USA}
\affil{$^3$Astrophysical Sciences Division, Bhabha Atomic Research Centre, Mumbai-400085, India}
\affil{$^4$Department of Astronomy and Astrophysics, Tata Institute of Fundamental Research, Homi Bhabha Road, Mumbai, India}
\email{vpaliya@g.clemson.edu}

\begin{abstract}
We present the results of a multi-frequency, time-averaged analysis of blazars included in the Candidate Gamma-ray Blazar Survey catalog. Our sample consists of 324 \gm-ray detected (\gl) and 191 non \gm-ray detected (\gq) blazars, and we consider all the data up to 2016 April 1. We find that both the \gl~and the \gq~blazar populations occupy similar regions in the {\it WISE} color-color diagram, and in the radio and X-ray bands \gl~sources are brighter. A simple one-zone synchrotron inverse-Compton emission model is applied to derive the physical properties of both populations. We find that the central black hole mass and the accretion disk luminosity ($L_{\rm disk}$) computed from the modeling of the optical-UV emission with a Shakura-Sunyaev disk reasonably matches with that estimated from the optical spectroscopic emission-line information. A significantly larger Doppler boosting in the \gl~blazars is noted, and their jets are more radiatively efficient. On the other hand, the \gq~objects are more MeV-peaked, thus could be potential targets for next-generation MeV missions. Our results confirm the earlier findings about the accretion-jet connection in blazars; however, many of the \gq~blazars tend to deviate from the recent claim that the jet power exceeds $L_{\rm disk}$ in blazars. A broadband study, considering a larger set of \gq~objects and also including BL Lacs, will be needed to confirm/reject this hypothesis and also to verify the evolution of the powerful high-redshift blazars into their low-power nearby counterparts.

\end{abstract}

\keywords{galaxies: active --- gamma-ray: galaxies--- galaxies: jets--- galaxies: high-redshift--- quasars: general}

\section{Introduction}{\label{sec:Intro}}
Blazars are radio-loud (RL) active galactic nuclei (AGN) that are known to host powerful relativistic jets closely aligned to the line of sight to the observer \citep[e.g.,][]{1995PASP..107..803U}. Blazars are known to emit radiation over the entire electromagnetic spectrum, from low-energy radio waves to very-high-energy \gm-rays, and they also exhibit flux and polarization variations \citep[][]{2002BASI...30..765S,2004MNRAS.348..176S,2010ApJ...715..362J,2011ApJ...730L...8A,2012MNRAS.424.2625B,2015ApJ...811..143P,2016A&A...591A..21M}. All these effects are believed to arise from the Doppler boosting that occurs due to the peculiar orientation of the jet (viewing angle $\theta_{\rm v} < 1/\Gamma$; where $\Gamma$ is the bulk Lorentz factor of the jet). Accordingly, for each observed blazar with a known $\Gamma$ there exist 2$\Gamma^2$ intrinsically similar sources with their jets pointed in other directions.

Based on the equivalent width (EW) of the optical emission lines, blazars are classified as flat spectrum radio quasars (FSRQs) and BL Lac objects, with FSRQs exhibiting broad emission lines (EW$>$5\AA). The observation of strong emission lines from FSRQs indicates the presence of a luminous broad line region (BLR), which in turn suggests a high and efficient accretion process that illuminates the BLR \citep[e.g.,][]{2012MNRAS.421.1764S}. In fact, the infrared-to-ultraviolet (IR to UV) spectrum of many FSRQs is found to be dominated by the thermal emission from the accretion disk \citep[e.g.,][]{2010MNRAS.402..497G}. The presence/absence of narrow emission lines in the optical spectrum of BL Lac objects, on the other hand, indicates a relatively low and inefficient accretion and/or the dominance of the non-thermal synchrotron radiation originating from the plasma moving along the relativistic jet.

The radio to \gm-ray spectral energy distribution (SED) of blazars exhibits a characteristic double hump structure. The low-energy peak lies at radio to optical-UV or X-ray frequencies and is attributed to  synchrotron emission, whereas the origin of the high-energy peak is often associated to the inverse Compton (IC) scattering off  low-energy photons by the energetic electrons present in the jet. When the low-energy target photons are provided by synchrotron emission, the IC mechanism is called synchrotron self Compton \citep[SSC; e.g.,][]{1985ApJ...298..114M}. This, along with the synchrotron process, is found to satisfactorily explain the broadband SEDs of BL Lac objects \citep[e.g.,][]{2010MNRAS.401.1570T,2014MNRAS.439.2933Y}. On the other hand, the high-energy \gm-ray emission from powerful FSRQs is conventionally explained by the IC process with low-energy photons originating outside the jet \citep[External Compton or EC; e.g.,][]{1987ApJ...322..650B}. In this case, the reservoirs of seed photons for IC mechanism could be the accretion disk \citep[][]{1993ApJ...416..458D}, the BLR \citep[][]{1994ApJ...421..153S}, and/or the dusty torus \citep[][]{2000ApJ...545..107B}. The SED of blazars is also successfully reproduced by hadronic processes \citep[e.g.,][]{2000CoPhC.124..290M}, though the resulting parameters (e.g., magnetic field) are found to be quite extreme \citep[e.g.,][]{2011ApJ...736..131A}.

It is well known since the Energetic Gamma Ray Experiment Telescope \citep[][]{1993ApJS...86..629T} era that the high-energy \gm-ray extragalactic sky is dominated by blazars \citep[]{1999ApJS..123...79H}. This has later been confirmed with the advent of the Large Area Telescope onboard \fermi~{\it Gamma-ray Space Telescope} \citep[\fermi-LAT;][]{2009ApJ...697.1071A}. In fact, \fermi-LAT  detected more than a thousand blazars in its first four years of operation \citep[][]{2015ApJ...810...14A}. The availability of good quality \fermi-LAT data, complemented by the lower-frequency monitoring from various observing facilities\footnote{https://fermi.gsfc.nasa.gov/ssc/observations/multi/programs.html}, has allowed population studies of blazars \citep[see, e.g.,][]{2009MNRAS.399.2041G,2010MNRAS.402..497G,2010MNRAS.401.1570T,2011ApJ...740...98M,2012ApJ...751..108A,2012A&A...541A.160G,2013ApJ...763..134F,2014ApJ...780...73A,2014Natur.515..376G,2014ApJS..215....5K,2014ApJ...788..104Z,2015AJ....150....8C,2015MNRAS.448.1060G,2016ApJS..226...20F,2016A&A...591A.130K,2016ApJS..224...26M,2017MNRAS.469..255G}.

Though \fermi-LAT has observed a significant \gm-ray emission from  a large number of blazars, an even larger number of such sources is yet to be detected in the \gm-ray band \citep[e.g.,][]{2015Ap&SS.357...75M}. Various explanations have been proposed to explain this, such as the differences in the Doppler boosting, apparent jet speed, apparent opening angle, very long baseline interferometry core flux densities, and brightness temperatures \citep[][]{2009A&A...507L..33P,2015ApJ...810L...9L}.

Motivated by the availability of good quality multi-wavelength (MW) data set for a large number of blazars, we systematically study their broadband properties using both observational and theoretical SED modeling approaches. Our primary goal is to investigate the fundamental properties of blazars such as the accretion-jet connection, the cosmological evolution of relativistic jets and also the differences in the physical characteristics of the known \gm-ray sources and non \gm-ray emitters. Compared to earlier studies, we focus on the observational results and then interpret them using a leptonic emission model. Here we present the results of our study on the blazars included in the Candidate Gamma-ray Blazars Survey catalog \citep[CGRaBS;][]{2008ApJS..175...97H}. Ours is probably the largest sample of blazars on which a physical SED model is applied. In Section \ref{sec2} we describe the steps adopted to select the objects, and in Section \ref{sec3} the details of the data reduction procedures are given. The observational results are presented in Section \ref{sec4}. We briefly describe the adopted one-zone leptonic emission model in Section \ref{sec5} and present the results associated with the SED modeling in Section \ref{sec6}. We discuss our findings in Section \ref{sec7} and summarize in Section \ref{sec8}.

We adopt a flat cosmology with $H_0=67.8$ km s$^{-1}$ Mpc$^{-1}$ and $\Omega_{\rm M}=0.308$ \citep[][]{2016A&A...594A..13P}.
 
\section{The Sample}\label{sec2}
\begin{table}
\caption{The CGRaBS sample studied in this work. Apart from them, we also include four CGRaBS blazars that were recently reported as \gm-ray emitting source in literature. See the text for details.\label{tab:sample}}
\begin{center}
\begin{tabular}{lcc}
\hline
 & \gl & \\
Catalog & FSRQs & BL Lacs \\
\hline															     			
3FGL                           & 235 & 75    \\
2FGL                           & 8    & 0       \\
1FGL                           & 2    & 0	    \\
\hline
 & \gq & \\
Catalog & FSRQs & BL Lacs \\
\hline
CGRaBS & 185  & 6  \\
\hline
\end{tabular}
\end{center}
\end{table}

 CGRaBS is a flux-limited \citep[8.4 GHz flux density $>65$ mJy, see also,][]{2007ApJS..171...61H} catalog of 1625 high-latitude ($|{\rm b}|>10^{\circ}$), flat radio spectrum objects prepared to serve for the MW followup of likely \gm-ray emitting AGNs \citep[][]{2008ApJS..175...97H}. To determine the \gm-ray detected CGRaBS blazars, we cross-match CGRaBS with all of the \fermi-LAT catalogs\footnote{The motivation behind cross-matching CGRaBS with all of the \fermi-LAT catalogs is to take into account the possibility that an object might have appeared in earlier catalogs but was left out in later ones, probably due to the detection significance of the source falling below the threshold.} \citep[3FGL, 2FGL, 1FGL, and 0FGL, sequentially;][]{2015ApJS..218...23A,2012ApJS..199...31N,2010ApJ...715..429A,2009ApJS..183...46A}. This is done by searching for a CGRaBS counterpart of the \gm-ray source within its 95\% positional uncertainty. We consider a source to be \gm-ray detected if it is included in any of the \fermi-LAT catalogs. We find a total of 467 matches and 1178 remain as no-LAT detected.
 Further, we look into the HEASARC archive for the availability of MW data, particularly X-rays. In our final sample, we keep only sources that have existing MW data set and have redshift information either given in \fermi-LAT  AGN catalogs or in CGRaBS. This exercise has led to the inclusion of 515 CGRaBS sources comprising 324 \gm-ray detected (or \gl) blazars and 191 non \gm-ray detected (hereafter \gq) objects. Table \ref{tab:sample} presents the breakdown of sources by  type for the entire sample. We also consider four CGRaBS blazars that were recently reported as new \gm-ray emitters. These include J0225+1846\footnote{We do not use the prefix CGRABS for the sake of clarity.} \citep[][]{2016ApJ...825...74P}, J0646+4451 and J2129$-$1538 \citep[][]{2017ApJ...837L...5A} and J1829$-$5813 \citep[][]{2014ATel.6067....1B}. 
 Moreover, we keep the FSRQ/BL Lac distinction as given in \fermi-LAT catalogs or in CGRaBS\footnote{If a source is classified as a FSRQ in the \fermi-LAT catalog and as a BL Lac in CGRaBS, we consider it as a FSRQ. This is because the \fermi-LAT catalogs are updated with more recent results. However, for \gq~sources we follow the CGRaBS classification.}. Since there are only a few BL Lac objects present in our sample ($\sim 15\%$), we refrain from performing a detailed comparison of their physical properties with FSRQs and defer it to a future publication where we will consider all \fermi-LAT detected blazars, including those FSRQs and BL Lac objects that are not present in this work (Paliya et al., in preparation). Instead, here we focus on comparing the MW properties of the \gl~and the \gq~CGRaBS sources. 
 
Our results are based on the time-averaged study of CGRaBS sources, as we have considered all of the available MW data (up to 2016 April 1) without producing for each source multiple SEDs with only contemporaneous data. Blazars are known to show fast variability; however, due to our very large sample such detailed analysis was not feasible. The non-simultaneity could affect the parameters of the individual objects; however, in an overall statistical sense the derived distributions of the results are reliable. We assume that the averaged data represent a typical activity state of the source. This assumption may not be valid for any individual blazar, but it is a reasonable choice when studying hundreds of objects.  The motivation is to characterize the population as a whole. Our approach is similar to that adopted in \citet[][]{2015MNRAS.448.1060G}, although we also reduce the MW data to understand the observational properties of CGRaBS quasars, and we update the \gm-ray results using the latest advancements in  the \fermi-LAT data, particularly the release of the Pass 8 dataset \citep[][]{2013arXiv1303.3514A}. Furthermore, our sample size is significantly larger and consists of both \gl~and \gq~blazars, whereas \citet[][]{2015MNRAS.448.1060G} have focused solely on \gl~objects.

\section{Data Reduction}\label{sec3}
\subsection{\fermi-LAT}

We consider the \fermi-LAT Pass~8 source class photons collected for the period 2008 August 5 to 2016 April 1 ($\sim$92 months) for all 515 blazars studied in this work. We follow the standard data reduction procedure\footnote{http://fermi.gsfc.nasa.gov/ssc/data/analysis/documentation/}, but with a few modifications \citep[see also,][]{2017ApJ...837L...5A}.  A sky model is defined considering all $\gamma$-ray sources included in 3FGL \citep[][]{2015ApJS..218...23A} lying within a region of interest (ROI) of 15$^{\circ}$ radius centered on the target quasar. The isotropic and Galactic diffuse emission models \citep[][]{2016ApJS..223...26A} are also considered. We perform a binned likelihood analysis to derive the best optimized spectral parameters of all of the sources and power-law normalization factors of the diffuse background models. The significance of the source detection is computed by means of the maximum likelihood test statistic TS=  2$\Delta \log \mathcal{L}$, where $\mathcal{L}$ denotes the likelihood function, between models with and without a point object at the position of the source of interest. The spectral models of all of the \fermi-LAT detected blazars are the same as those considered in the LAT catalogs, whereas for the remaining \gq~sources we adopt a simple power-law model to compute the upper limits.

We use a minimum energy of 60 MeV and set the energy upper limit (E$_{\rm max}$) to 300 GeV. To account for the poorer energy resolution of the \fermi-LAT at lower energies we enable the energy dispersion correction for all of the objects except the background diffuse models. Moreover, Pass 8 introduces the classification of photons by point-spread-function (PSF) type, sub-classified into four quartiles by quality of directional reconstruction, with the lowest (PSF0) and the highest (PSF3) quartiles having the worst and the best, respectively, angular reconstruction. We perform a component-wise analysis for each PSF and finally perform a joint fitting using the SUMMED likelihood method of the Science Tools\footnote{http://fermi.gsfc.nasa.gov/ssc/data/analysis/software/}.

In order to search for  faint \gm-ray emitters that are present in the data but not in the 3FGL catalog, we adopt an iterative approach. This step is necessary to update the background models in order to derive the accurate spectral parameters for the sources of interest. We generate a residual TS map for each ROI and scan it for unmodeled excess emissions (TS$\geq$25). Once found, the locations of such unmodeled objects are optimized and inserted in the model file with a power-law spectrum. We repeat this procedure until the TS map stops showing any excess emission.  

For \gq~blazars, we calculate the 3$\sigma$ \fermi-LAT flux sensitivity limit in the direction of the source, for the period covered in the analysis and assuming a photon index of 2.4. For all sources, flux upper limits are derived at 95\% confidence for energy bins with TS$<9$ while generating \gm-ray spectra of the sources. All errors associated with the \fermi-LAT data analysis are 1$\sigma$ statistical uncertainties, unless specified. The entire data analysis is performed using the publicly available python package {\tt fermiPy} \citep[][]{2017arXiv170709551W}.

\subsection{\swift}
We use all of the data from the three instruments onboard the \swift~satellite \citep[][]{2004ApJ...611.1005G}: the Burst Alert Telescope \citep[BAT, 15$-$150 keV;][]{2005SSRv..120..143B}, the X-Ray Telescope \citep[XRT, 0.3$-$10 keV;][]{2005SSRv..120..165B}, and the Ultraviolet Optical Telescope \citep[UVOT][]{2005SSRv..120...95R}, which can observe in six filters, $V, B, U, UVW1, UVM2$, and $UVW2$.

We use publicly available 70-month \swift-BAT survey \citep[][]{2013ApJS..207...19B} spectra and the instrument response file to cover the hard X-ray (15$-$150 keV) part of the SEDs of the blazars included in it.

 We generate 0.3$-$10 keV XRT spectrum using the online tool ``\swift-XRT data product generator\footnote{http://www.swift.ac.uk/user\_objects/}" \citep[][]{2009MNRAS.397.1177E}. The source spectra are appropriately rebinned (20 or 1 counts per bin, depending on the brightness of the source), and we perform the spectral fitting in XSPEC. The faint sources are fitted with a simple power-law model following the C-statistic \citep[][]{1979ApJ...228..939C}. For the remaining objects we follow a $\chi^2$ fitting procedure. In the $\chi^2$ method, we fit two models, power law and log parabola, and choose the best-fitted model based on the outcome of the {\tt f-test}\footnote{We prefer a log-parabola model over the power-law, if the {\tt f-test} probability of null hypothesis is $<10^{-4}$.}. We consider the Galactic neutral Hydrogen column densities from \citet[][]{2005AA...440..775K}. The uncertainties are estimated at the 90\% confidence level.
 
We combine UVOT snapshots using the tool {\tt uvotimsum} and extract source magnitudes with {\tt uvotsource}. For the latter, we extract the source counts from a circular region of 5$^{\prime\prime}$ radius centered at the quasar. The background is chosen as a circular region of the radius of 30$^{\prime\prime}$ from a nearby region free from  source contamination. The extracted magnitudes are corrected for the Galactic reddening \citep[][]{2011ApJ...737..103S} and converted to energy flux using the conversion factors of \citet[][]{2011AIPC.1358..373B}.

\subsection{XMM-Newton and Chandra}
The \xmm~\citep[][]{2001A&A...365L...1J} data (0.3$-$10 keV) are analyzed using the Science Analysis Software version 15.0.0. The task {\tt epproc} is used to generate EPIC-PN event files. We remove the high flaring background using {\tt evselect}. To extract the source and the background spectra, we define the respective regions as circles of 40$^{\prime\prime}$ radius each with the source region being centered at the quasar and the background from the same chip but avoiding contaminating objects. The response files are generated using the tool {\tt rmfgen} and {\tt arfgen}. We bin the spectra using {\tt specgroup} to have a minimum of 20 counts per bin and use XSPEC for spectral fitting.

We reduce the \chandra~\citep[][]{2000SPIE.4012....2W} Advanced CCD Imaging Spectrometer data (ACIS, 0.3$-$7 keV) using the package Chandra Interactive Analysis of Observations (CIAO, version 4.9) and the associated calibration database (v 4.7.3). The data are first reprocessed with the tool {\tt chandra\_repro}. The source and the background spectra are extracted from the cleaned and calibrated event file using the tool {\tt specextract}. A circular region of 3$^{\prime\prime}$ radius centered at the source of interest is chosen to extract source counts, and the background is selected as a circle of 10$^{\prime\prime}$ radius from a nearby source-free region. The data are binned to have at least 20 or 1 count per bin (depending on the brightness of the source), and we perform the fitting in XSPEC.

\section{Observational Characteristics}\label{sec4}
\subsection{Redshift distribution}
\begin{figure}[t]
\includegraphics[scale=0.7]{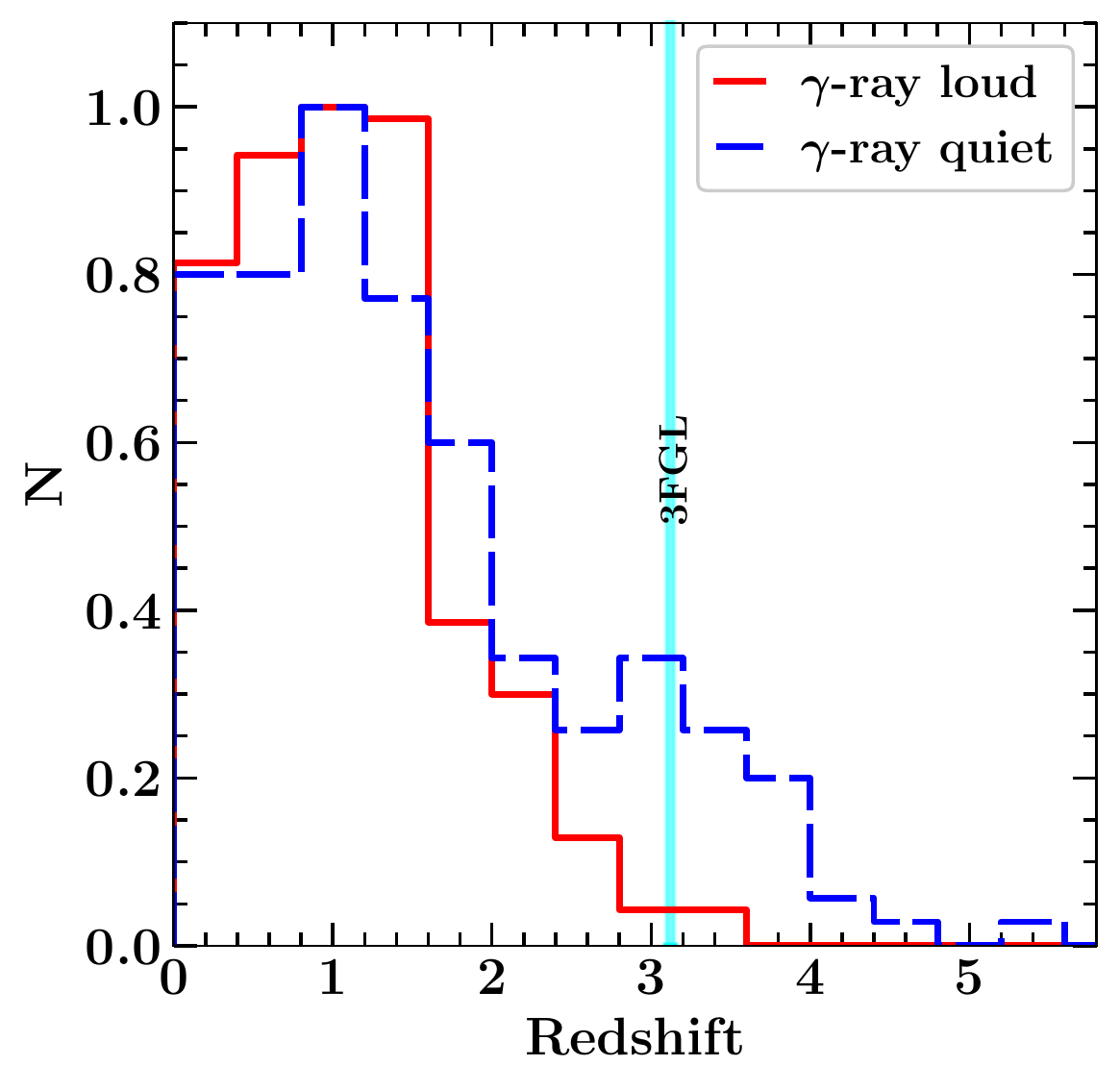}
\caption{The redshift distribution of \gl~(red solid) and \gq~(blue dashed) quasars. For an equal comparison, we have normalized the distributions, i.e., the number of the sources at the peak of the distribution has been set equal to unity. The solid cyan line represents the redshift of the most distant known \gm-ray emitting blazar in the 3FGL catalog, i.e., $z=3.1$. New \gl~blazars lie beyond this value \citep[see also,][]{2017ApJ...837L...5A}.}\label{fig:redshift}
\end{figure}

\begin{figure*}[t]
\hbox{
\includegraphics[scale=0.7]{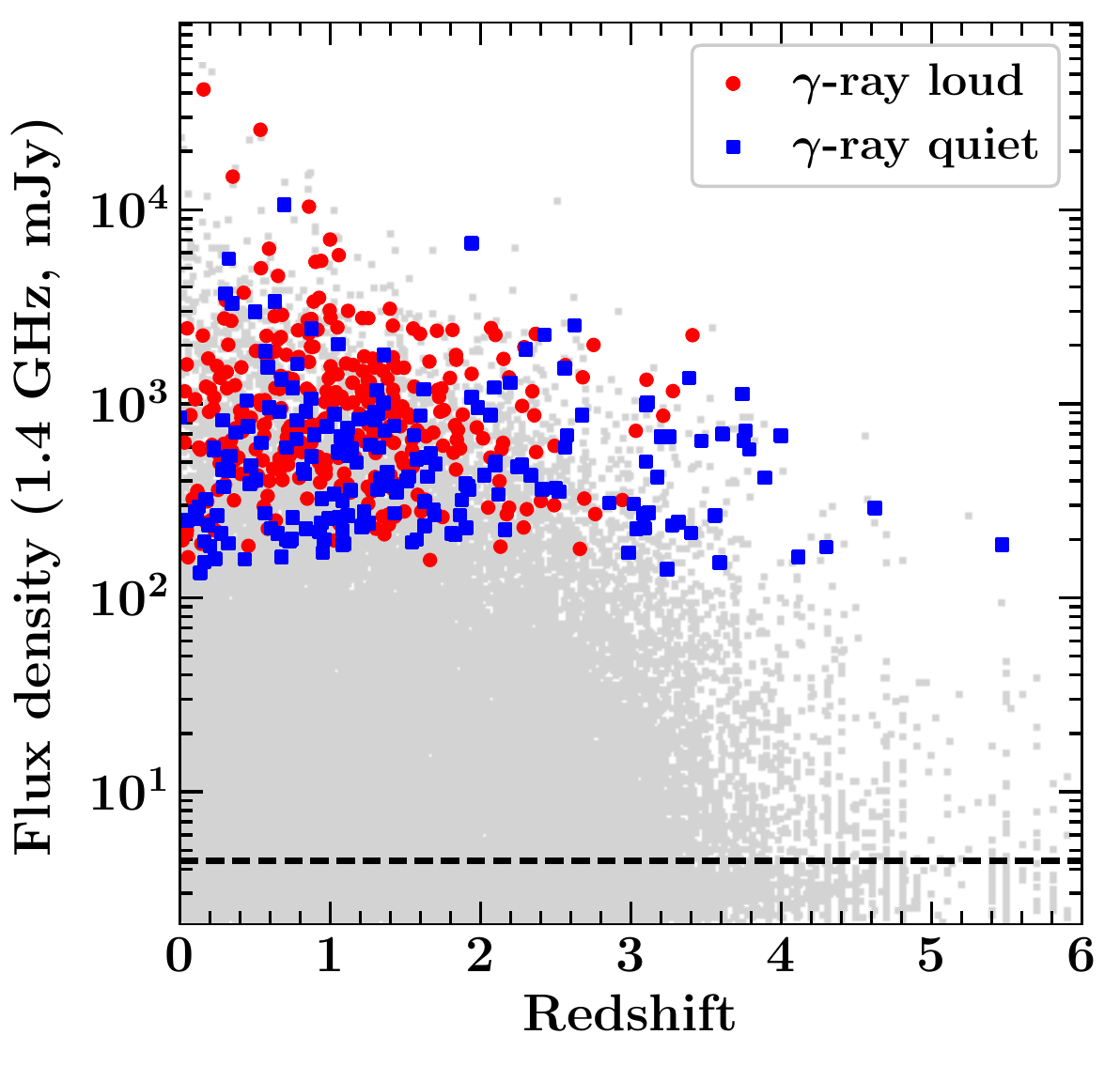}
\includegraphics[scale=0.7]{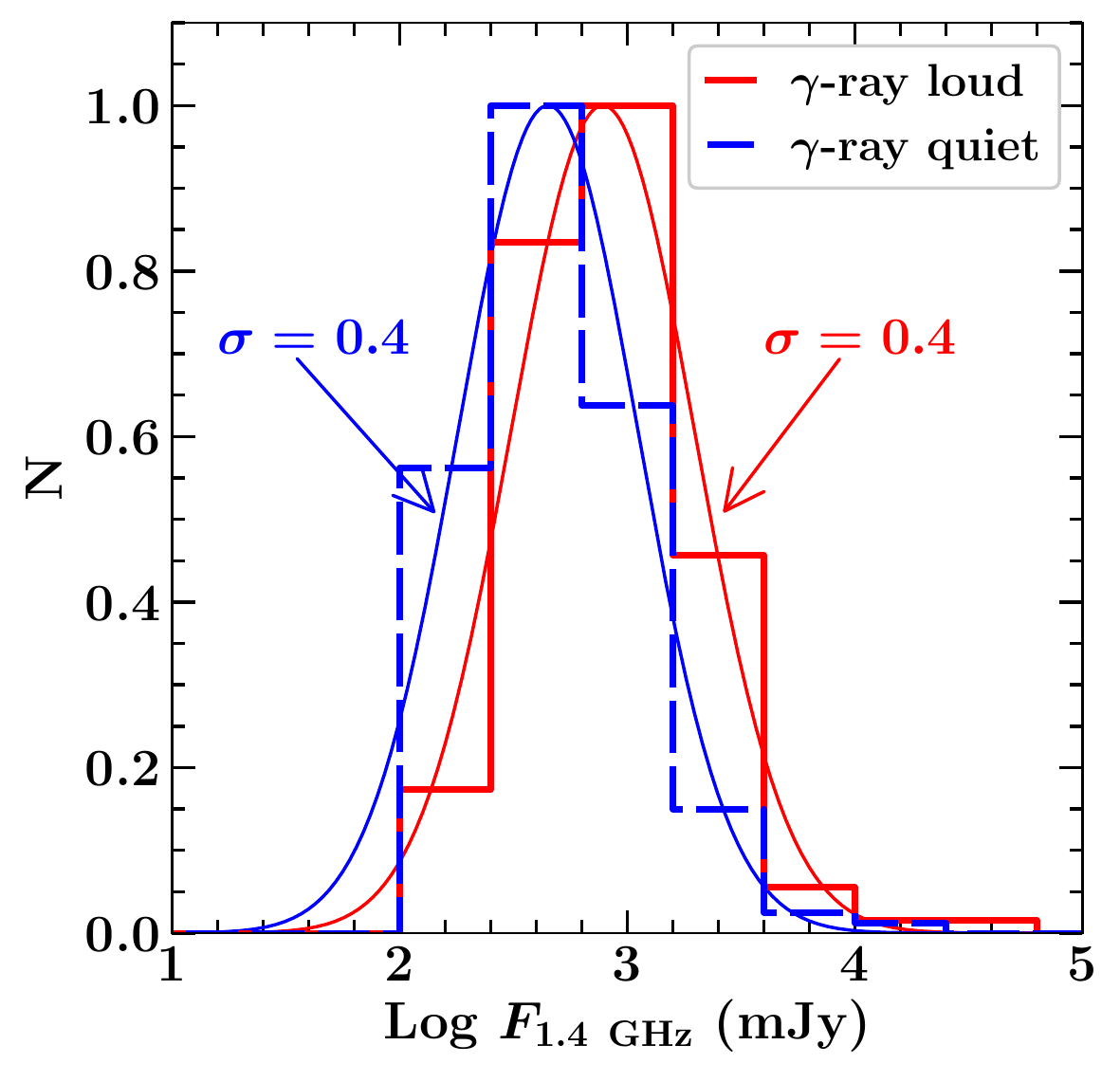}
}
\caption{Left: The 1.4 GHz flux density of \gl~(red) and \gq~(blue) blazars as a function of their redshift. The gray data points correspond to the radio flux densities of the sources included in the Million Quasar Catalog (MQC). The horizontal dashed line represent the median flux density of MQC sources. Right: The radio flux density distributions of \gl~and \gq~sources. The distributions are fitted with a Gaussian function whose width is quoted.}\label{fig:radio}
\end{figure*}

We show the redshift distribution of both \gl~and \gq~blazars in Figure \ref{fig:redshift}. The distributions are normalized for an equal comparison. As can be seen, the sample of \gq~blazars spans a larger redshift range and the most distant \gq~object has a redshift of 5.47 \citep[J0906+6930;][]{2004ApJ...610L...9R}. The relatively low redshifts of \gl~sources could still be due to the relatively high flux threshold of \fermi-LAT. However, it is important to note that there are two sources in our sample, J0646+4451 ($z=3.41$) and J2129$-$1538 ($z=3.28$), from which significant \gm-ray emission is detected. Both these objects are located farther away than the most distant blazar reported in 3FGL (3FGL J0540.0$-$2837, $z=3.1$). They were recently found by \citet[][]{2017ApJ...837L...5A} as new \gl~blazars.

\subsection{Radio Properties}
Both \gl~and \gq~objects present in our sample are bright radio sources. In the left panel of Figure \ref{fig:radio}, we compare the 1.4 GHz flux densities ($F_{1.4~{\rm GHz}}$) of our sources with those included in the Million Quasar Catalog \citep[MQC;][]{2015PASA...32...10F}\footnote{We derive $F_{1.4~{\rm GHz}}$ from $F_{8.4~{\rm GHz}}$ provided in the CGRaBS catalog, assuming $\alpha=0~(F\propto\nu^\alpha)$. For MQC objects, we extract $F_{1.4~{\rm GHz}}$ from the NRAO VLA Sky Survey \citep[NVSS;][]{1998AJ....115.1693C}. This is done by searching for the NVSS counterpart of the MQC source within the positional uncertainty reported in the NVSS catalog.}. All CGRaBS sources have $F_{1.4~{\rm GHz}}>100$ mJy. They are well above the median radio flux density of MQC objects ($\sim$5 mJy, shown with a black dashed line). Comparing the \gl~and \gq~sources (the right panel), we find that the \gl~blazars are brighter ($\langle {F_{\rm 1.4~GHz}} \rangle=776$ mJy) than the \gq~sources ($\langle {F_{\rm 1.4~GHz}} \rangle=447$ mJy). Fitting the shown distributions with a Gaussian function returns an equal dispersion of 0.4 dex.

\subsection{WISE IR-colors}
\begin{figure}[t]
\includegraphics[scale=0.7]{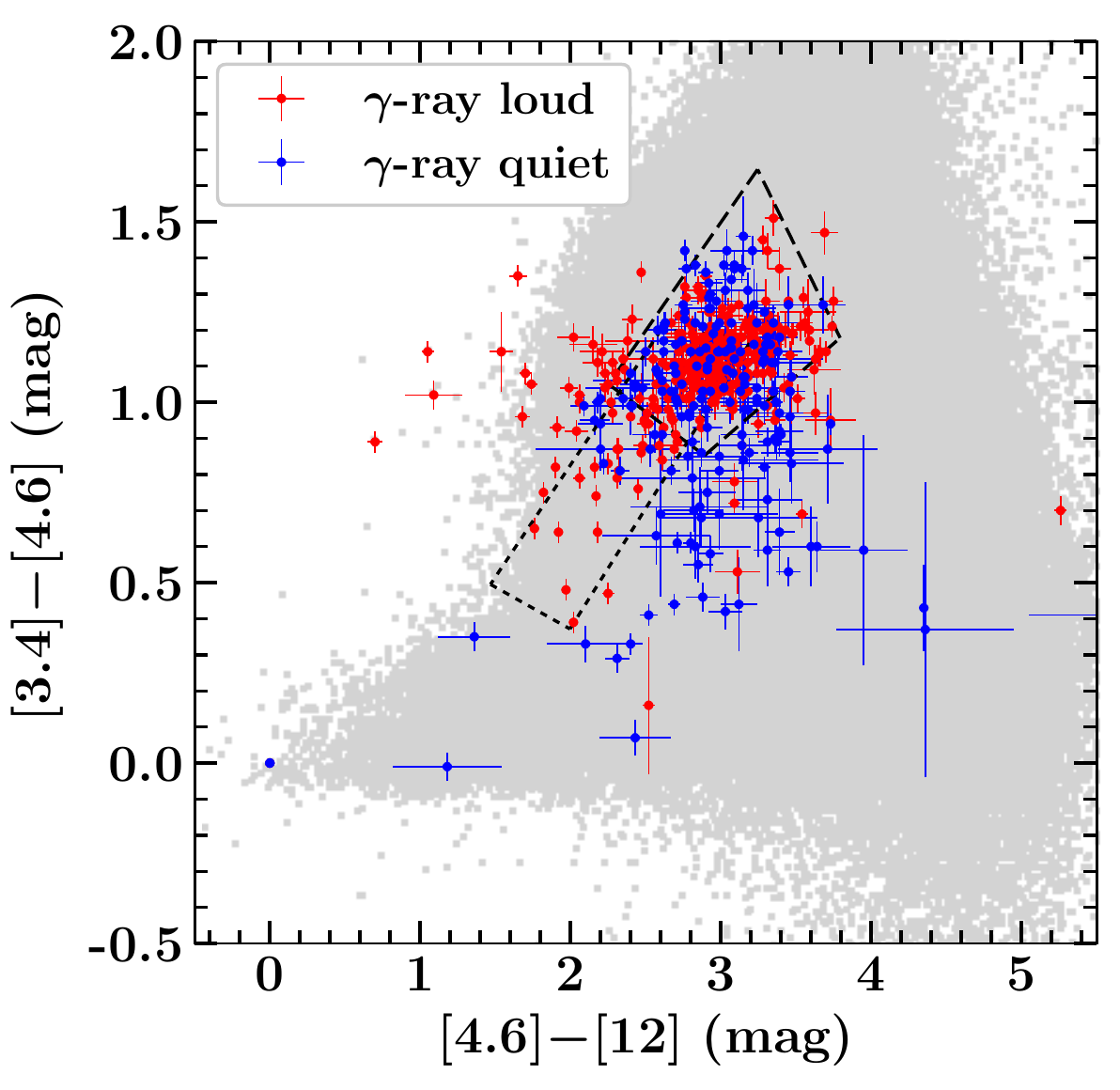}
\caption{The {\it WISE} [3.4]$-$[4.6]$-$[12] $\mu$m color-color diagram for CGRaBS quasars (red and blue for \gl~and \gq~objects, respectively) and MQC sources (gray). The black dashed and dotted lines correspond to the {\it WISE} Gamma-ray strips for FSRQs and BL Lac objects, respectively \citep[][]{2012ApJ...750..138M}.}\label{fig:wise}
\end{figure}

\begin{figure*}
\includegraphics[scale=0.7]{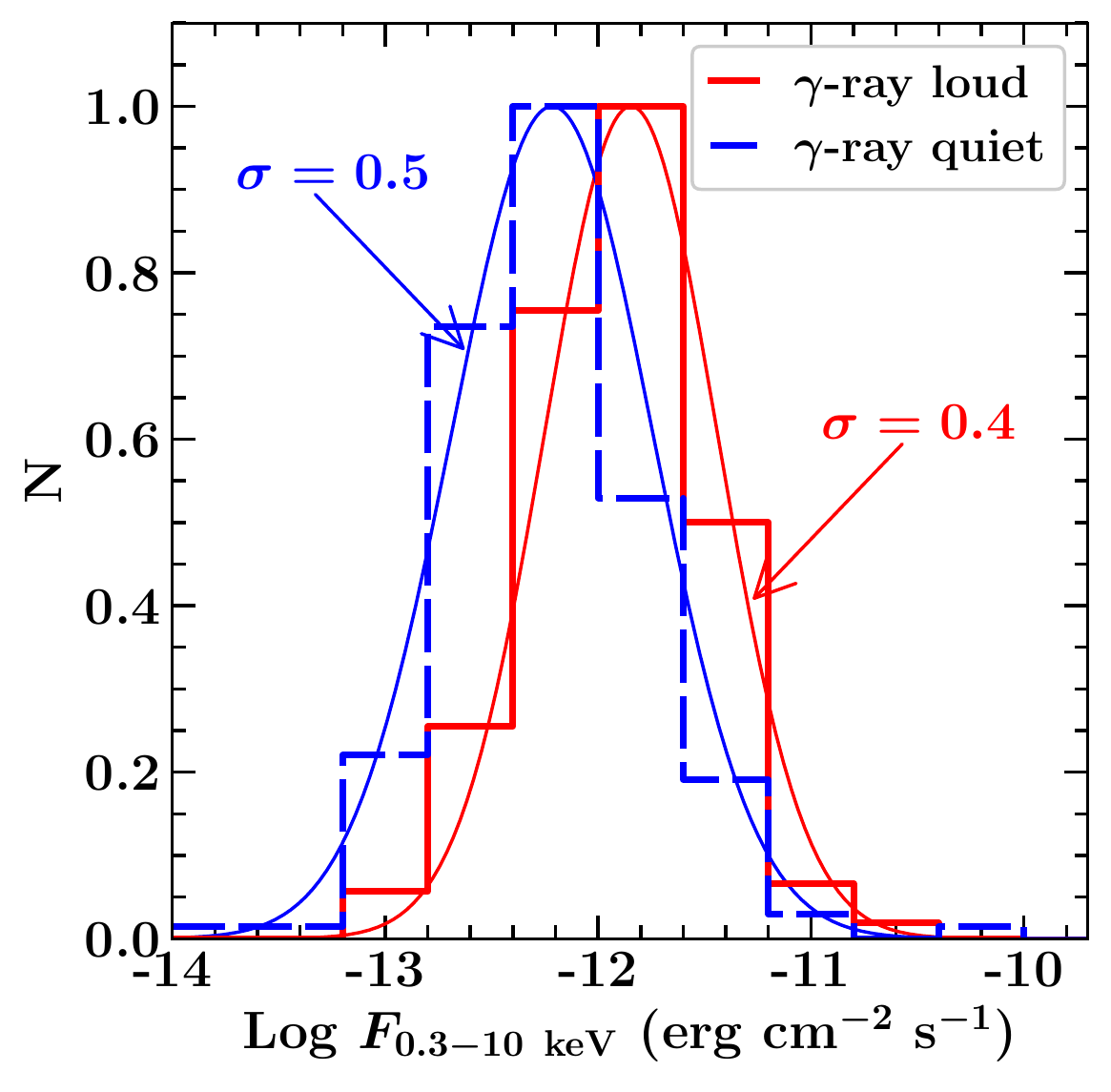}
\includegraphics[scale=0.7]{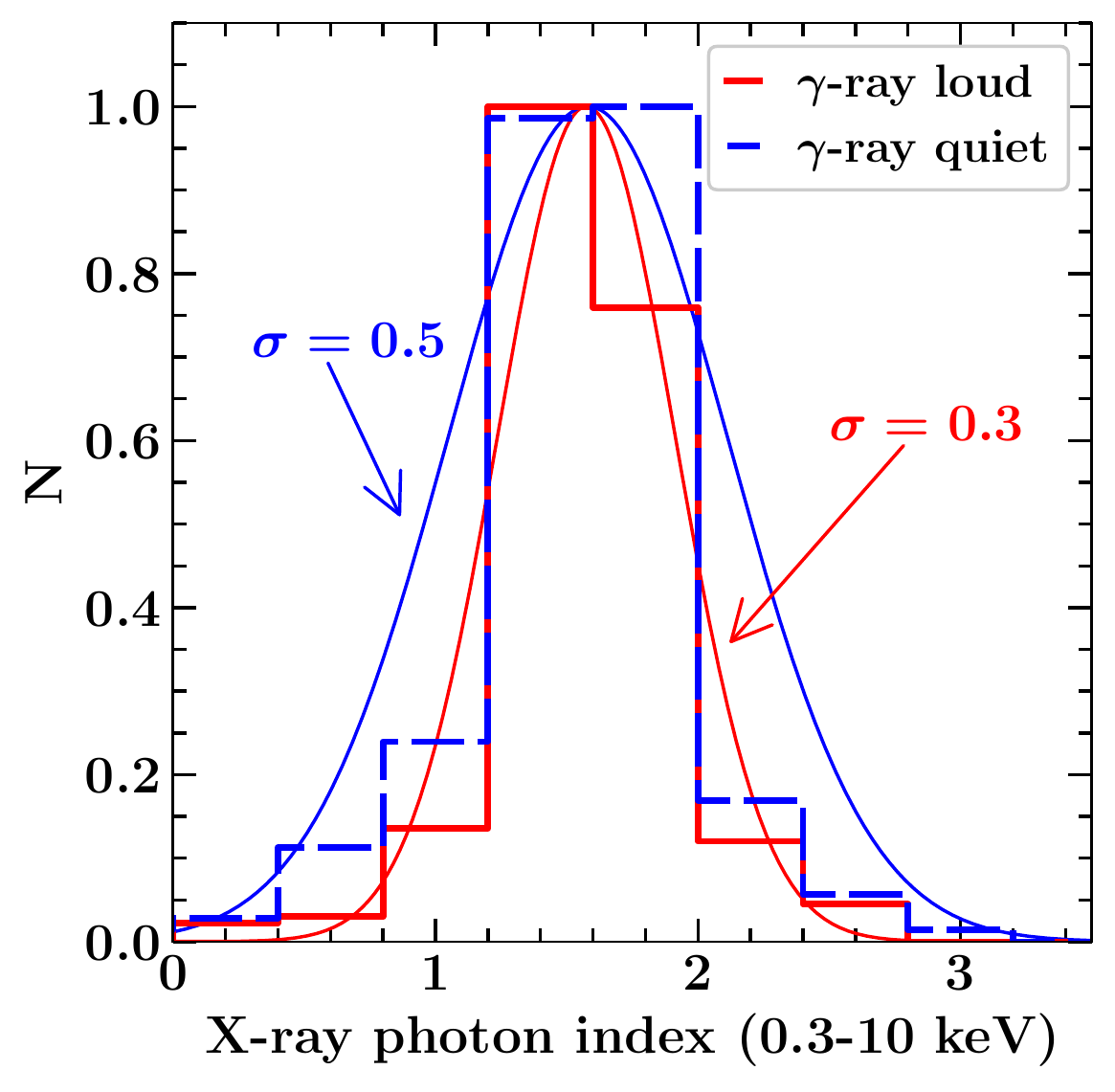}
\caption{The distribution of the observed 0.3$-$10 keV X-ray flux (left) and  photon index (right) for the \gl~and  \gq~blazars. We show the results only for those sources whose X-ray spectra are fitted with an absorbed power-law model, for an equal comparison. See the text for details.}\label{fig:x-ray}
\end{figure*}

Recently, it has been noticed that \fermi-LAT detected blazars occupy a distinct region in the {\it Wide-field Infrared Survey Explorer} \citep[{\it WISE};][]{2010AJ....140.1868W} mid-IR color-color diagram,  named the {\it WISE Gamma-ray Strip} \citep[WGS;][]{2011ApJ...740L..48M}. In Figure \ref{fig:wise}, we show the {\it WISE} 3.4$-$4.6 versus 4.6$-$12 $\mu$m color-color plot for CGRaBS objects and also include MQC sources for a comparison\footnote{We determine the {\it WISE} counterparts within 6$^{\prime\prime}$ of the radio position.}. A majority of the \gl~blazars are found to occupy a narrow region in this diagram (see red points and WGS for FSRQs and BL Lac objects), confirming earlier results. Interestingly, \gq~blazars are also found to be centered in the same region of the diagram, however with a larger scatter. This can be understood in terms of the origin of the IR emission. As also noticed in \citet[][]{2012ApJ...748...68D}, the observed correlation between 3.4$-$4.6 $\mu$m and 4.6$-$12 $\mu$m colors is tighter for the objects in which both IR and \gm-rays originate from the same electron population via synchrotron and SSC/EC mechanisms. On the other hand, the IR emission in FSRQs could be dominated by thermal radiation from the torus and hence may not correlate with the \gm-ray emission. Indeed, out of 191 \gq~blazars 185 are FSRQs, indicating the thermal origin of the IR emission as a possible explanation for the observed scatter.

\subsection{X-ray emission}
There are \swift-XRT observations for a major fraction ($>90\%$) of CGRaBS quasars. For bright sources (e.g., J1256$-$0547 or 3C 279), we also find \xmm~and/or {\it Chandra} data, though we prefer to use \swift-XRT as it allows us to use the simultaneous optical-UV observations from the UVOT. In Table \ref{tab:gamma_XRT} and \ref{tab:non-gamma_XRT}, we present the results of the \swift-XRT data analysis for the \gl~and the \gq~blazars, respectively. Furthermore, for 18 objects (2 \gl~and 16 \gq), we could find only \chandra~or \xmm~observations, which we use to extract the spectral information in the X-ray band. We report the associated fitting parameters in Table \ref{tab:chandra}.

We compare the X-ray properties of the \gl~and the \gq~blazars in Figure \ref{fig:x-ray}. This is done for the absorbed power-law modeled objects, which constitute a major fraction ($>93\%$) of the sample, and including all soft X-ray instruments, i.e., \swift-XRT, \chandra-ACIS, and \xmm~EPIC-PN. For an equal comparison, we extrapolate \chandra~analysis results to 10 keV. In the 0.3$-$10 keV energy range, the \gl~sources ($\langle {F_{\rm 0.3-10~keV}} \rangle=1.4\times10^{-12}$ \ergflux) are brighter than the \gq~objects ($\langle {F_{\rm 0.3-10~keV}} \rangle=6\times10^{-13}$ \ergflux); however, the spectral slopes ($\langle \Gamma_{\rm 0.3-10~keV}\rangle=1.58$ for both) of the distributions are similar (Figure \ref{fig:x-ray}), with a slightly larger scatter for \gq~blazars.

\section{Modeling the Broadband Emission}\label{sec5}
The MW SEDs of blazars are often modeled with a one-zone synchrotron-IC model \citep[e.g.,][]{2009ApJ...692...32D,2009MNRAS.397..985G,2009ApJ...704...38S}. We adopt the leptonic emission model of \citet[][]{2009MNRAS.397..985G} to interpret the broadband SEDs of CGRaBS blazars and briefly describe it here. The emission region is considered to have a spherical shape, located at a distance $R_{\rm diss}$ from the central black hole of mass $M_{\rm BH}$. It moves along the jet axis with a bulk Lorentz factor $\Gamma$. Under the assumption of a conical jet with semi-opening angle $\psi=0.1$ radian, the size of the emission region is constrained from $R_{\rm diss}$. We assume the initial acceleration phase of the jet by considering $\Gamma\sim\sqrt{R_{\rm diss}/3R_{\rm Sch}}$ up to a final value and constant after that \citep[][]{2004ApJ...605..656V,2007MNRAS.380...51K,2009MNRAS.397..985G}, where $R_{\rm Sch}$ is the Schwarzschild radius. 

The emission region is assumed to be filled with an electron population that follows a smooth broken power-law energy distribution of the following type

 \begin{equation}
 S(\gamma)  \, = S_0\, { (\gamma_{\rm b})^{-p} \over
(\gamma/\gamma_{\rm b})^{p} + (\gamma/\gamma_{\rm b})^{q}}.
\end{equation}
where $S_0$ is the normalization constant, $\gamma_b$ represents the peak of the distribution, i.e., the break Lorentz factor, and $p$ and $q$ are the power-law slopes before and after $\gamma_b$. The relativistic electrons emit synchrotron and IC radiation in the presence of a uniform but tangled magnetic field. We consider various sources of low energy seed photons that originate both internally and externally to the jet. This includes IC scattering off the synchrotron photons (SSC) and photons radiated from the accretion disk (EC-disk), the BLR (EC-BLR), and the torus (EC-torus). The accretion disk is assumed as a geometrically thin and optically thick disk \citep[][]{1973A&A....24..337S}. Its emission profile has a multi-temperature blackbody shape of the following type \citep[][]{2002apa..book.....F}

 \begin{equation}
 \label{eq:disk_flux}
 F_\nu = \nu^3\frac{4\pi h \cos \theta_{\rm v} }{c^2 {D_l}^2}\int_{R_{\rm in}}^{R_{\rm out}}\frac{R\,{\rm d}R}{e^{h\nu/kT(R)}-1}
\end{equation}
where $h$ is the Planck constant, $\theta_{\rm v}$ is the angle between the jet axis and the line of sight, $D_l$ is the luminosity distance, $k$ is the Boltzmann constant, $c$ is the speed of light, and $R_{\rm in}$ and $R_{\rm out}$ are the inner and outer disk radii, assumed as 3$R_{\rm Sch}$ and 500$R_{\rm Sch}$, respectively. The radial dependence of the temperature is given as follows

\begin{equation}
\label{eq:temp_profile}
T(R)\, =\, {  3 R_{\rm Sch}  L_{\rm disk }  \over 16 \pi\eta_{\rm acc}\sigma_{\rm SB} R^3 }  
\left[ 1- \left( {3 R_{\rm Sch} \over  R}\right)^{1/2} \right]^{1/4},
\end{equation}
where $\sigma_{\rm SB}$ is the Stefan-Boltzmann constant and $\eta_{\rm acc}$ is the accretion efficiency, adopted here as 10\%.

We assume that certain fractions of the $L_{\rm disk}$ are reprocessed by the BLR ($f_{\rm BLR}=0.1$) and the torus \citep[$f_{\rm torus}=0.5$,][]{2009MNRAS.397..985G}. The radiation from the BLR and the torus are adopted to follow a simple blackbody emission profile peaking at the Hydrogen Lyman-$\alpha$ line frequency and at the characteristic temperature ($T_{\rm torus}$) of the torus, respectively. Both the BLR and the torus are assumed as spherical shells located at distances $R_{\rm BLR}=10^{17}~L_{\rm d, 45}^{1/2}$ cm and $R_{\rm torus}=2.5\times10^{18}~L_{\rm d, 45}^{1/2}$ cm, respectively, from the central black hole, where $L_{\rm d,45}$ is the accretion disk luminosity in units of 10$^{45}$ \lum. We also consider the presence of the X-ray corona lying close to the accretion disk. It reprocesses 30\% of the disk radiation. Its spectral shape is considered as $L_{\rm cor}(\nu)  \propto \nu^{-1}  \exp (- h\nu / 150\, {\rm keV} )$. We calculate the radiative energy densities of these components in the comoving frame following the prescriptions of \citet[][]{2009MNRAS.397..985G} and use them to derive the EC fluxes. 

{\it Jet powers:}
We estimate the jet power carried by electrons ($P_{\rm ele}$), Poynting flux ($P_{\rm mag}$),  radiation ($P_{\rm rad}$), and  protons ($P_{\rm kin}$) as follows \citep[][]{2008MNRAS.385..283C}

 \begin{equation}\label{pjet}
 P_{k}={2\pi}R_{\rm size}^{2}\Gamma^{2}{\beta_c}U'_{k},
\end{equation}
where the factor of 2 accounts for two-sided jets and $R_{\rm size}$ is the size of the emission region. $U'_{k}$ are the comoving frame energy density of the magnetic field ($k={\rm mag}$), relativistic electrons ($k={\rm ele}$), and cold protons ($k={\rm kin}$) and can be estimated using the following equations

\begin{equation}
U'_{\rm mag}=B^2/8\pi,
\end{equation}
 \begin{equation}
U'_{\rm ele}=m_{e}c^2\int{S(\gamma)\gamma}d\gamma,
\end{equation}
\begin{equation}
U'_{\rm kin}= m_{p}c^2\int{S(\gamma)}d\gamma,
\end{equation}

where $B$ is the magnetic field and $m_e$, $m_p$ are the electron and proton masses, respectively. The radiative power is derived as follows \citep[][]{2014Natur.515..376G}

\begin{equation}
P_{\rm rad,~EC} \, = \, 2 \,{4\Gamma^2\over 3\delta^4}~L_{\rm bol} 
\end{equation}

\begin{equation}
P_{\rm rad,~syn/SSC} \, = \, 2 \,{16\Gamma^4\over 5\delta^6}~L_{\rm bol}
\end{equation}
where $L_{\rm bol}$ is the bolometric jet luminosity in the observed frame and the factor of 2 accounts for the two-sided jet. To calculate the kinetic power of the jet, we assume an equal number density of electrons and cold protons \citep[e.g.,][]{2008MNRAS.385..283C}.

{\it Black hole mass and the disk luminosity:}
Two crucial parameters in the blazar SED modeling, particularly in FSRQs, are $M_{\rm BH}$ and $L_{\rm disk}$. In general, these can be constrained either from  single-epoch optical spectroscopy \citep[assuming a virialized BLR, e.g.,][]{2012ApJ...748...49S} or from fitting the big blue bump at optical-UV energies by a standard \citet[][]{1973A&A....24..337S} disk model \citep[][]{2013MNRAS.431..210C,2013A&A...560A..28C}. There are uncertainties associated with both  approaches, e.g., the typical errors in the virial spectroscopic black hole mass calculation is $\sim$0.4 dex \citep[][]{2006ApJ...641..689V,2011ApJS..194...45S} and the disk fitting method does not take into account the jet emission. However, when there is a sufficient coverage at optical-UV energies and the big blue bump is visible, the uncertainty in the disk fitting method is quite small \citep[a factor of $\sim$2,][]{2013MNRAS.431..210C}. Both approaches have been thoroughly discussed in \citet[][]{2015MNRAS.448.1060G} and we refer an interested reader to this article for details.

We adopt the following steps to determine $M_{\rm BH}$ and $L_{\rm disk}$ for CGRaBS blazars.
\begin{enumerate}
\item Whenever we find a big blue bump at IR-UV energies, we reproduce it with a standard accretion disk model and derive both $M_{\rm BH}$ and $L_{\rm disk}$. We are able to apply this method to 178 \gl~and 176 \gq~blazars. They are flagged with the keyword `D' in Table \ref{tab:sed_param_loud} and \ref{tab:sed_param_quiet}.
\item If the bump is not observed or if there are insufficient data points at optical-UV energies, we search in the literature for  availability of the optical spectrum \citep[][]{2011ApJS..194...45S,2012ApJ...748...49S,2012RMxAA..48....9T}. Using the broad emission line (H$_{\beta}$, Mg~{\sc ii}, and C~{\sc iv}) information and the empirical relations of \citet[][]{2011ApJS..194...45S}, we determine the mass of the black hole. From the line luminosities, we compute the BLR luminosity following the scaling relations of \citet[][]{1991ApJ...373..465F} and \citet[][]{1997MNRAS.286..415C}. Under the assumption that the BLR reprocesses 10\% of $L_{\rm disk}$, we derive the luminosity of the accretion disk. When  parameters for more than one line are known, we take the geometric mean of the derived quantities. For 50 \gl~and 5 \gq~sources, we determine their $M_{\rm BH}$ and $L_{\rm disk}$ via optical spectroscopic measurements. In Table \ref{tab:sed_param_loud} and \ref{tab:sed_param_quiet}, they are flagged with the keyword `O'.
\item When the bump is not visible at optical-UV wavelengths and the optical spectral information is also not available, we adopt an empirical relation between the \gm-ray luminosity and the BLR luminosity, as suggested by \citet[][]{2012MNRAS.421.1764S}. It is found that the \gm-ray luminosity is a good tracer of the BLR luminosity and the following empirical relation holds \citep[albeit with a large scatter,][]{2012MNRAS.421.1764S}

\begin{equation}
L_{\rm BLR}\sim4 L_{\gamma}^{0.93}.
\end{equation}
where, $L_{\rm BLR}$ is the BLR luminosity. We then derive $L_{\rm disk}$ from the knowledge of the BLR luminosity. In such sources, we assume a typical $M_{\rm BH}$ of 5 $\times$ 10$^{8}~M_{\odot}$. We compute $L_{\rm disk}$ in 83 \gl~blazars using the empirical method. These objects are flagged with `E' in Table \ref{tab:sed_param_loud}.
\item For \gq~blazars, the previous method could not be used due to lack of \gm-ray information. In such sources, we assume an appropriate value of $L_{\rm disk}$ while keeping in mind not to overproduce the observations \citep[i.e., about a factor $\sim10-20$ below the optical-UV data points, e.g.,][]{2010MNRAS.402..497G} and also to maintain sub-Eddington $L_{\rm disk}$ and adopt $M_{\rm BH}$ = 5 $\times$ 10$^{8}~M_{\odot}$. In 10 \gq~blazars (i.e., $<2\%$ of the sample size), this approach is adopted to get an estimation of $L_{\rm disk}$. These objects can be identified with the flag `A' in Table \ref{tab:sed_param_quiet}.
\end{enumerate}

{\it SED modeling guidelines:}
Our SED modeling code does not perform a statistical fitting. We merely reproduce the observations following a `fit-by-eye' approach. Once we have information about the $L_{\rm disk}$ and $M_{\rm BH}$, as described above, there are eight free parameters in the modeling: $p$, $q$, $B$, $R_{\rm diss}$, $\Gamma$, $S_0$, $\gamma_b$, and $\gamma_{\rm max}$. The size of the emission region is constrained from $R_{\rm diss}$. Along with this, the following parameters are kept fixed: $\psi$, $\theta_{\rm v}$, $\gamma_{\rm min}$, $T_{\rm torus}$, and $f_{\rm BLR/torus/cor}$ (with rare exceptions). Though our modeling program does not compute uncertainties in the physical parameters, depending on the quality of the MW data, the SED parameters can be fairly well constrained from the observations \citep[e.g.,][]{1998ApJ...509..608T}. Below, we briefly elaborate our adopted choices to constrain the free parameters from the observations.

 The leptonic model used here fails to reproduce the low frequency (sub-mm to radio) data due to self-absorption of the synchrotron emission. In the low synchrotron peaked objects, however, radio observations provide a clue about the typical flux level of the synchrotron radiation. For many sources, the synchrotron peak lies in the self-absorbed regime. In these objects, the synchrotron peak is the self-absorption frequency. Moreover, the measurement of $L_{\rm disk}$, either from the disk modeling or from the spectroscopic line measurements, regulates the size of the BLR and the torus, and hence the corresponding radiative energy densities of these components within $R_{\rm BLR}$ and $R_{\rm torus}$, respectively.

\begin{figure}[t!]
\includegraphics[scale=0.7]{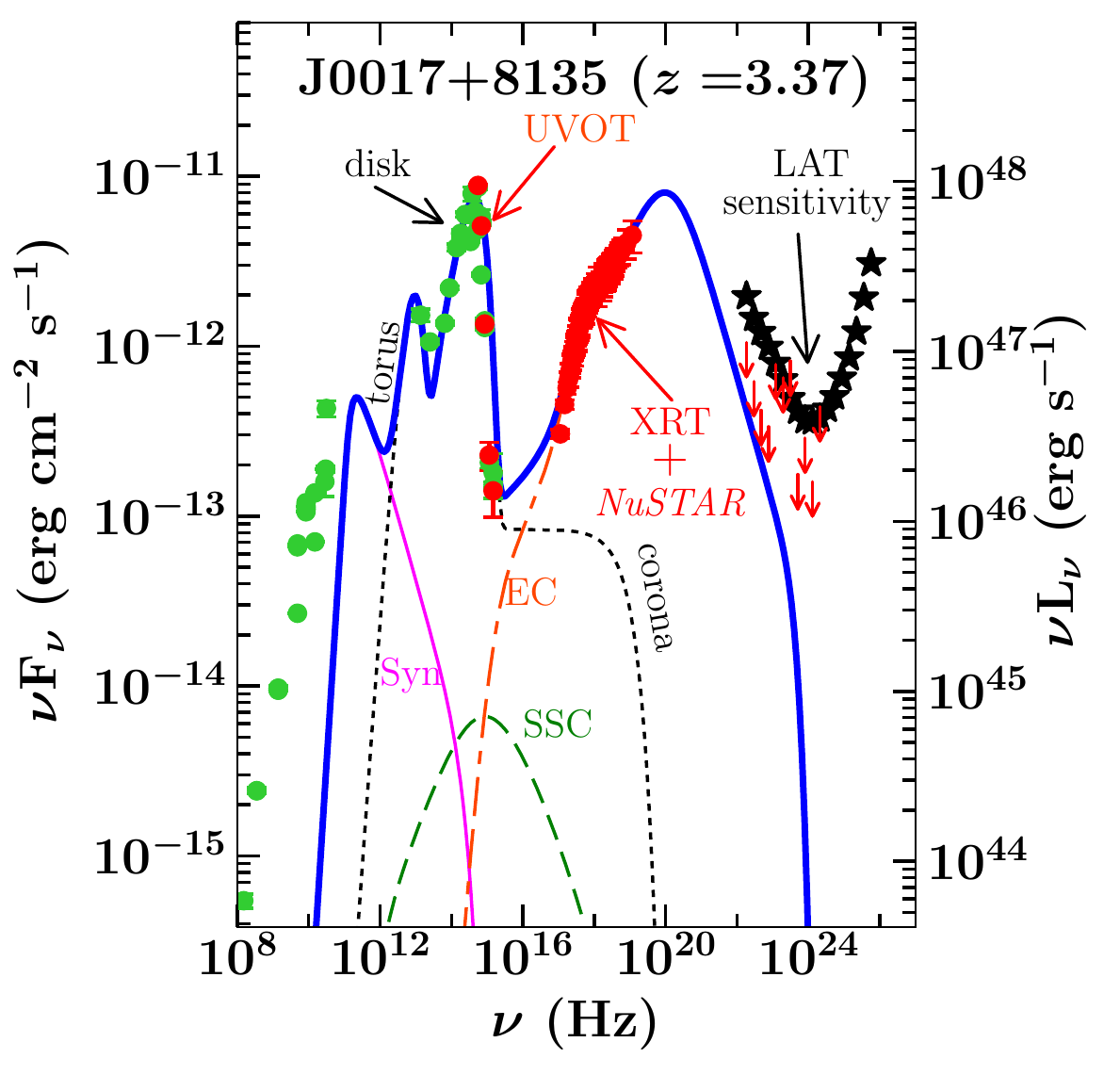}
\caption{The modeled SED of a \gq~blazar. Red filled circles represent the data analyzed by us, while the lime-green filled circles are archival observations. The black dotted line corresponds to the thermal emission from the torus, the accretion disk, and the X-ray corona. The pink thin solid line denotes the synchrotron emission. The green long-dashed and orange dash-dash-dot lines represent the SSC and EC processes, respectively. The blue thick solid line is the sum of all of the radiative components. Black stars represent the 3$\sigma$ \fermi-LAT sensitivity for the duration covered in this work and toward the direction of the source. Red downward arrows correspond to the 2$\sigma$ upper limits. Both sets of information are extracted assuming a photon index of 2.4. The \nustar~data in this source are taken from \citet[][]{2016ApJ...825...74P}. [{\it The modeled SED plots for all other \gm-ray quiet blazars are shown in Figs. xxx$-$xxx in the electronic version.}]\label{fig:SED_quiet}}
\end{figure}

 In powerful \gl~blazars, the shape of the \gm-ray spectrum directly constrains the high-energy slope of the underlying broken power-law electron energy distribution ($q$).  This slope in turn controls the slope of the high-energy tail of the synchrotron component. The lack of  \gm-ray information for the \gq~blazars hampers the determination of $q$. Since a majority of these sources have disk-dominated optical-UV SEDs, it is also not possible to constrain $q$ from the synchrotron process. However, the \fermi-LAT sensitivity limits (shown with black stars in Figure \ref{fig:SED_quiet}) provide us a hint that $q$ has to be steep to avoid the \fermi-LAT detection threshold. In FSRQs, the low-energy slope $p$ can be constrained from the X-ray observations. In high synchrotron peaked (HSP) blazars, though, the X-ray spectrum exhibits a steep falling shape, originating from the high-energy tail of the synchrotron component. In such objects, the \gm-ray spectrum has a rising shape ($\Gamma_{\gamma}<2$), and that slope can be used to get an estimation of $p$.

\begin{figure}[t]
\includegraphics[scale=0.7]{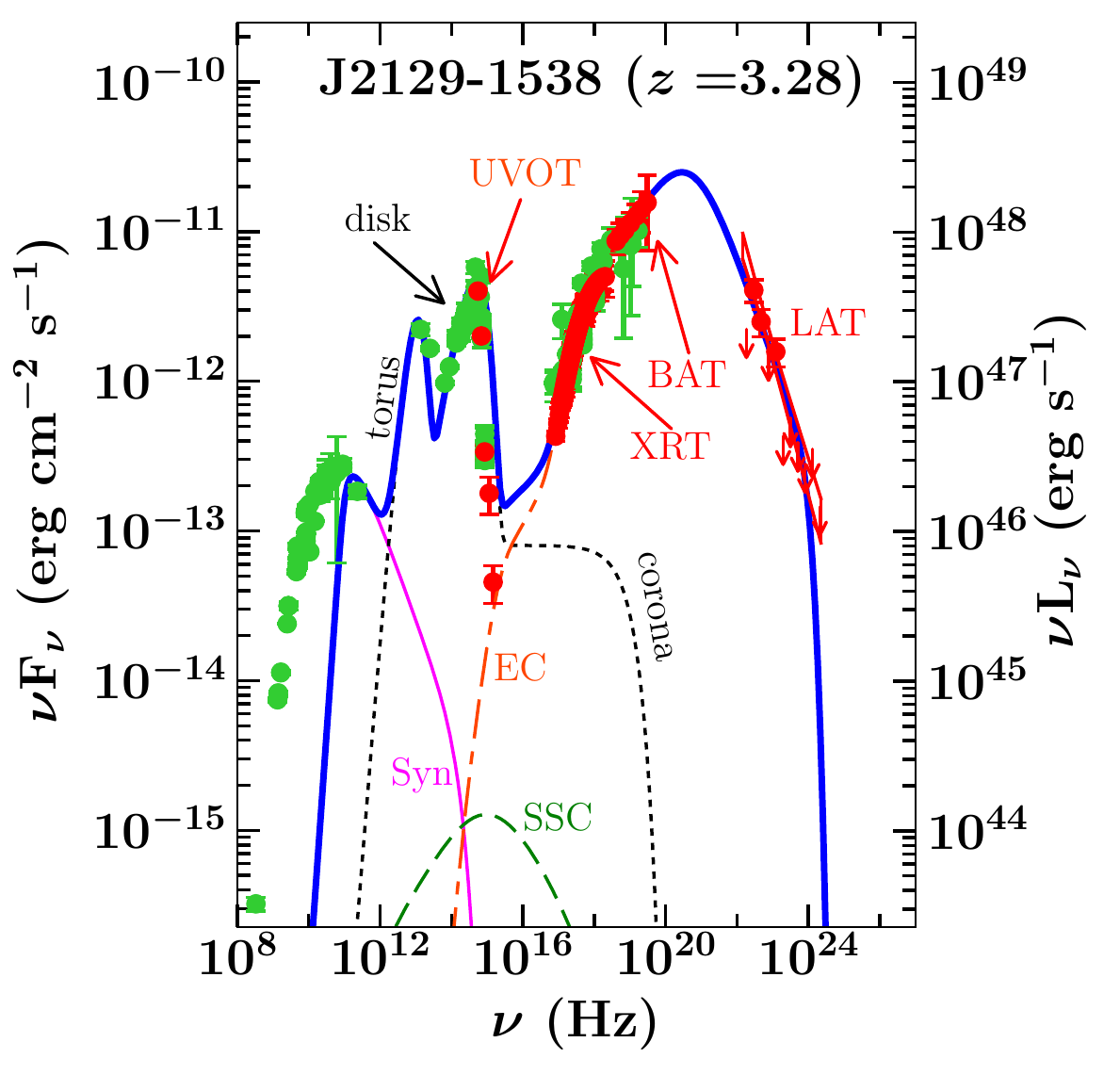}
\caption{The modeled SED of a \gl~blazar. Other information are same as in Figure \ref{fig:SED_quiet}. [{\it The modeled SED plots for all of the other \gm-ray loud blazars are shown in Figs. xxx$-$xxx in the electronic version.}]\label{fig:SED_loud}}
\end{figure}

A rising X-ray spectrum (photon index at 0.3$-$10 keV $\lesssim1.5-1.6$) in FSRQs hints at a prevailing EC mechanism instead of the SSC process. This is because, in these two processes, not only are the seed photons for IC scattering  different but also the participating electrons are of very different energies. In the EC-dominated scenario, the X-ray spectrum  originates from the low-energy electron population, whereas relatively high-energy electrons lying close to the peak of the distribution contribute to the observed soft X-ray spectrum via the SSC mechanism \citep[see, a detailed discussion in][]{2016ApJ...826...76A}. In other words, a flat X-ray spectrum demands the (softer) SSC component to lie below the observations. This regulates the size of the emission region and the magnetic field. An increase in the magnetic field decreases both the SSC and the EC fluxes for a given flux level of the synchrotron emission. This is due to the fact that with increasing magnetic field, fewer electrons are required to produce the same synchrotron flux. Furthermore, we can also fine tune the bulk Lorentz factor $\Gamma$ from the X-ray and \gm-ray SEDs. For larger $\Gamma$, or the Doppler factor $\delta$, fewer electrons are required to make the same flux level of the synchrotron radiation, thus decreasing the synchrotron photon energy density and therefore lowering the SSC flux. However, since the external photon energy densities becomes larger, an increase in $\delta$ leads to the enhancement of the EC flux \citep[][]{2009ApJ...692...32D,2009MNRAS.397..985G}.

\begin{figure*}[t]
\hbox{
\includegraphics[scale=0.7]{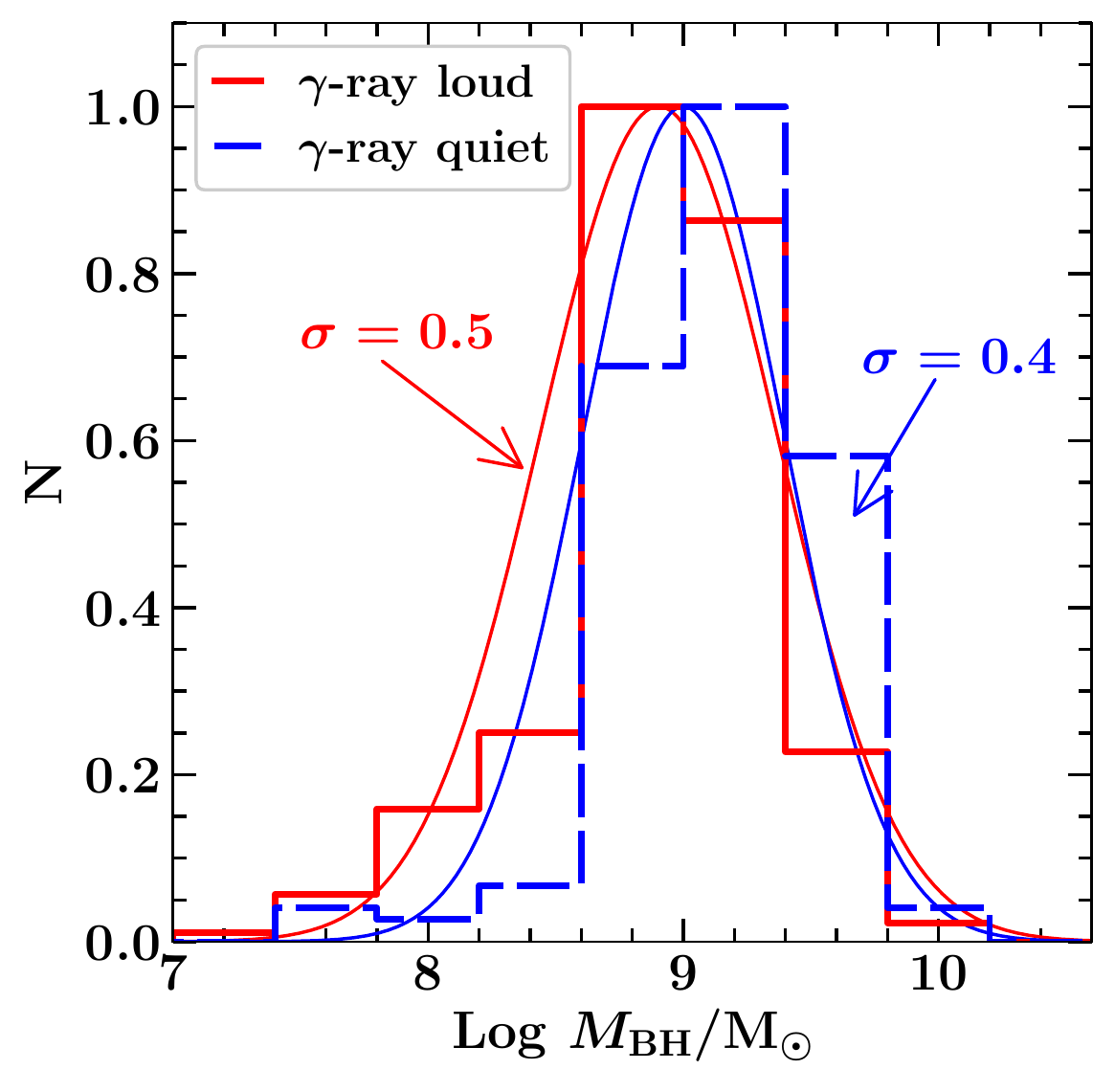}
\includegraphics[scale=0.7]{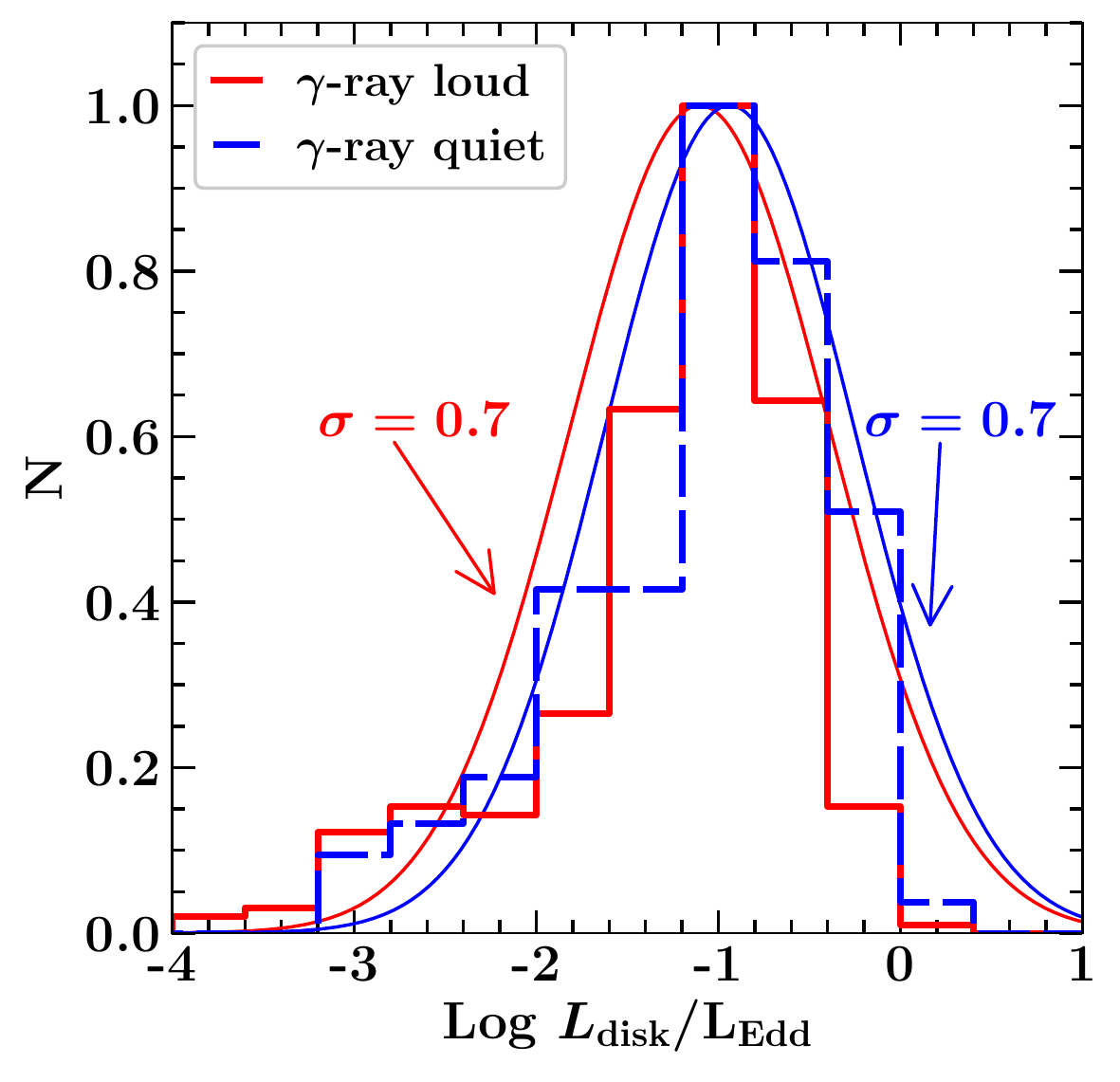}
}
\hbox{
\includegraphics[scale=0.7]{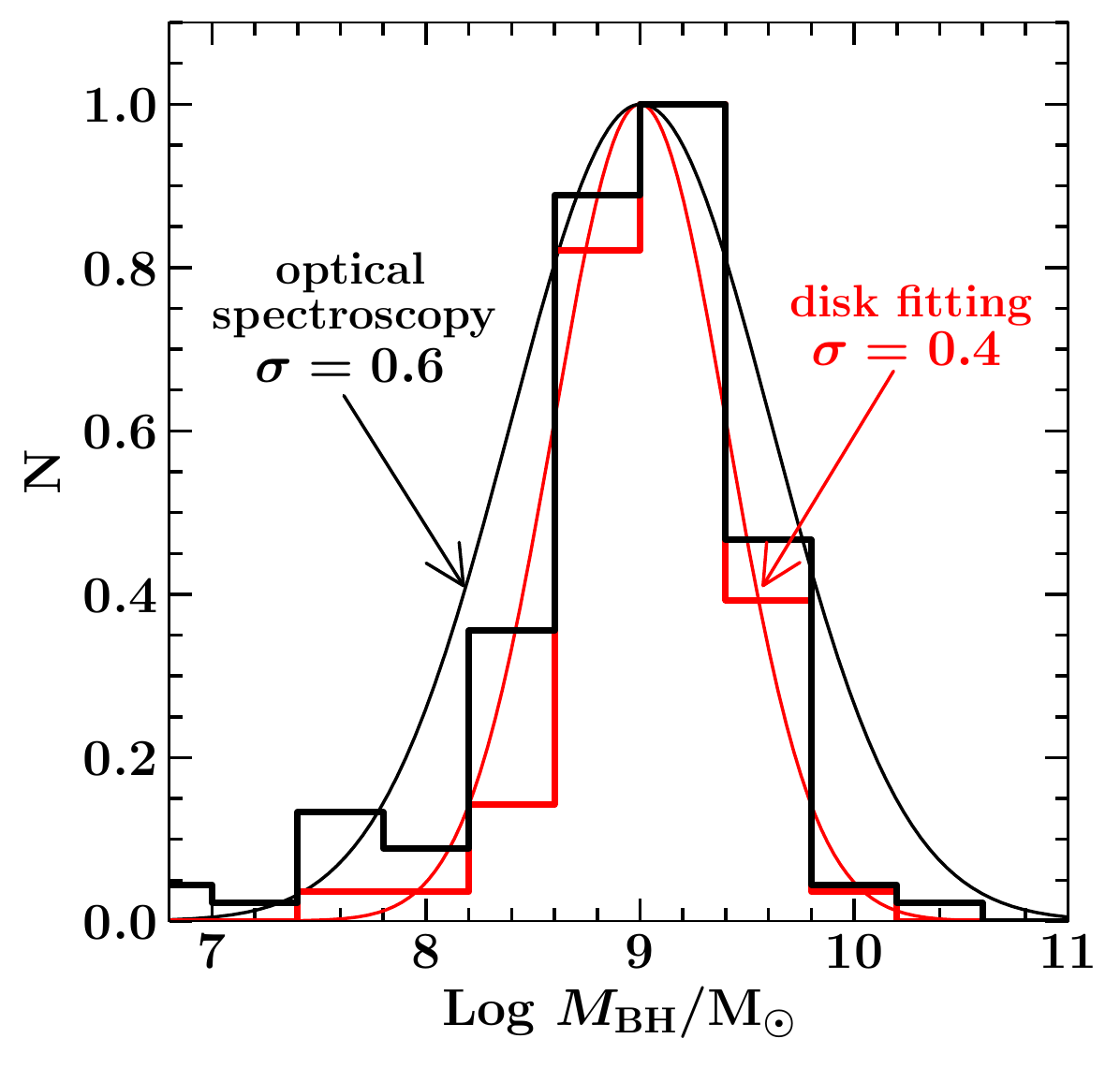}
\includegraphics[scale=0.7]{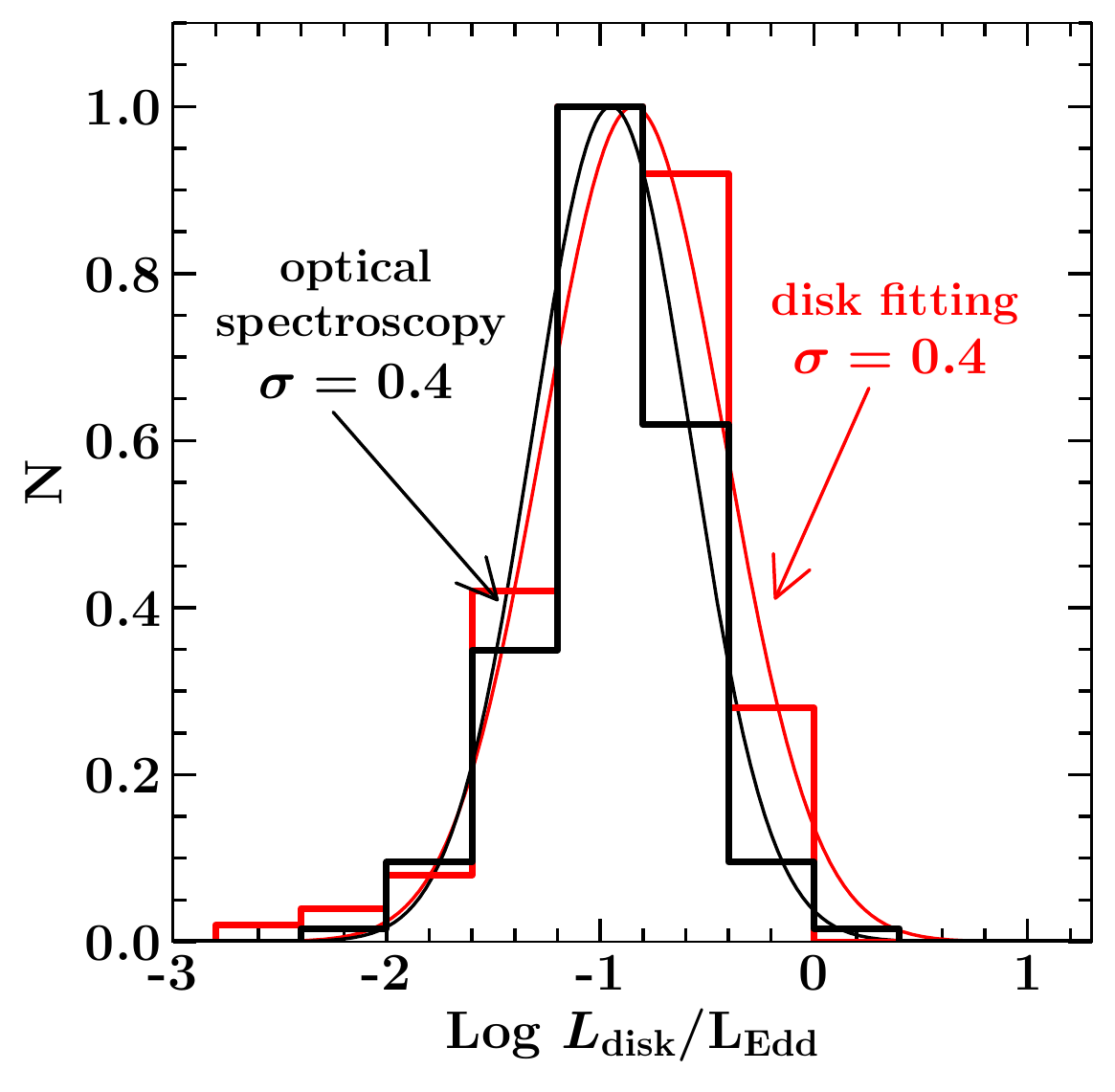}
}
\caption{Top: Distributions of $M_{\rm BH}$ and $L_{\rm disk}$ (in Eddington units) for the \gl~(red solid) and the \gq~(blue dashed) blazars. For an equal comparison the distributions are normalized with the peak set equal to one and we have excluded those sources whose $M_{\rm BH}$ is assumed as 5$\times10^8$ \Msun. Bottom: Comparison of $M_{\rm BH}$ and $L_{\rm disk}$ (in Eddington units) distributions shown only for those sources that have $M_{\rm BH}$ and $L_{\rm disk}$ computed from the optical spectroscopic (black solid) and disk fitting (red solid) methods. The distributions are fitted with  lognormal functions, whose dispersions are indicated.\label{fig:Mbh_Ld_sed_spec}}
\end{figure*}

The Compton dominance (CD), which is the ratio of the high-to-low energy SED peak luminosities, provides us  information about the relative prevalence of the external radiation energy densities compared to the  magnetic energy density. Since in our model both energy densities are a function of $R_{\rm diss}$ \citep[see,][]{2009MNRAS.397..985G,2009ApJ...704...38S}, we can derive the location of the emission region from the observed CD. A large CD ($>1$) indicates the dominance of the external photon field, originating from the BLR/torus, and thus suggests the emission region to lie inside the BLR or outside it but still inside the torus. Another constraint comes from the observation of the fast flux variability from many blazars, indicating a compact emission region, which in turn demands the emission region to lie closer to the central black hole \citep[see also, e.g.,][for alternative arguments]{2012MNRAS.420..604N,2014ApJ...780...87M}. On the other hand, comparatively, the EC-torus-dominated SED has a lower CD than the EC-BLR-dominant SED. The high-energy peak located at  lower frequencies ($\sim$MeV energies) also indicates the emission region to be located outside the BLR but inside the torus since a low IC peak probably hints at a smaller characteristic frequency of the seed photons participating in the EC process. A precise determination of CD in \gq~blazars is rather difficult due to lack of a \gm-ray spectrum. However, along with the shape and the flux level of the soft X-ray spectrum, the \fermi-LAT sensitivity limit constrains it reasonably well. Whenever the hard X-ray observations are available (see, e.g., Figure \ref{fig:SED_quiet}), CD is even further constrained.

\section{Physical Characteristics}\label{sec6}
\begin{figure}[t]
\includegraphics[scale=0.7]{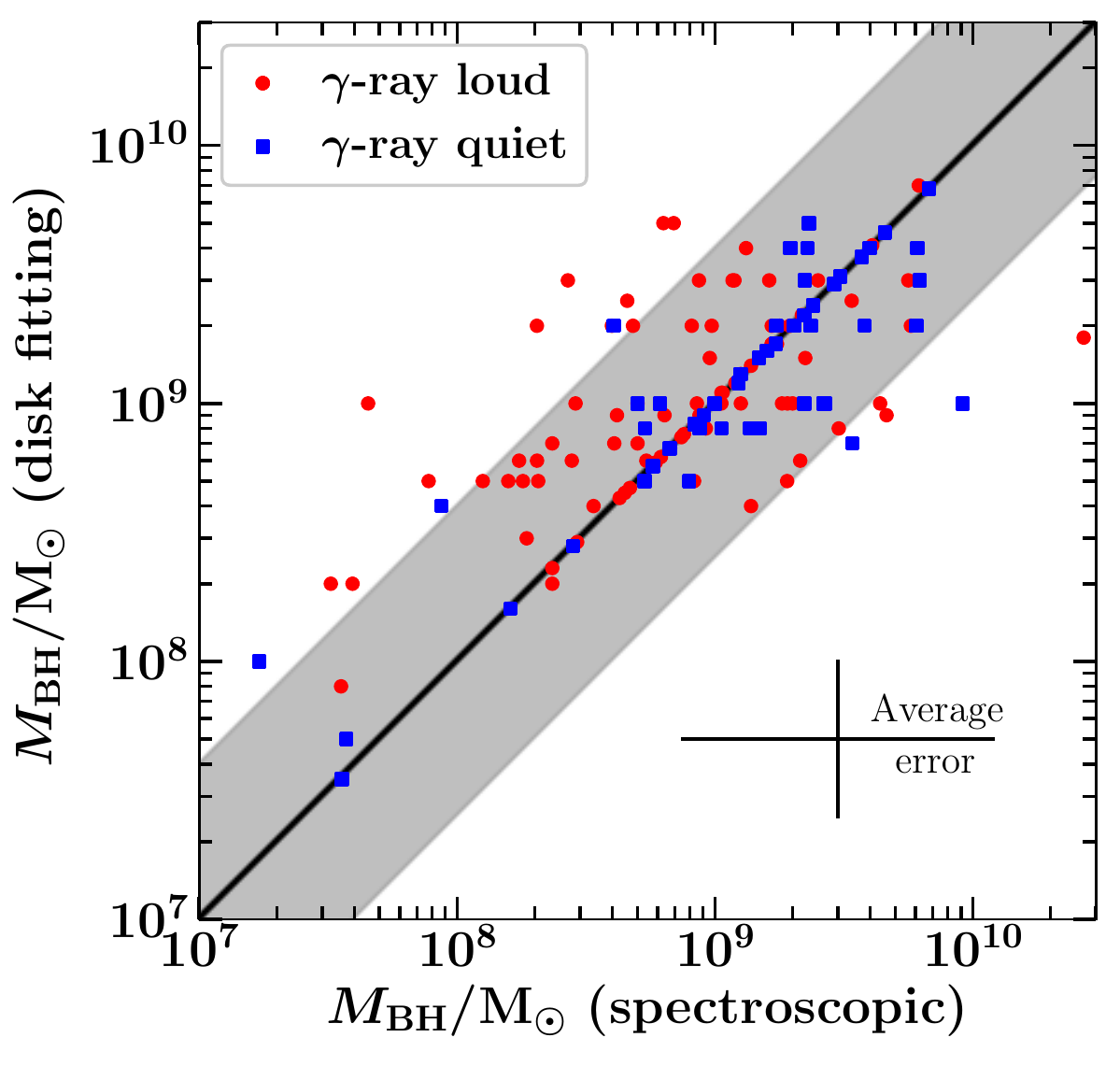}
\caption{The $M_{\rm BH}$ computed from the optical spectroscopic line information versus that estimated using the disk fitting method. The black solid line is the equality line and the gray shaded region corresponds to a factor of 4 uncertainty in the $M_{\rm BH}$ derived from the optical spectroscopic method \citep[e.g.,][]{2006ApJ...641..689V}.\label{fig:mbh1_mbh2}}
\end{figure}

The broadband emissions of both \gq~and \gl~blazars are reproduced using the simple one-zone leptonic emission model described in Section \ref{sec5}. In Figure \ref{fig:SED_quiet} and \ref{fig:SED_loud}, we show the modeled SEDs of these two populations, respectively. The parameters associated with the SED modeling are given in Table \ref{tab:sed_param_loud} and \ref{tab:sed_param_quiet}. We provide the derived jet powers in Table \ref{tab:jet_loud} and \ref{tab:jet_quiet}. The SEDs of 13 \gl~blazars (see Table \ref{tab:13_syn}) are well reproduced using synchrotron and SSC emission mechanisms, i.e., without invoking the EC process. All of them are HSP BL Lac objects with synchrotron-dominated SEDs, and they are characterized by featureless optical spectra \citep[e.g.,][]{1990PASP..102.1120F,1996MNRAS.281..425M}. Since these objects are modeled without needing the information about $L_{\rm disk}$ and $M_{\rm BH}$, we do not consider them when we discuss the physical properties of the \gl~blazars associated with $L_{\rm disk}$ and $M_{\rm BH}$.
 
\subsection{Black Hole Mass and the Accretion Disk Luminosity}
In the top panel of Figure \ref{fig:Mbh_Ld_sed_spec}, we compare $M_{\rm BH}$ and $L_{\rm disk}$ for the \gl~(red) and the \gq~(blue) blazars. For $M_{\rm BH}$ distributions, we have considered only those objects whose black hole masses are measured either from the optical spectrum or from the disk fitting method, as explained in the previous section. The fitted distribution peaks at $\langle {M_{\rm BH}} \rangle \approx 1\times10^9~M_{\odot}$ and $8\times10^8~M_{\odot}$, for the \gq~and the \gl~objects, respectively. The fact that the \gq~blazars host slightly more massive ($\sim2.5\sigma$ significance) black holes compared to \gl~sources can be explained by noting that they are located at larger redshifts (Figure \ref{fig:redshift}),  possibly a selection effect of detecting only the heaviest black holes at  high redshifts. On the other hand, the accretion disk luminosities in the \gl~and the \gq~blazars show a similar distribution and peak around 10\% of the Eddington luminosity ($L_{\rm Edd}$). This implies that, in absolute units (i.e., in \lum), \gq~objects host more luminous accretion disks compared to \gl~blazars.

In our sample, there are a total of 138 blazars (86 \gl~and 52 \gq) that have $M_{\rm BH}$ and $L_{\rm disk}$ measurements from both the disk fitting and  optical spectroscopy methods. Therefore, it is interesting to check the consistency of the results derived from these two different approaches. In the bottom panel of Figure \ref{fig:Mbh_Ld_sed_spec}, we plot $M_{\rm BH}$ and $L_{\rm disk}$/$L_{\rm Edd}$ distributions estimated from the optical spectroscopic (black) and the disk modeling (red) techniques. The $M_{\rm BH}$ computed from the disk fitting matches well with the virial measurements ($\langle {M_{\rm BH} \rangle=1\times 10^9}~M_{\odot}$ for both). Fitting the distribution with a log normal function returns a dispersion of 0.4 dex for the disk fitting method and 0.6 dex for the virial masses. Similarly, the distributions of the accretion disk luminosities derived from the disk modeling and the optical emission line approaches show a good agreement. We also show the $M_{\rm BH}$ estimated from the disk fitting method as a function of the $M_{\rm BH}$ derived from the virial approach, for both \gl~and \gq~blazar populations, in Figure \ref{fig:mbh1_mbh2}. Though there is a large dispersion, overall both results match reasonably. These observations support the argument that the disk modeling could be a robust tool to calculate $M_{\rm BH}$ and $L_{\rm disk}$ in powerful FSRQs.

\subsection{The Particle Distribution}
\begin{figure*}[t]
\hbox{
\includegraphics[scale=0.5]{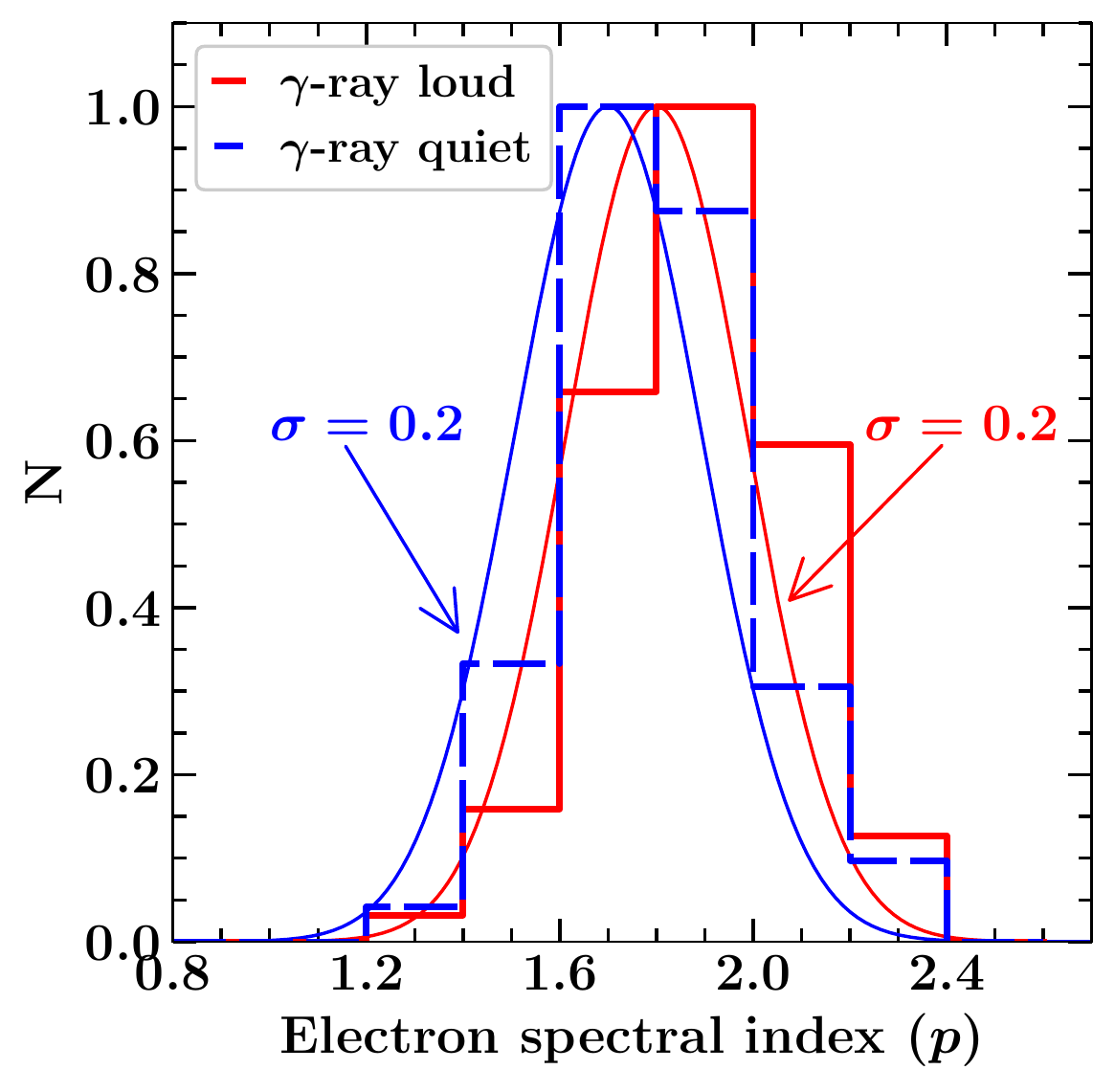}
\includegraphics[scale=0.5]{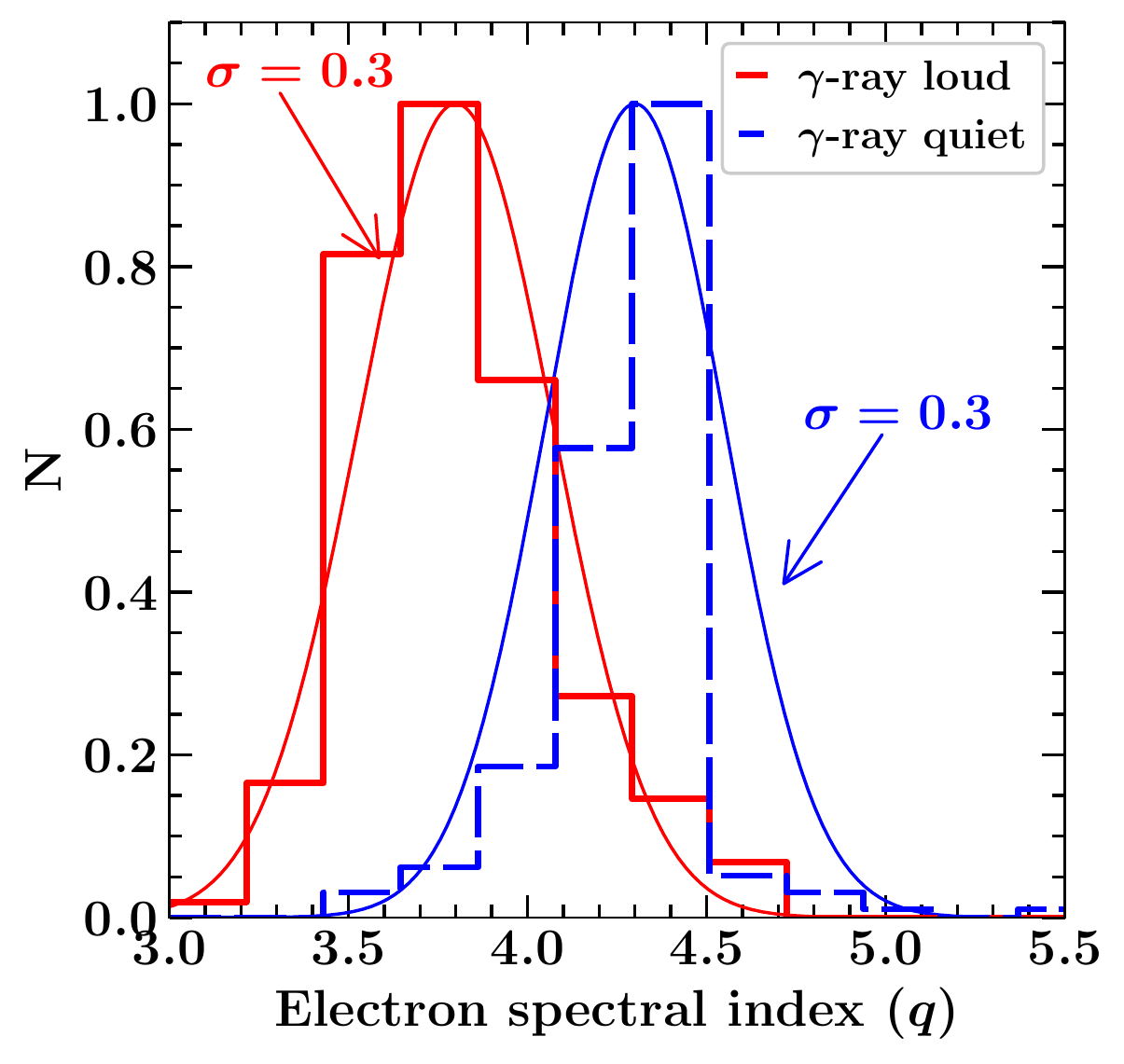}
\includegraphics[scale=0.5]{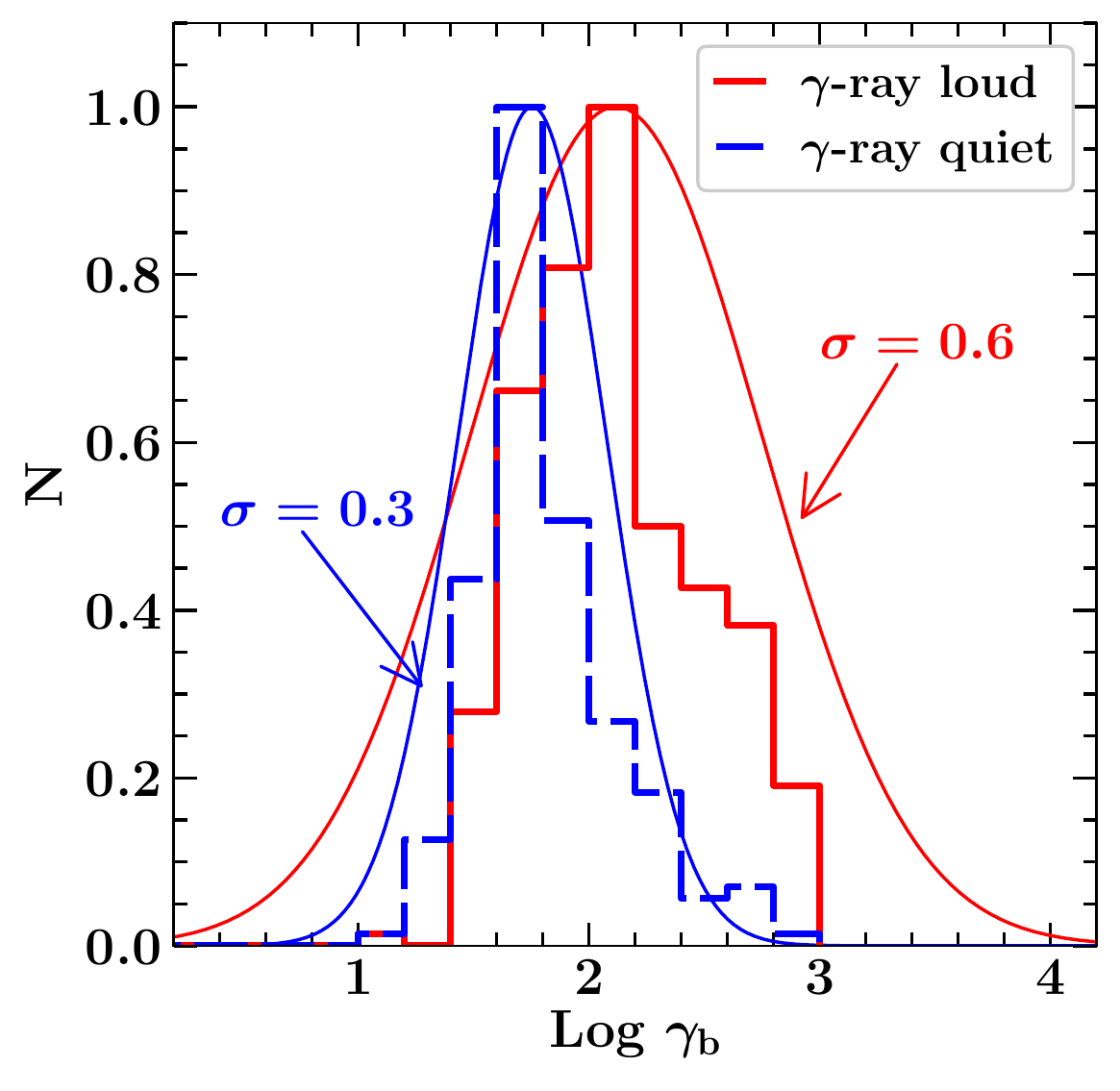}
}
\hbox{
\includegraphics[scale=0.5]{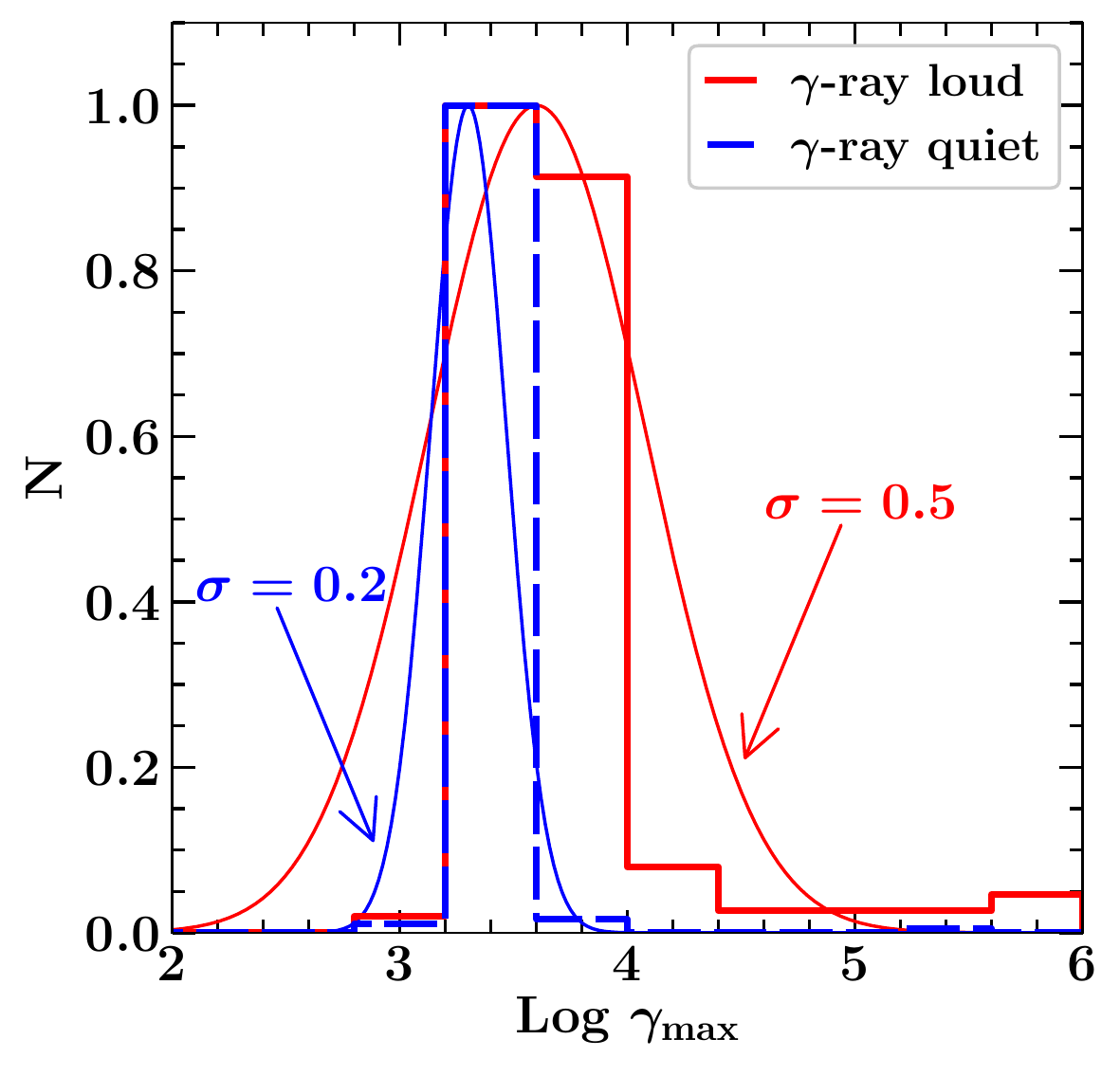}
\includegraphics[scale=0.5]{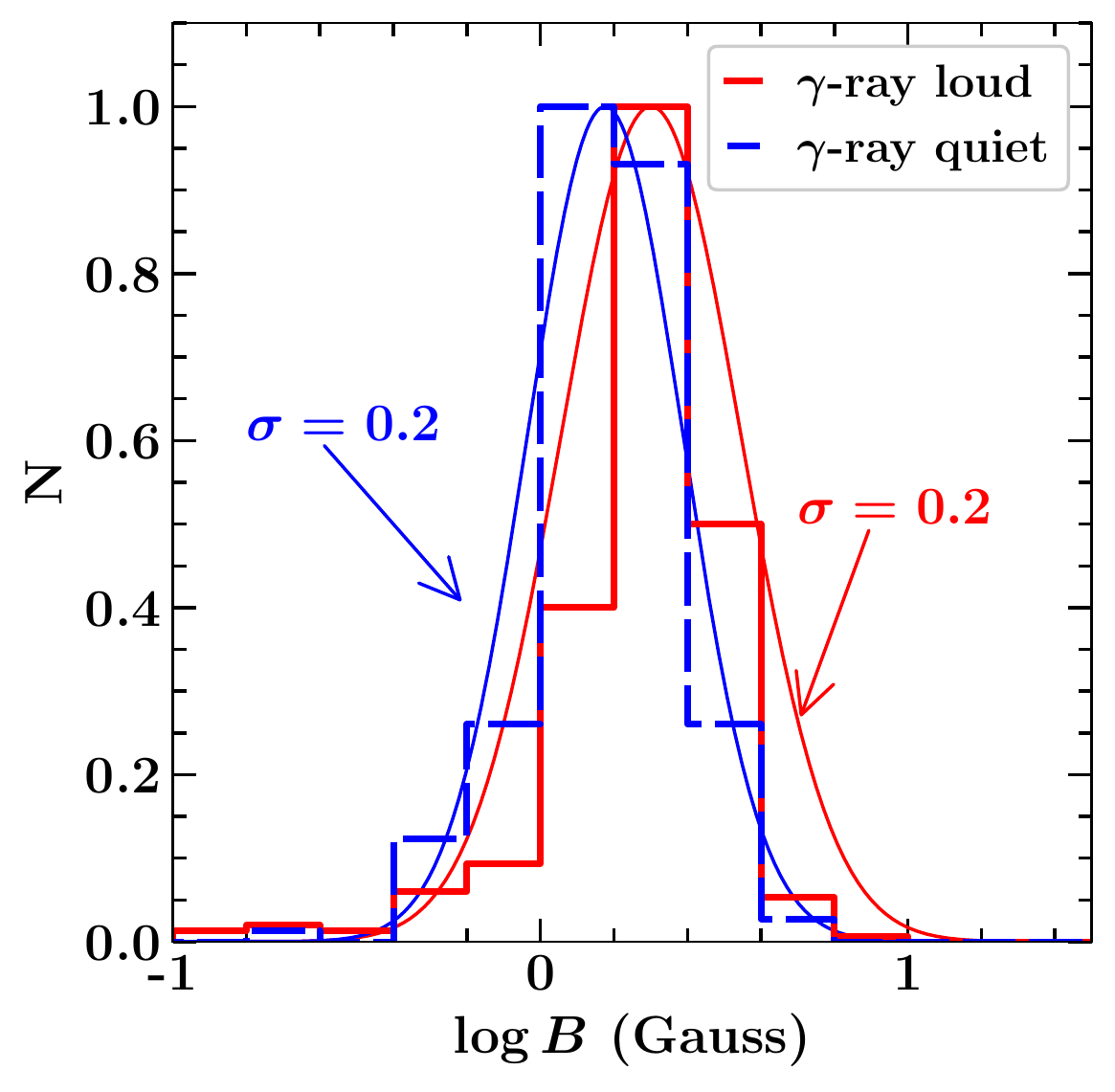}
\includegraphics[scale=0.5]{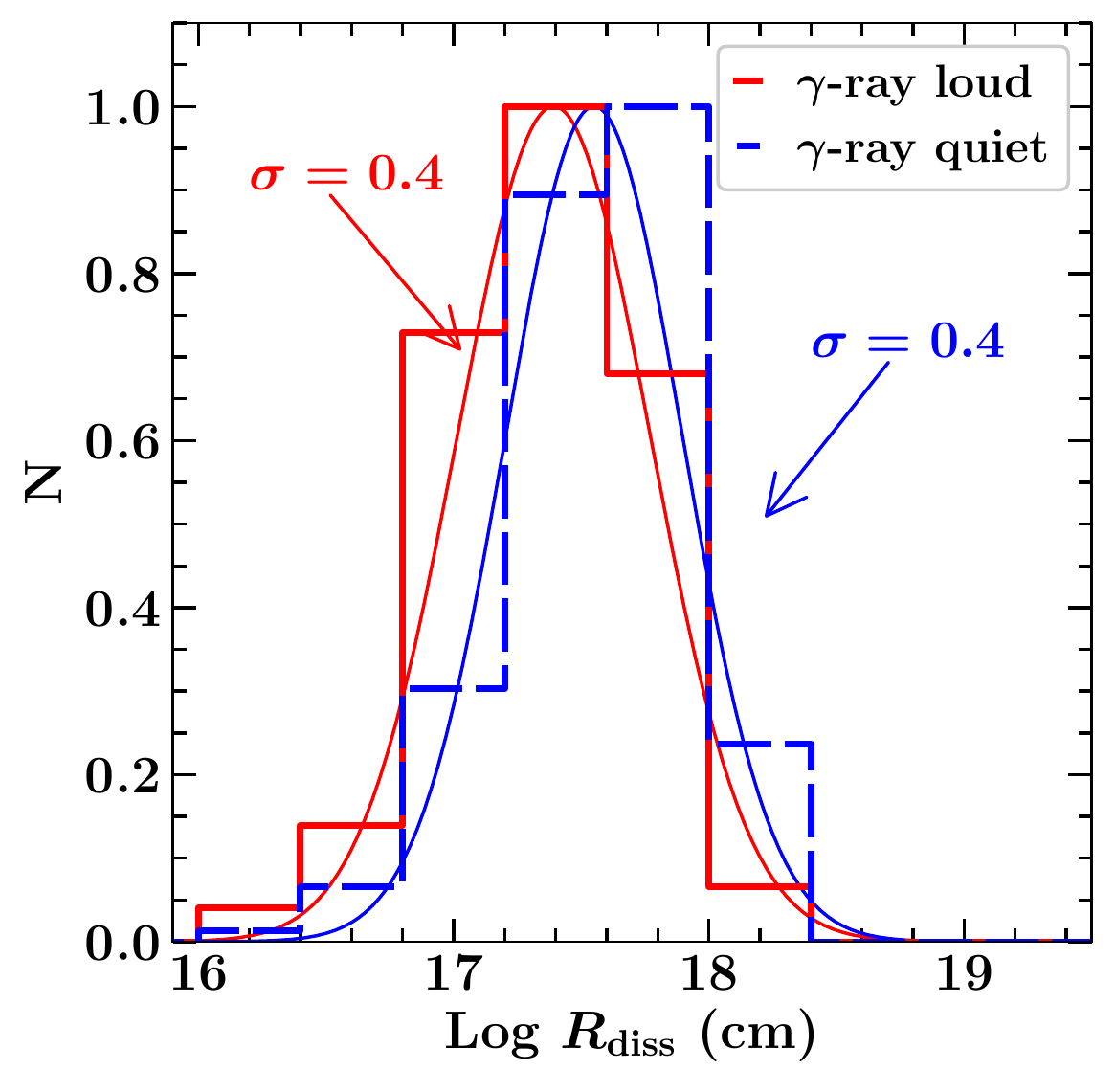}
}
\hbox{
\includegraphics[scale=0.5]{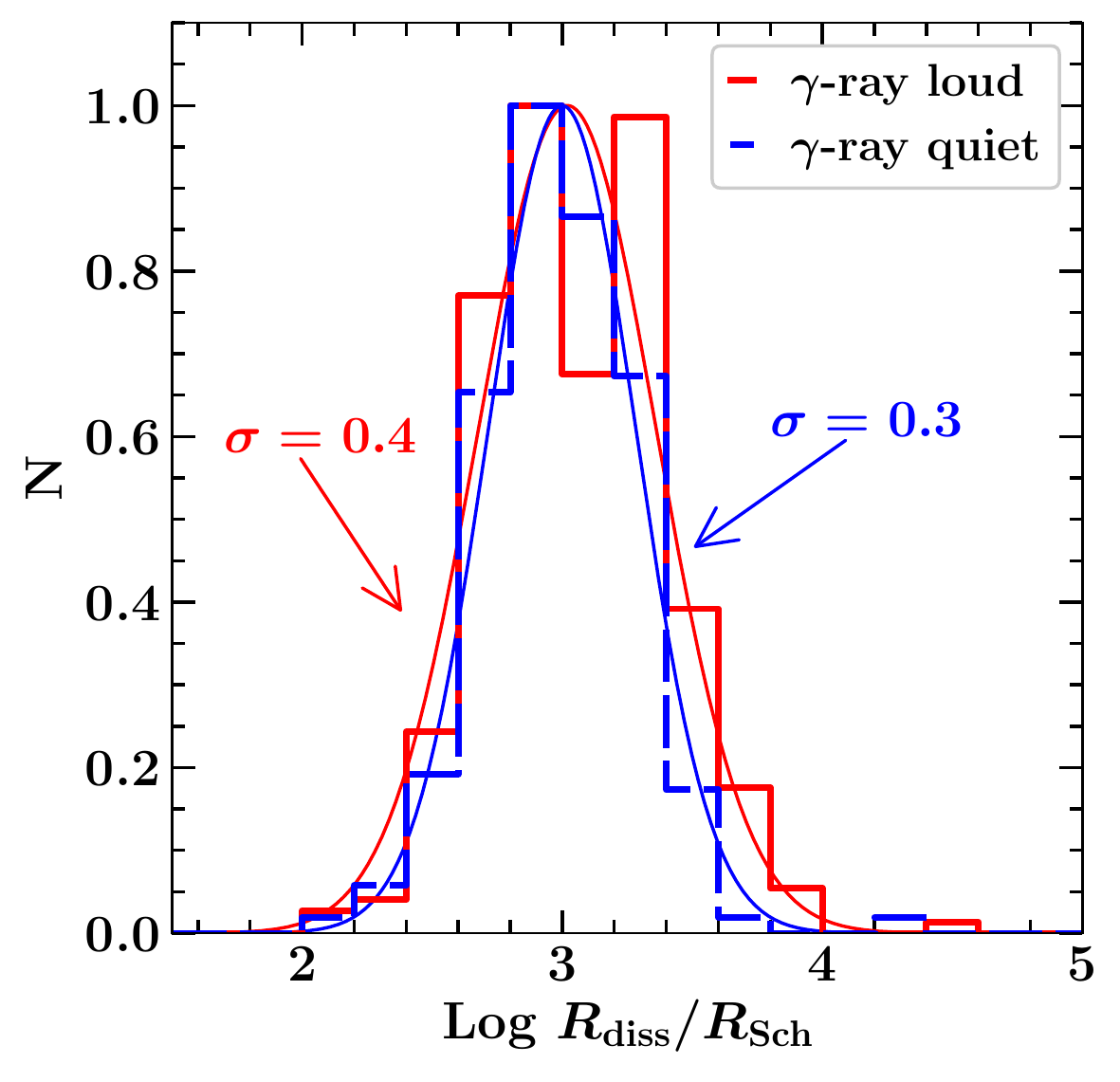}
\includegraphics[scale=0.5]{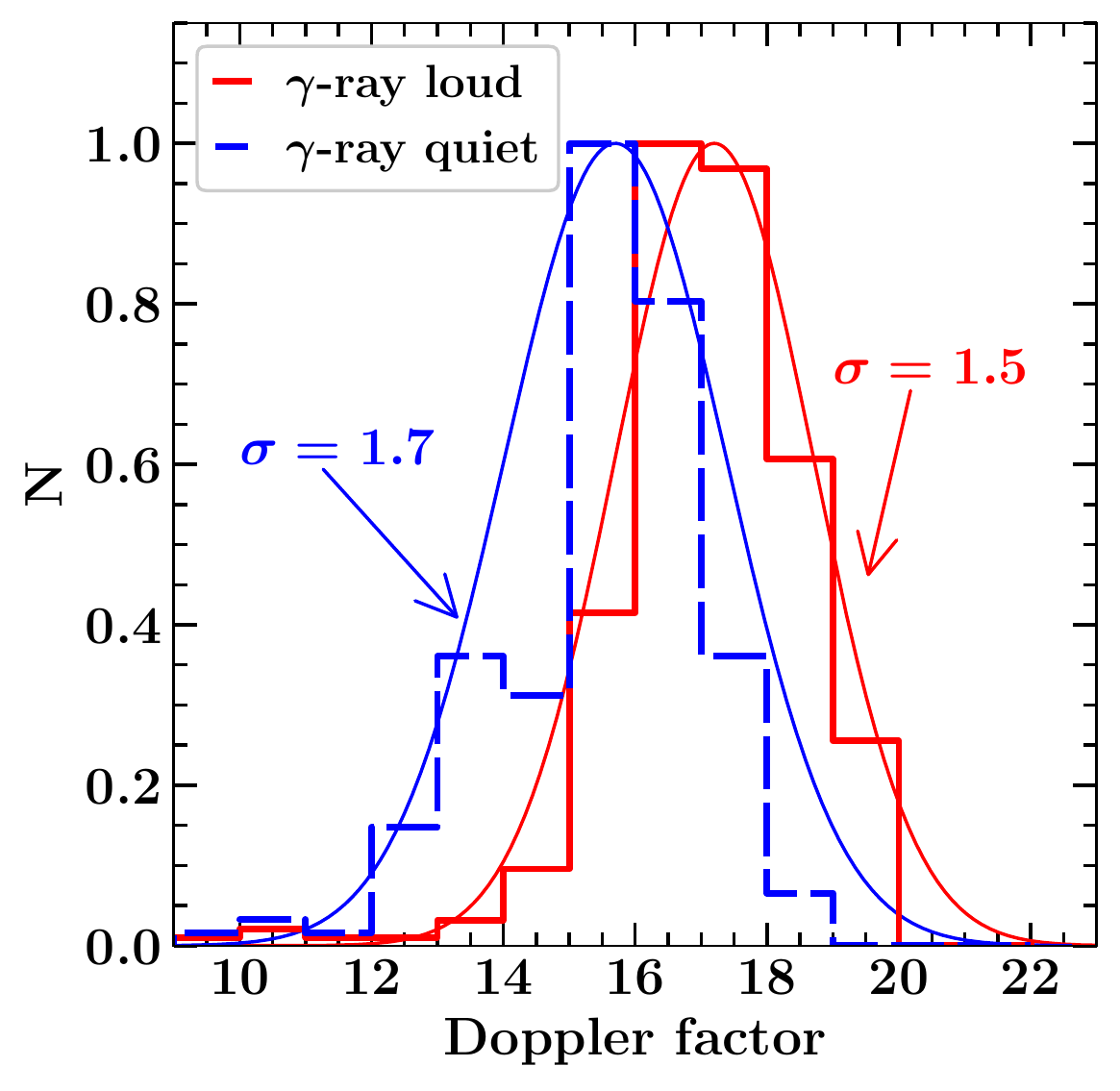}
\includegraphics[scale=0.5]{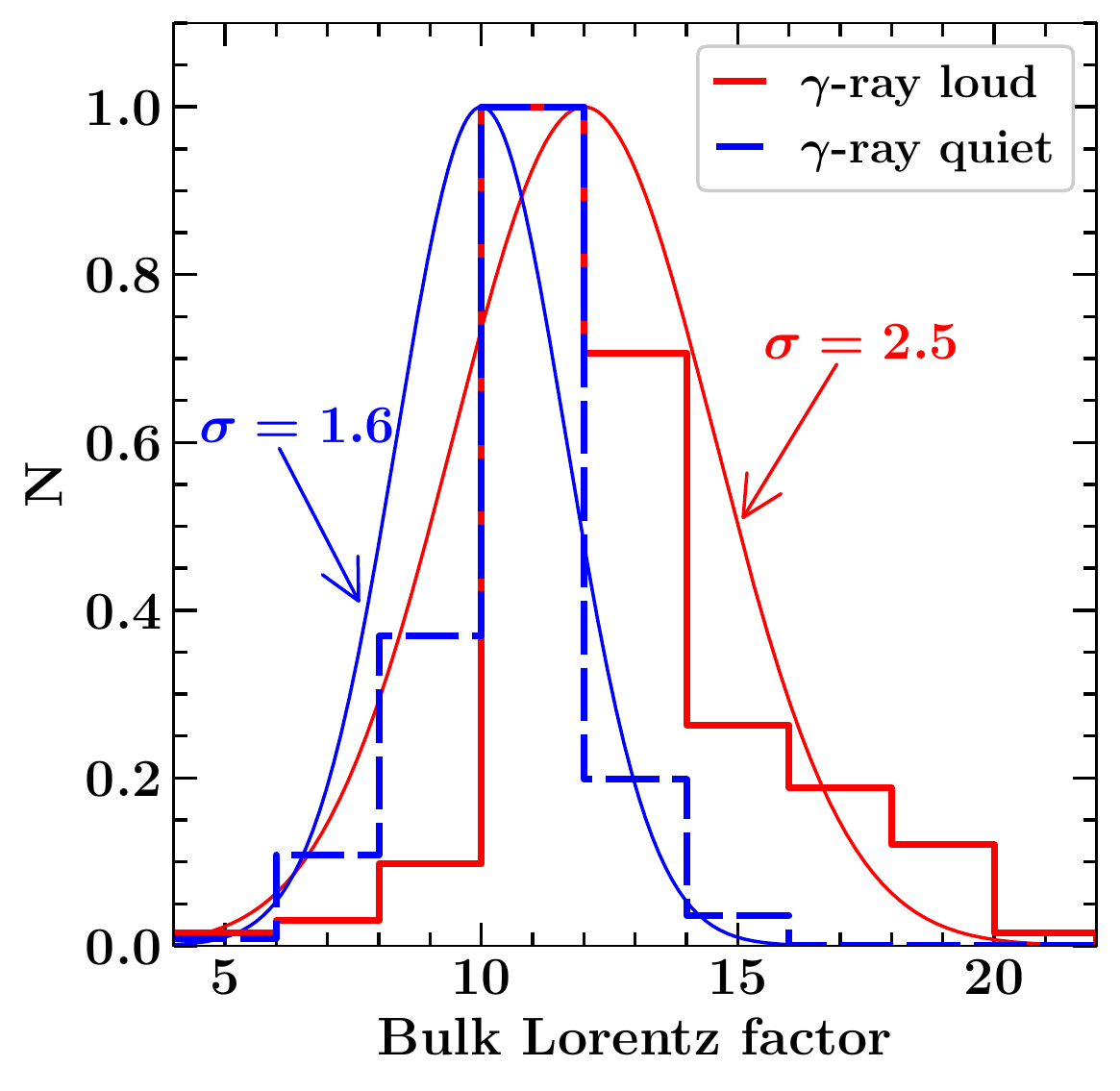}
}
\caption{Distributions of the particle spectral indices before ($p$, top left) and after ($q$, top middle) the break energy ($\gamma_{\rm b}$, top right), the maximum Lorentz factor ($\gamma_{\rm max}$, central left), the magnetic field (central middle), distance of the emission region from the central black hole in cm (central right) and in $R_{\rm Sch}$ (bottom left), the Doppler factor (bottom middle) and the bulk Lorentz factor (bottom right).\label{fig:sed_param}}
\end{figure*}

The average properties of the electron energy distribution (EED) can be seen in the top left and middle panels of Figure \ref{fig:sed_param}. For both \gl~and \gq~blazars, the low- and high-energy spectral indices of the EED ($\langle p \rangle=1.8,~1.7$ and $\langle q \rangle=3.8,~4.3$, respectively) have a narrow distribution. In powerful FSRQs, the synchrotron peak is typically located at radio to sub-mm frequencies and due to synchrotron self-absorption, it is impossible to characterize the spectral shape of the underlying EED from the synchrotron emission alone. In such blazars, the EED can be constrained using X-ray and \gm-ray data, which (in the EC scenario) constrain the spectral shapes of the low-energy and high-energy electron population, respectively. The availability of the \swift-BAT or the \nustar~data for a few sources also helps in putting further constraints on the low-energy index $p$. Note that $p$ regulates the total number of radiating electrons that are present in the emission region, and therefore the total number of protons, which controls the kinetic power of the jet. On the other hand, the distribution of the high energy index $q$ shows a clear bimodality, with the high-energy electron population of the \gl~blazars exhibiting a relatively hard spectrum compared to the \gq~objects. This can be understood in terms of the non-detection of the \gq~sources in the \gm-ray band. In such sources, a steep falling EC spectrum is necessary to avoid the detection limit of the \fermi-LAT (see, e.g., Figure \ref{fig:SED_quiet}). The shape of the optical-UV spectrum can also provide a good constraint to $q$ if it is synchrotron-dominated. However, out of 191 \gq~blazars we find an accretion-disk-dominated optical-UV SED in 176 sources, implying $q$ is rather unconstrained from the synchrotron emission.

We show the distribution of the break Lorentz factor ($\gamma_{\rm b}$) and the maximum Lorentz factor ($\gamma_{\rm max}$) of the EED in the top right and middle left panels of Figure \ref{fig:sed_param}. In both plots, the distribution of the \gq~blazars is relatively low peaked ($\langle \gamma_{\rm b}\rangle \sim56$ and $\langle \gamma_{\rm max}\rangle \sim2000$) and has narrower dispersion, when fitted with a log normal function. The wider distributions for the \gl~blazars are probably due to the presence of a few HSP BL Lac objects such as J1653+3945 (or Mrk 501). These sources have their synchrotron peak located at very high frequencies ($>10^{16}$ Hz), thus implying the peak of the underlying EED, i.e., $\gamma_{\rm b}$, to have a large value. Furthermore, $\gamma_{\rm max}$ is rather insensitive to the modeling for the steeply falling \gm-ray spectrum FSRQs. However, a hard rising \gm-ray spectrum in HSP sources indicates a large $\gamma_{\rm max}$. This is the reason that the distribution of $\gamma_{\rm max}$ for the \gl~population extends to very large values and accordingly is centered at a larger $\gamma_{\rm max}$ of $\sim4000$ (Figure \ref{fig:sed_param}).

\subsection{Magnetic Field and the Location of the Emission Region}
The distribution of the magnetic field for the \gl~blazars peaks around $\langle B \rangle=2.0$ Gauss, whereas, for \gq~sources it is centered at a low value of 1.5 Gauss (see middle plot in the middle panel of Figure \ref{fig:sed_param}). Fitting them with a log normal function returns a narrow width of 0.2 dex for both distributions. 

The logarithmic distribution of the location of the emission region in absolute units (cm) peaks at $\langle R_{\rm diss} \rangle=2.5\times10^{17}$ cm and $3.5\times10^{17}$ cm for the \gl~and the \gq~blazars, respectively (middle right panel of Figure \ref{fig:sed_param}). Fitting the distributions with a log normal function gives an equal dispersion of 0.4 dex for both of them. On the other hand, considering the dissipation distance in units of $R_{\rm Sch}$, we find no difference in the distributions of \gl~and \gq~sources. On average, the emission region is found to be located inside the BLR for a majority of sources.

\subsection{Doppler and the Bulk Lorentz Factors}
The distributions of $\delta$ and $\Gamma$ are shown in the bottom middle and right panels of Figure \ref{fig:sed_param}. On average, the \gl~blazars exhibit a larger $\delta$ and $\Gamma$ ($\langle \delta \rangle=17.2,~\langle \Gamma \rangle=12$) compared to the \gq~blazars ($\langle \delta \rangle=15.7,~\langle \Gamma \rangle=10$). Recently, based on the radio study of 1.5 Jy MOJAVE AGNs, \citet[][]{2015ApJ...810L...9L} have reported that the \gm-ray detected blazars have significantly larger Doppler boosting factors compared to sources not detected by \fermi-LAT \citep[see also,][and references therein, for earlier results]{2010A&A...512A..24S}. The similarity of the results derived from two different approaches, from the SED modeling and the 15 GHz radio measurements, is rather striking. We emphasize here that the lower $\delta$ of the \gq~blazars may not be due to a large $\theta_{\rm v}$ or misalignment of the jet \citep[see also,][]{2011ApJ...740...98M}. It is primarily due to small $\Gamma$ and can be explained as follows. In a majority of the FSRQs studied here, the X-ray to \gm-ray SEDs are explained by the EC process (e.g., Figure \ref{fig:SED_quiet} and \ref{fig:SED_loud}). The EC mechanism has a stronger dependence on $\theta_{\rm v}$, and due to anisotropy of the external radiation field in the comoving frame of the emission region, it has an additional boosting \citep[][]{1995ApJ...446L..63D}. A hard-rising EC emission at X-rays indicates a small $\theta_{\rm v}$ and vice-versa. In other words, the X-ray spectrum in EC-dominated blazars can be used as a tool to determine $\theta_{\rm v}$ \citep[e.g.,][]{2013ApJ...777..147S}. We find the X-ray spectral shapes of both the \gl~and the \gq~blazars to be similar (Figure \ref{fig:x-ray}), thus indicating a similar $\theta_{\rm v}$, which is further confirmed from the SED modeling (Table \ref{tab:sed_param_loud} and \ref{tab:sed_param_quiet}).

\subsection{Jet powers}
\begin{figure}[t]
\includegraphics[scale=0.75]{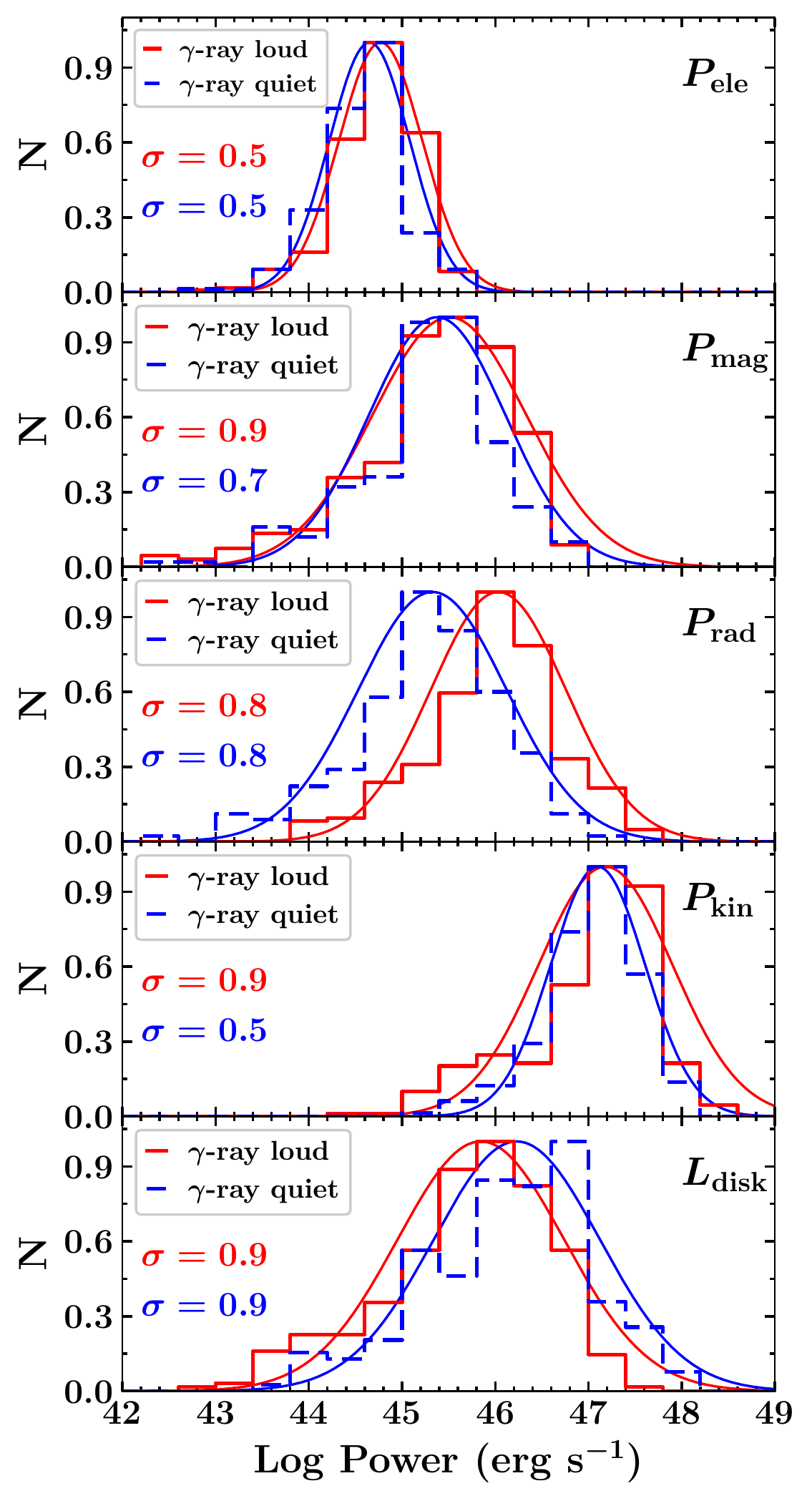}
\caption{Distributions of various jet powers considering two-sided jets. In the bottom panel, we also show the distribution of $L_{\rm disk}$ for a comparison with the jet power. The histograms are fitted with a log normal function and the corresponding widths are quoted. The average values of the distributions, for the \gl~and the \gq~blazars, respectively, are as follows: $\langle \log~P_{\rm ele}\rangle=44.76, 44.65$, $\langle \log~P_{\rm mag}\rangle=45.48, 45.38$, $\langle \log~P_{\rm rad}\rangle=45.99, 45.33$,  $\langle \log~P_{\rm kin}\rangle=47.15, 47.10$, and $\langle \log~L_{\rm disk}\rangle=45.85, 46.23$. All units are in \lum.\label{fig:jet_power}}
\end{figure}
\begin{figure}[t]
\includegraphics[scale=0.72]{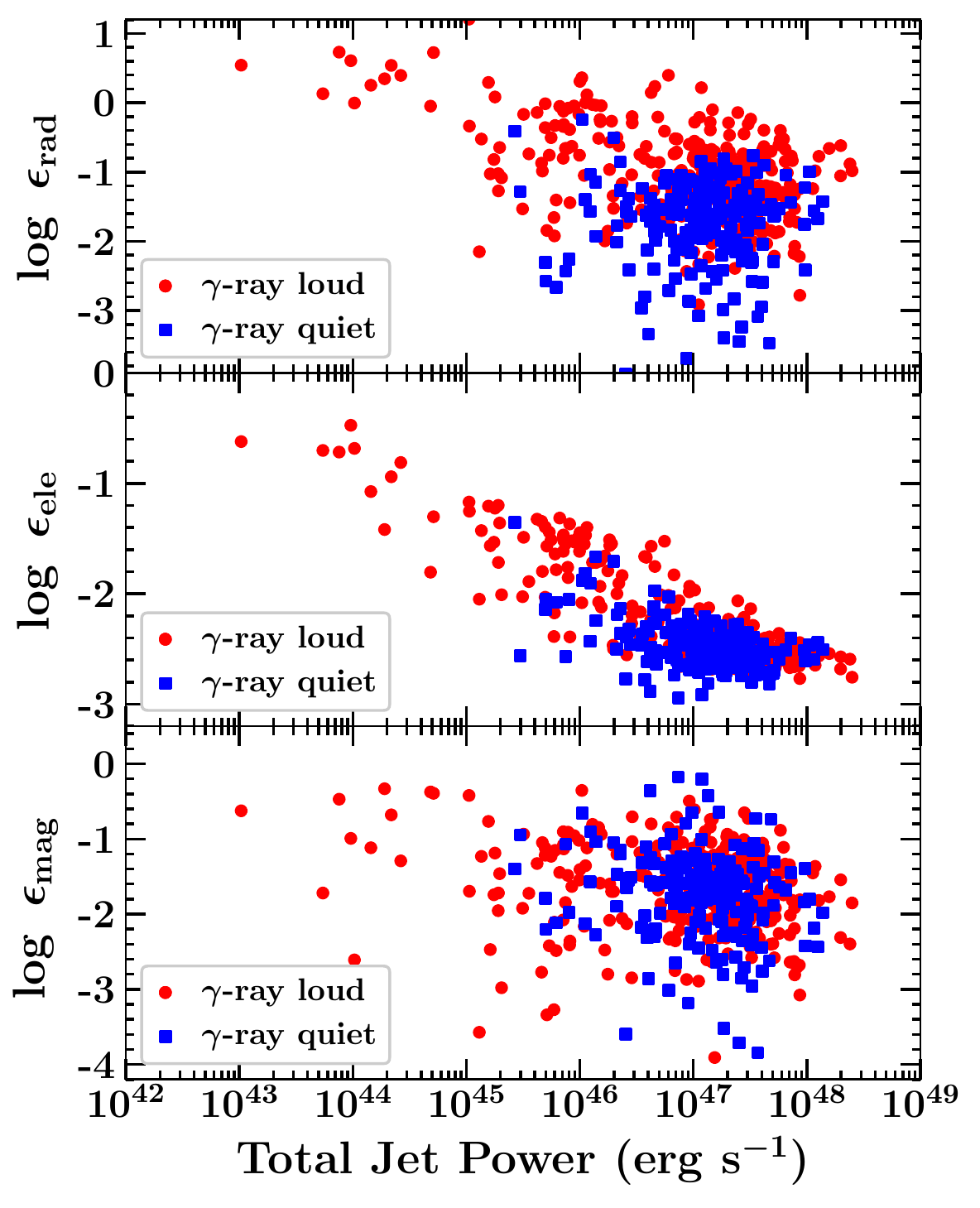}
\caption{The fraction of the total jet power transformed into radiation ($\epsilon_{\rm rad}$, top), relativistic electrons ($\epsilon_{\rm ele}$, middle), and Poynting flux ($\epsilon_{\rm mag}$, bottom).\label{fig:jet_effi}}
\end{figure}

Figure \ref{fig:jet_power} represents the distributions of various jet powers. As can be seen, the \gl~and the \gq~blazars share a similar distribution of the power carried by relativistic electrons, Poynting flux, and cold protons. However, the \gl~sources exhibit a larger $P_{\rm rad}$ compared to the \gq~objects. This is probably due to \gm-ray emitting jets being more radiatively efficient. Also, $P_{\rm ele}$ is substantially less than $P_{\rm rad}$. This is the direct consequence of the rapid cooling of electrons in a time substantially smaller than the light crossing time of the emission region, so that more power is released in the form of radiation than remains in electrons. The fact that $P_{\rm mag}$ is slightly lower than $P_{\rm rad}$ indicates that the magnetic field alone is not responsible for the observed radiation. This leaves us to consider $P_{\rm kin}$ as a plausible reservoir for $P_{\rm rad}$. In fact, most of the jet power remains in the form of dynamically dominant protons which  produce the large scale radio structures. Only a small fraction ($\sim1-10\%$, see also the top panel of Figure \ref{fig:jet_effi}) of it gets converted to radiation. We have assumed an equal number density of electrons and cold protons. This assumption is crucial and often leads to the total jet power in powerful FSRQs to exceed $L_{\rm Edd}$ \citep[e.g.,][]{2014Natur.515..376G}. Considering the presence of the electron-positron pairs would reduce the budget of the jet power \citep[see, e.g.,][]{2016ApJ...831..142M,2017MNRAS.465.3506P}; however, their number cannot be large \citep[$\lesssim$10-15 pairs per proton, e.g.,][]{2000ApJ...534..109S,2008MNRAS.385..283C} to avoid the Compton rocket effect \citep[][]{1981ApJ...243L.147O}. If present in a substantial fraction, these pairs would produce a large amount of  soft X-radiation (so-called the ``Sikora bump"), which is yet to be observed. Alternatively, instead of a uniform one-zone emission, if one considers the broadband emission to originate from a spine-sheath structured jet, the total power of the jet will come down \citep[][]{2016MNRAS.457.1352S}. Moreover, as can be seen in Figure \ref{fig:jet_power}, $P_{\rm mag}$ is tiny compared to $P_{\rm kin}$ and hints at a weak magnetization of the emission region. These observations, therefore, argue against the Poynting-flux-dominated scenario (e.g., \citealt{2012MNRAS.423.3083M}, but see, \citealt{2014ApJ...796L...5N,2015MNRAS.449..431J,2015MNRAS.451..927Z}).

In Figure \ref{fig:jet_effi}, we show the fraction of the total jet power ($P_{\rm jet} = P_{\rm ele} + P_{\rm mag} + P_{\rm kin}$) converted to radiation ($\epsilon_{\rm rad}$), carried by relativistic electrons ($\epsilon_{\rm ele}$), and magnetic field ($\epsilon_{\rm mag}$). It can be seen in the top panel of this plot that there are a few low-power blazars that have $\epsilon_{\rm rad}\gtrsim1$. These are primarily BL Lac objects that are known to exhibit $P_{\rm rad}\sim P_{\rm kin}$. In these sources, almost all of the available jet power is used to produce the radiation \citep[see,][for a relevant discussion]{2010MNRAS.402..497G}.

\section{Discussion}\label{sec7}

By applying a simple leptonic emission model, we are able to study a few fundamental physical properties of blazars, which are briefly discussed in the following sub-sections. In order to determine the
 strength of the correlations, we compute the Spearman's rank-correlation coefficient ($\rho_{\rm s}$) and the probability of no correlation, PNC. Since we do not have error estimation for the derived 
 SED parameters, we quantify the strength of the correlation by performing a Monte Carlo simulation following a bootstrapping approach that takes into account the dispersion of the plotted quantities. 
 We create $10^4$ data sets, each consisting of N data pairs $x_i$ and $y_i$ (N is the sample size) and compute $\rho_{\rm s}$ and PNC for each data set. A pair ($x_ i$, $y_i$) is randomly
 chosen from the original data set such that some of the original pairs may appear more than once in a given data set or not at all. Assuming that the returned values follow a Gaussian distribution, we
 estimate the correlation coefficient and the 1$\sigma$ uncertainty by deriving the average and the standard deviation of the calculated values. Quoted PNC values are the average of the calculated PNC in each
 simulation. Note that it can be argued that the observed correlations could be due to intrinsic correlations of the input SED parameters. However, in our work, the chosen modeling parameters are 
 independent and they are constrained only by the observations, thus supporting the connection of the observed correlations with the physical behavior of the sources. In various correlation plots, the reported average errors are 1$\sigma$ standard deviation of the plotted parameters for the whole population, i.e., including both \gl~and \gq~blazars, unless specified.

\subsection{Accretion-Jet Connection in Blazars}
\begin{figure}[t]
\includegraphics[scale=0.7]{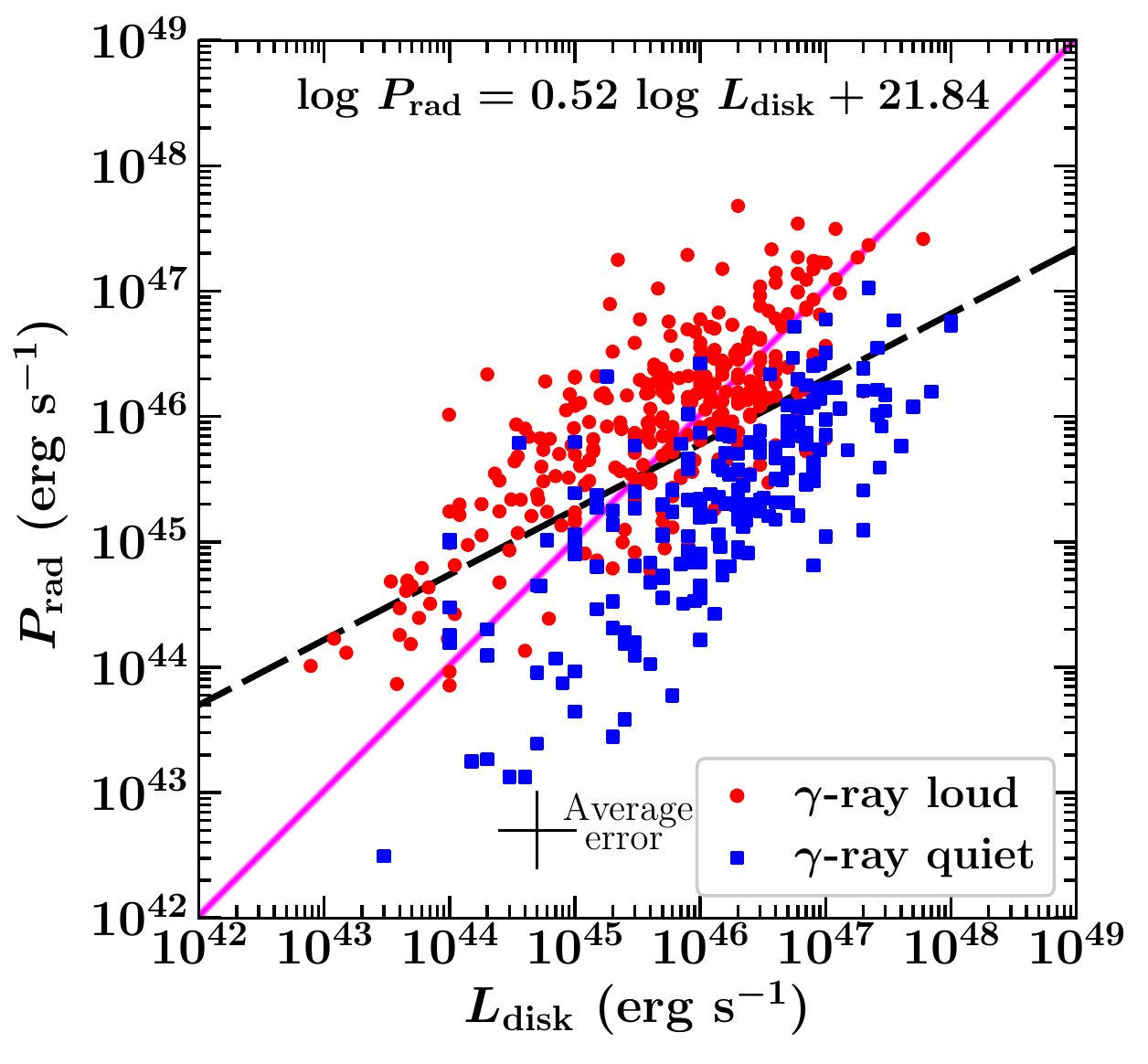}
\includegraphics[scale=0.7]{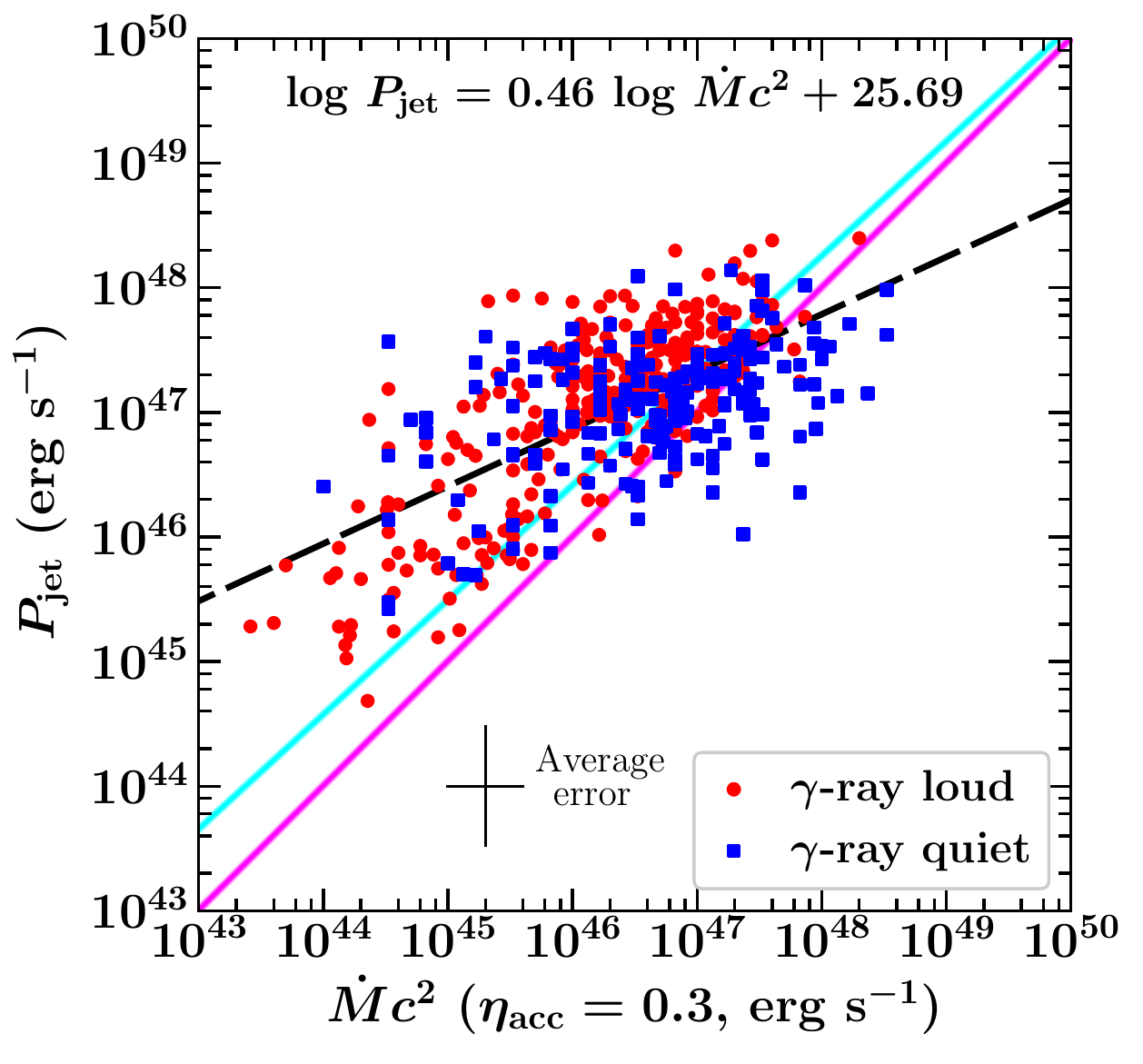}
\caption{The disk-jet connection in blazars. Top: The plot of $P_{\rm rad}$ versus the $L_{\rm disk}$ for the \gl~(red) and the \gq~(blue) blazars. Bottom: The jet power versus the total accretion power, assuming an accretion efficiency $\eta_{\rm acc}=0.3$. In both plots, the pink solid line represents the one-to-one correlation and the black dashed line denotes the best fit. In the bottom panel, the cyan solid line shows the correlation reported by \citet[][]{2014Natur.515..376G}. The uncertainties  in both $P_{\rm rad}$ (corresponding to the uncertainty in $\Gamma^2$) and $L_{\rm disk}$ are a factor of 2, the same as that in $\dot{M}c^2$. On the other hand, the average uncertainty in $P_{\rm jet}$ is a factor of 3 \citep[e.g.,][]{2014Natur.515..376G}. \label{fig:disk_jet}}
\end{figure}

There are evidences for a positive correlation between the jet power and the accretion luminosity in jetted AGNs \citep[e.g.,][]{1991Natur.349..138R}. More importantly, it has been claimed, both from theoretical and observational arguments, that the former exceeds the latter \citep[][]{2008MNRAS.385..283C,2011MNRAS.418L..79T,2014Natur.515..376G}. We test these hypotheses on CGRaBS quasars.

In the top panel of Figure \ref{fig:disk_jet}, we show the variation of $P_{\rm rad}$ as a function of $L_{\rm disk}$. The results of the correlation analysis are provided in Table \ref{tab:statistics}. Overall, we find a strong positive correlation ($\rho_{\rm s}=0.63\pm0.03$, PNC$<1\times10^{-5}$), which remains valid if we consider the \gl~and the \gq~blazar populations separately (Table \ref{tab:statistics}). However, since both  $L_{\rm disk}$ and $P_{\rm rad}$ depend on redshift, we also perform a partial correlation test by following the prescriptions of \citet[][]{1992A&A...256..399P}. Even after excluding the common redshift dependence, both $L_{\rm disk}$ and $P_{\rm rad}$ still correlate, although the significance becomes a bit weaker ($\rho_{\rm par}=0.19\pm0.08$, PNC$<1\times10^{-5}$). Interestingly, a major fraction of the \gl~blazars lie above the best fit (black dashed line) and except for a few sources, almost all of the \gq~objects occupy the low $P_{\rm rad}$ regime. This is probably due to the fact that the \gl~blazars are more radiatively efficient than the \gq~blazars, although they share a similar range of $L_{\rm disk}$. 

\begin{figure*}[t]
\hbox{
\includegraphics[scale=0.7]{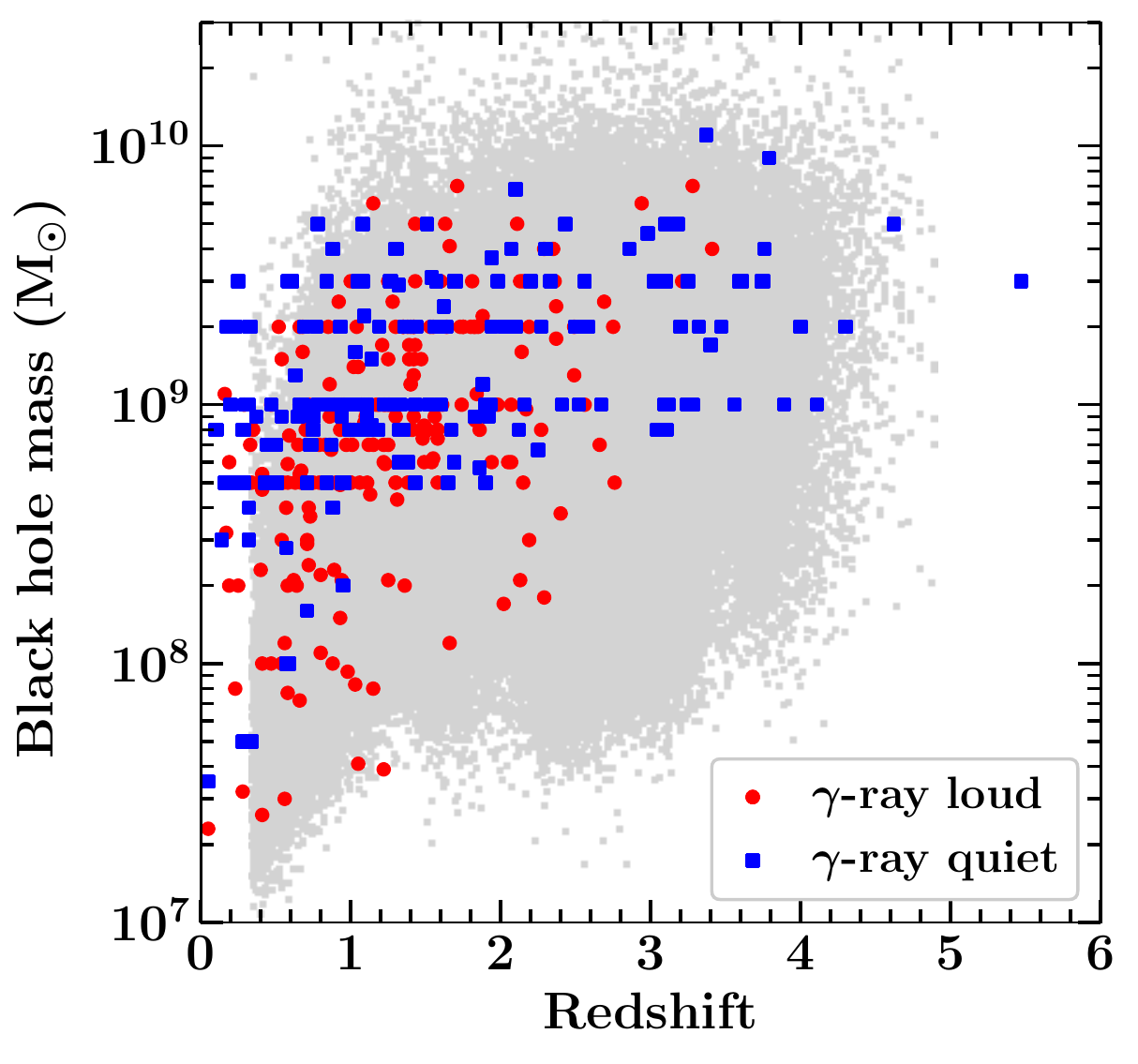}
\includegraphics[scale=0.7]{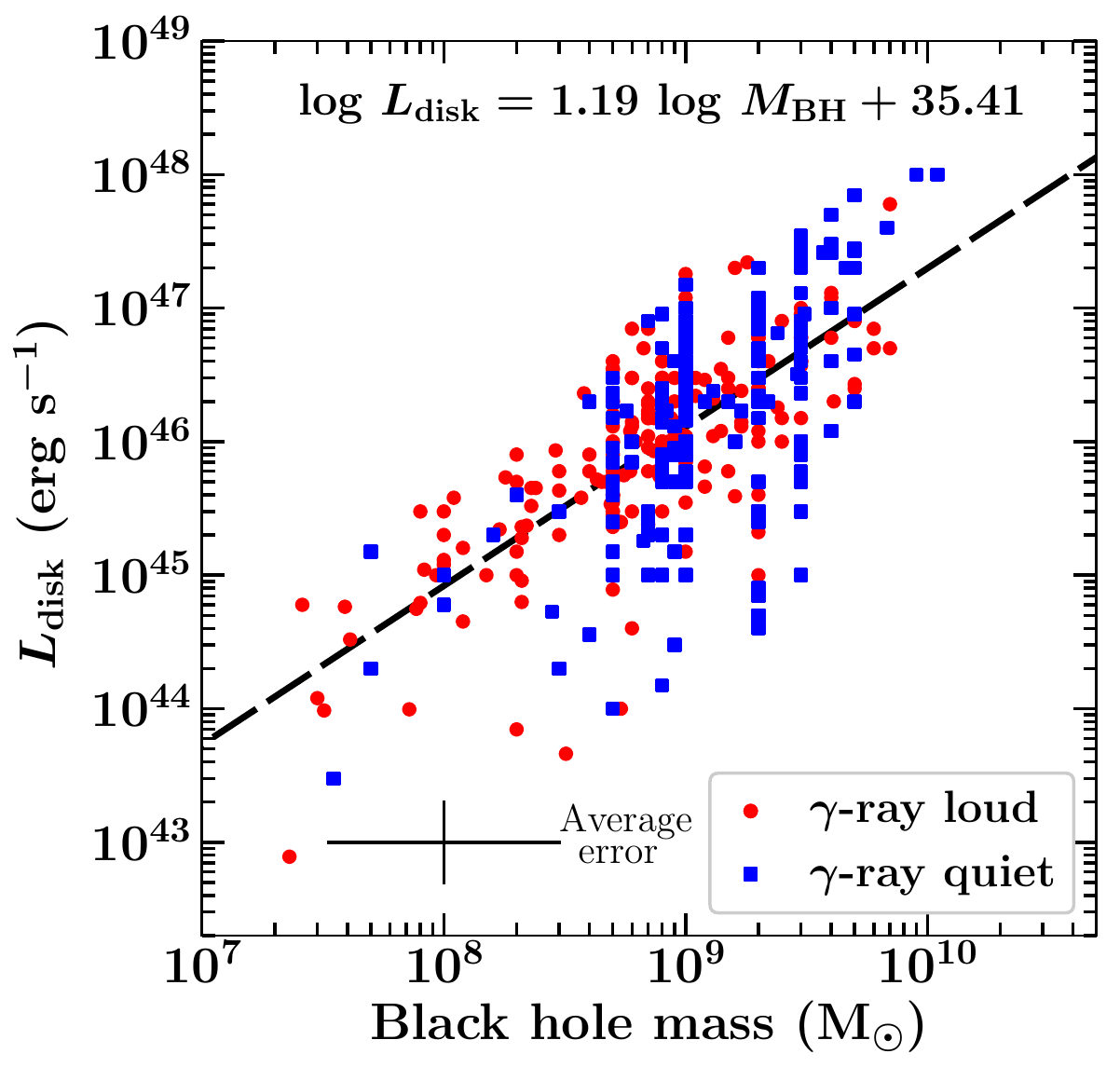}
}
\hbox{
\includegraphics[scale=0.7]{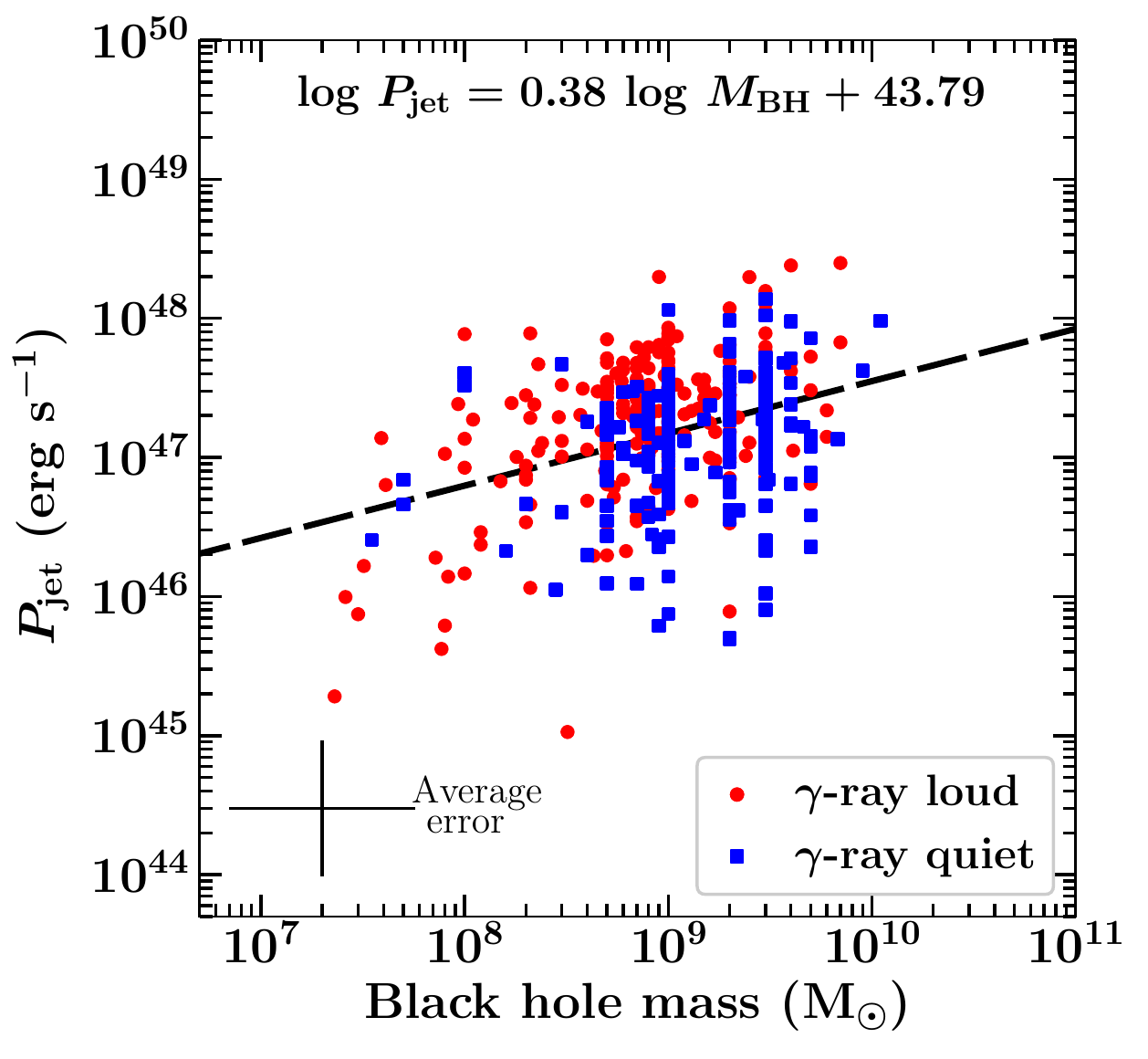}
\includegraphics[scale=0.7]{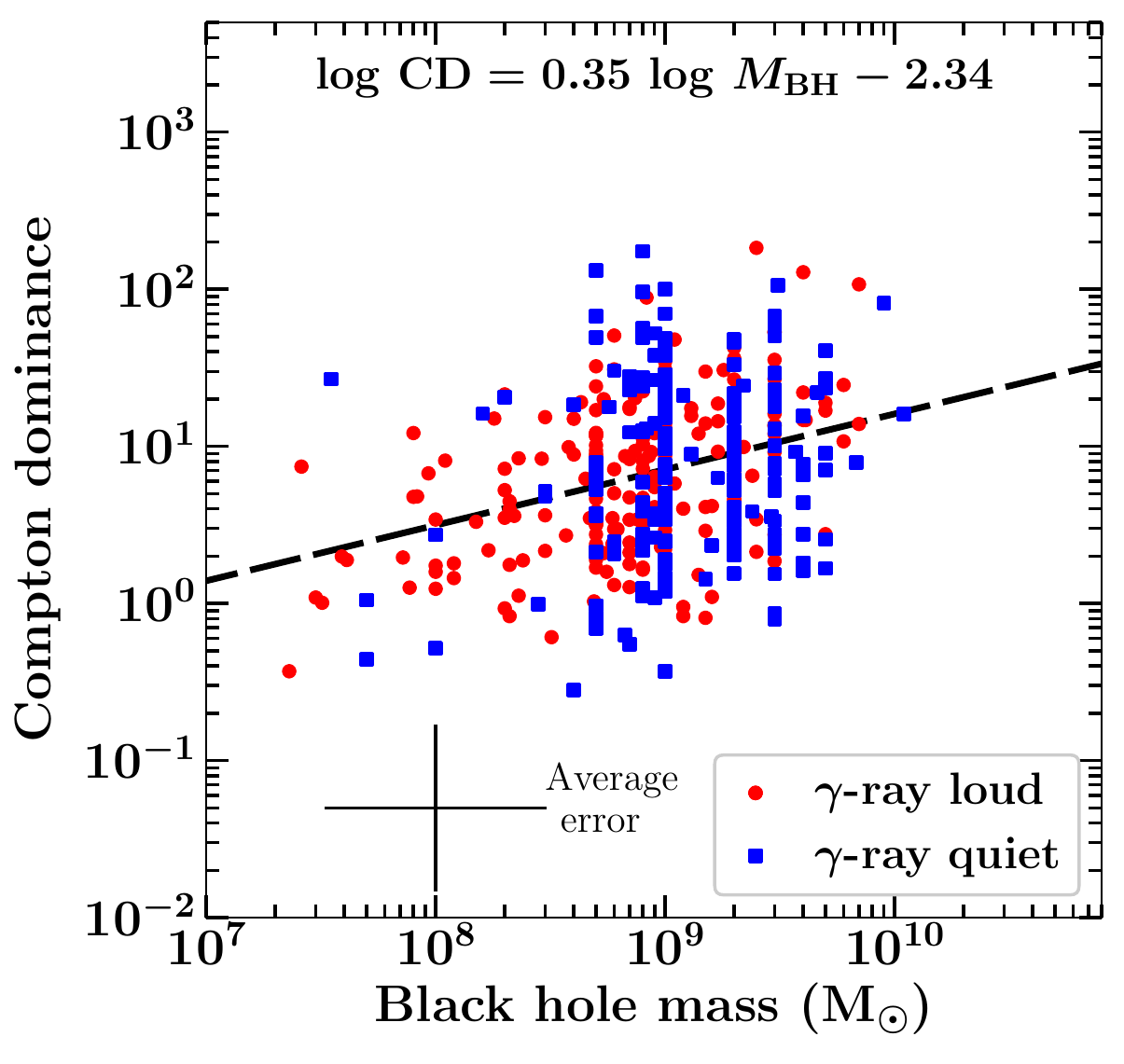}
}
\caption{Top left: The redshift dependence of $M_{\rm BH}$ for the \gl~(red) and the \gq~blazars (blue). The gray dots correspond to SDSS quasars \citep[][]{2017ApJS..228....9K}. The accretion luminosity $L_{\rm disk}$ (top right), the total jet power $P_{\rm jet}$ (bottom left), and the Compton dominance (bottom right) are shown as a function of the derived black hole masses. In all of the plots, the black dashed line is the best fit. The quoted uncertainty in the black hole mass is a factor of 3, which is the average of that typically reported for the disk fitting and virial estimation methods \citep[e.g.,][]{2006ApJ...641..689V,2013A&A...560A..28C}. \label{fig:disk_jet_bh_mass}}
\end{figure*}

We show the variation of the total jet power, $P_{\rm jet}$, as a function of the total available accretion power ($\dot{M}c^2$) in the bottom panel of Figure \ref{fig:disk_jet}. Note that for a maximally rotating black hole, the efficiency of accretion is $\eta_{\rm acc}=0.3$ \citep[][]{1974ApJ...191..507T}, and we have $\dot{M}c^2=L_{\rm disk}/\eta_{\rm acc}$. The results of the correlation analysis suggest a strong positive correlation, which remains valid even after removing the common redshift dependence (Table \ref{tab:statistics}). Therefore, we confirm the earlier findings about the accretion-jet connection, considering both \gl~and \gq~blazar populations. However, comparing our results with the ones derived by \citet[][]{2014Natur.515..376G}, we notice an interesting observation. In our sample, there are many \gq~blazars that have larger $L_{\rm disk}$ but host a relatively moderate power jet. As can be seen in Figure \ref{fig:disk_jet}, almost all \gl~blazars lie above the one-to-one correlation (pink solid line) but there are many \gq~blazars that are located below. Accordingly, the slope of the best fitted line in our case is softer. In other words, the fact that the jet power exceeds the accretion power is probably true for the \gl~sources; however, we may see a deviation from this trend when considering a larger \gq~blazar population.

\subsection{Black Hole Mass Dependence}
\begin{figure*}[t]
\hbox{
\includegraphics[scale=0.7]{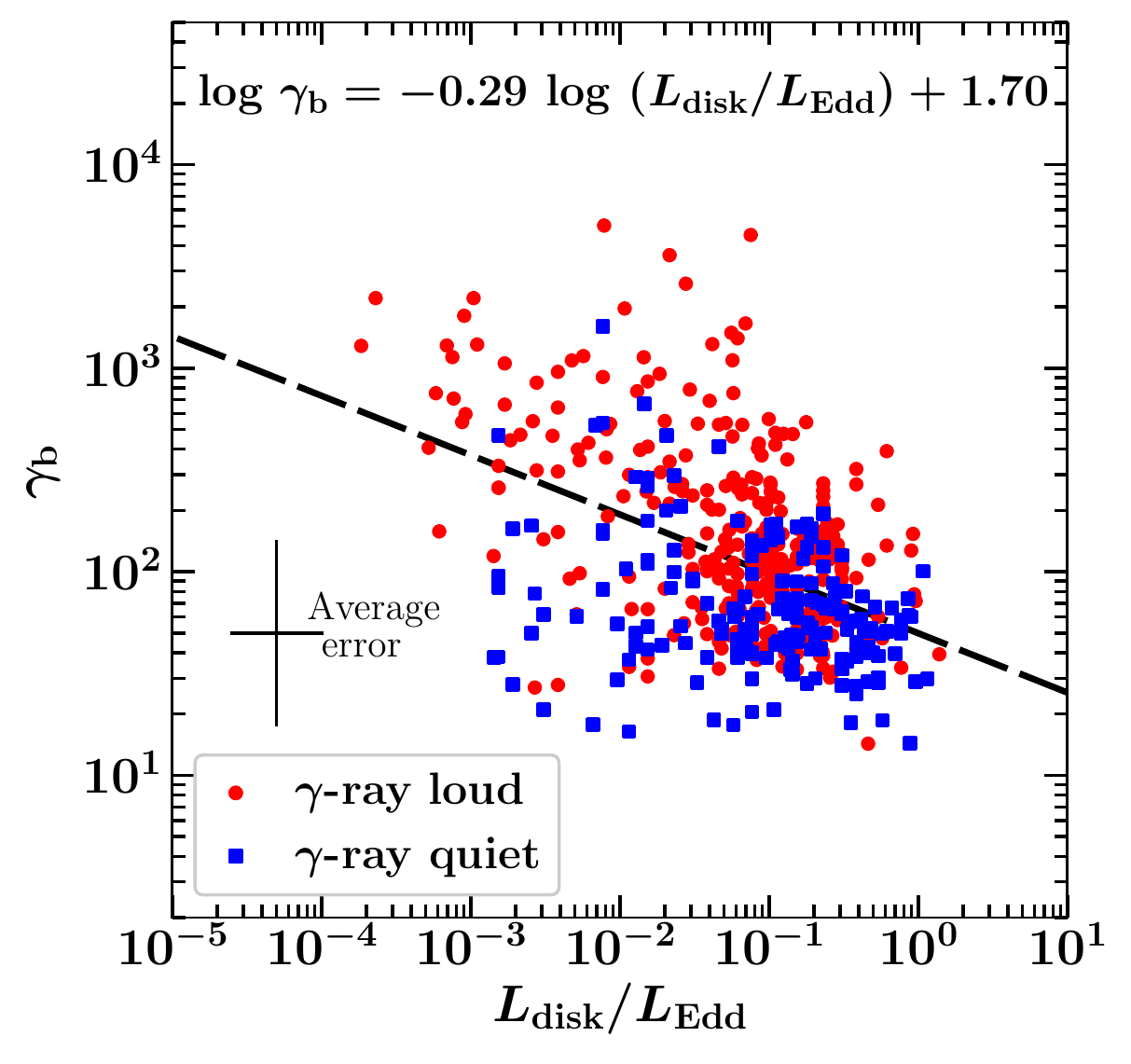}
\includegraphics[scale=0.7]{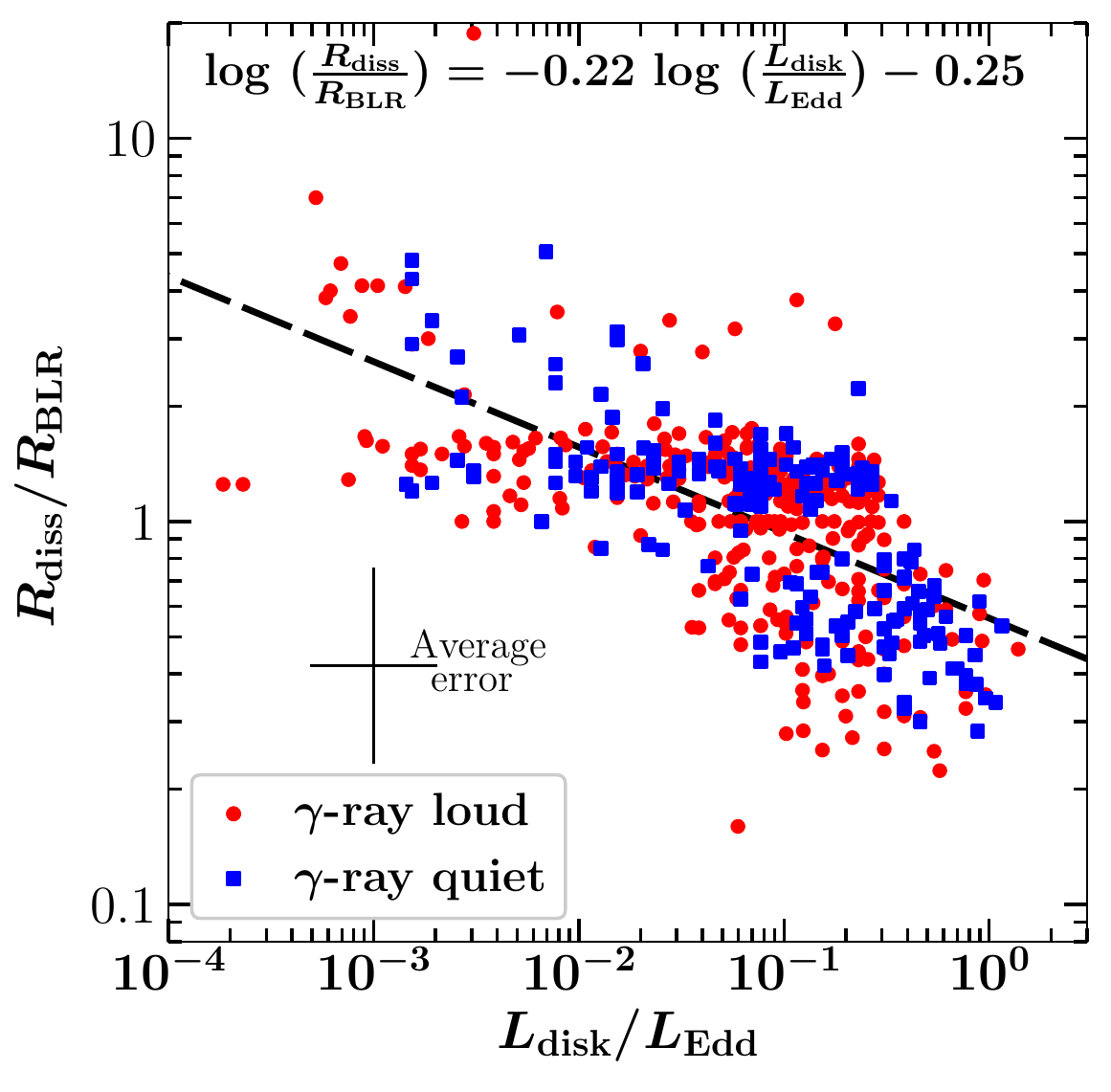}
}
\hbox{
\includegraphics[scale=0.7]{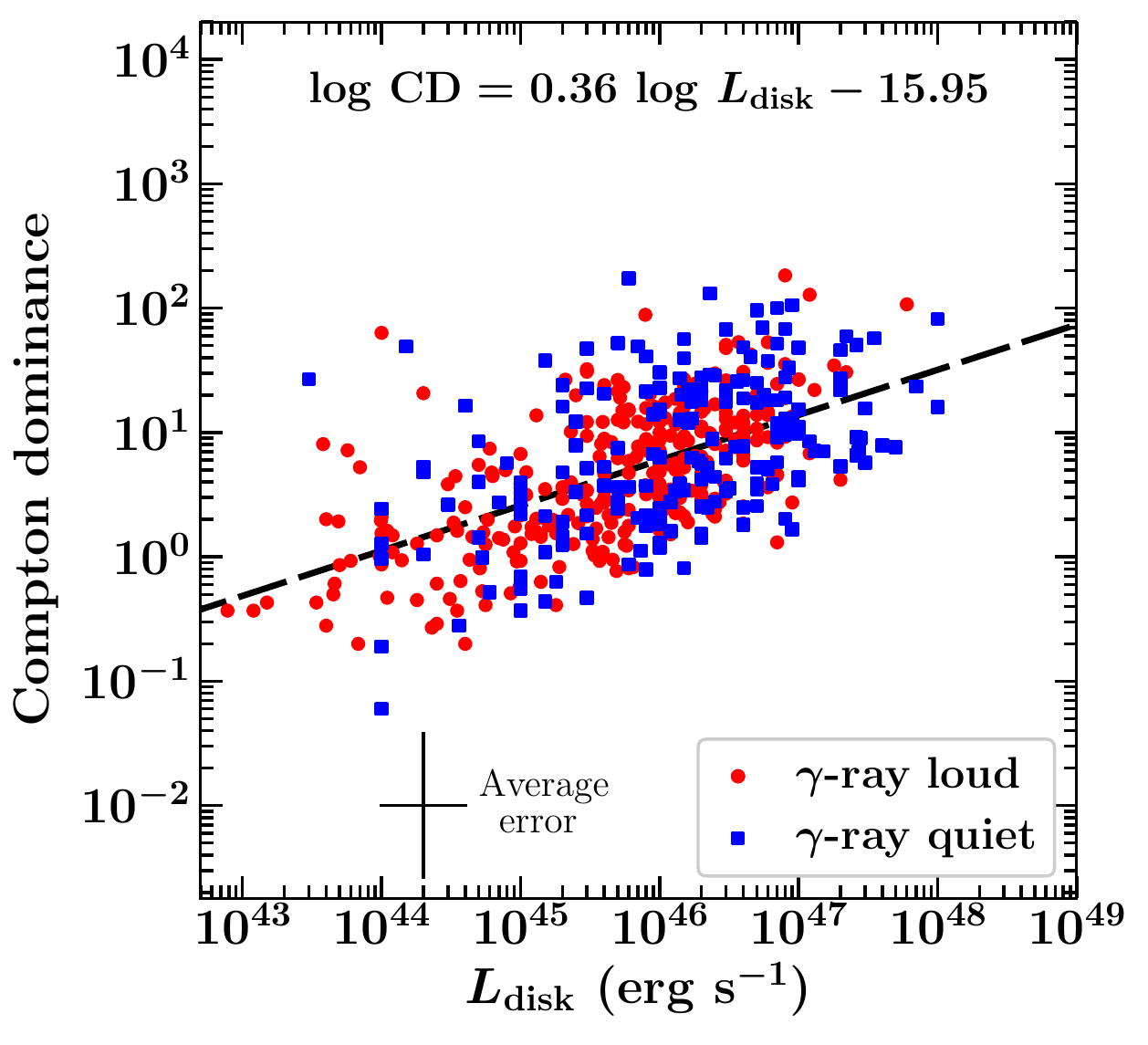}
\includegraphics[scale=0.7]{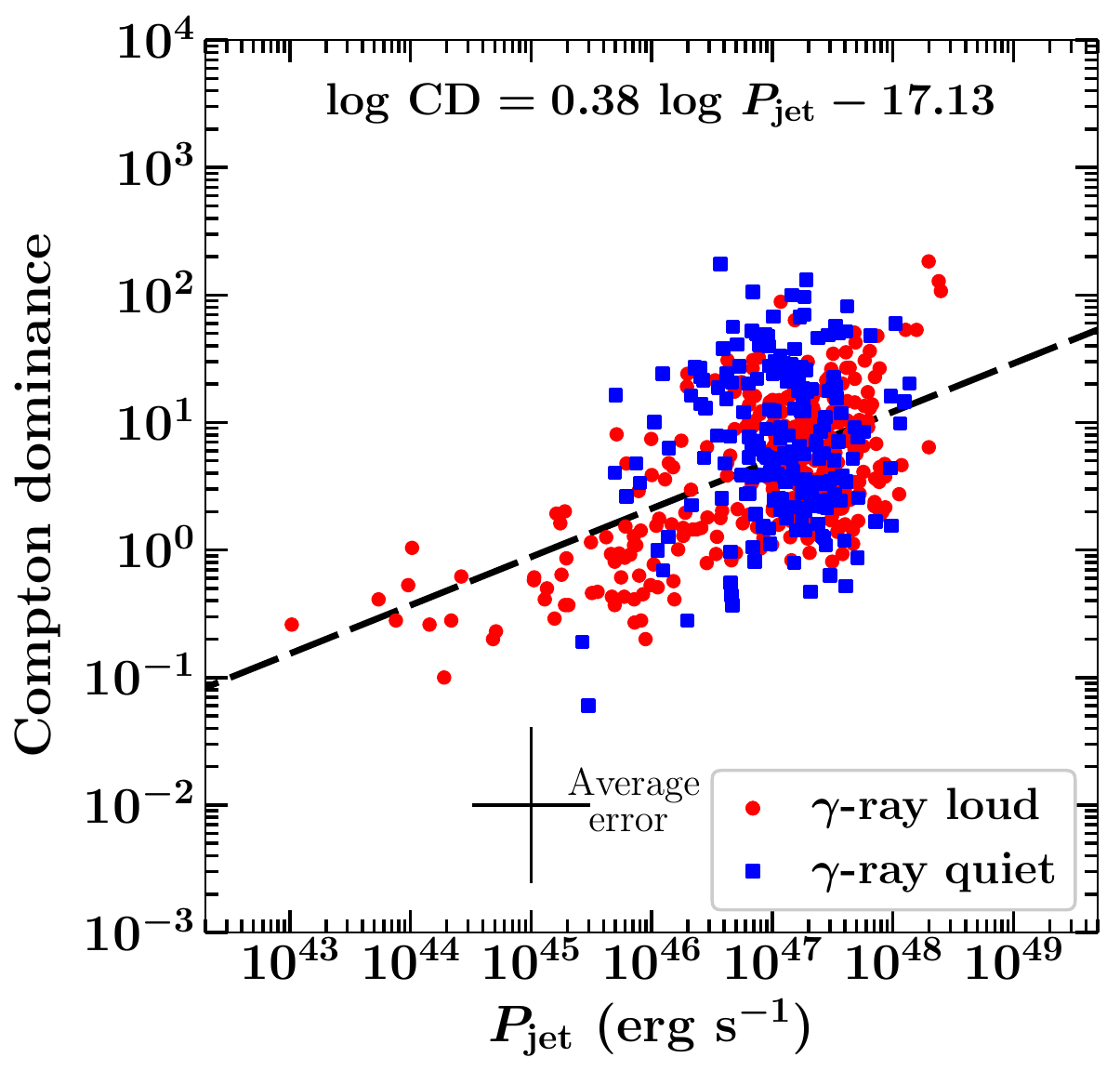}
}
\caption{Top: The variations of $\gamma_{\rm b}$ (left) and the dissipation distance (right, in units of the BLR radius) as a function of the accretion luminosity (in Eddington units). The pink solid lines represent the inner and outer boundaries of the BLR. The emission region is located inside or close to the outside of the inner boundary of the BLR for objects with a large $L_{\rm disk}$. Bottom: The Compton dominance as a function of $L_{\rm disk}$ (left) and $P_{\rm jet}$ (right). A significant positive correlation is found. In all plots, the black dashed line represents the best fit.\label{fig:blazar_sequence}}
\end{figure*}

In Figure \ref{fig:disk_jet_bh_mass}, we show the redshift dependence of $M_{\rm BH}$ derived in this work. Note that we consider only those objects whose $M_{\rm BH}$ are computed either from the disk fitting or the optical spectroscopic methods. For a comparison, we also show $M_{\rm BH}$ values for SDSS quasars as reported in \citet[][]{2017ApJS..228....9K}.  The results of the correlation analysis suggest a positive correlation between $M_{\rm BH}$ and redshift (Table \ref{tab:statistics}), which is probably a selection effect. Furthermore, we plot $L_{\rm disk}$, $P_{\rm jet}$, and CD as a function of the derived masses in Figure \ref{fig:disk_jet_bh_mass}. All three correlate positively with $M_{\rm BH}$. However, since $L_{\rm disk}$, $P_{\rm jet}$ and $M_{\rm BH}$ are a function of the redshift, we also perform partial correlation tests to exclude the common redshift dependence. As can be seen in Table \ref{tab:statistics}, even after subtracting the redshift effect, $L_{\rm disk}$ shows a positive correlation with $M_{\rm BH}$ ($\rho_{\rm s}=0.39\pm0.06$, PNC$<1\times10^{-5}$), whereas $P_{\rm jet}$ is very weakly correlated ($\rho_{\rm s}=0.17\pm0.09$, PNC$<1\times10^{-5}$), considering the entire sample. 
We also find a positive correlation between CD and $M_{\rm BH}$. However, since CD, being the ratio of two luminosities, is a redshift-independent quantity whereas $M_{\rm BH}$ is not, we adopt a semi-partial correlation analysis to exclude the redshift dependence of $M_{\rm BH}$. The results of this exercise lead to the conclusion that CD does not correlate with $M_{\rm BH}$ (see Table \ref{tab:statistics}). Although massive black-hole-hosted blazars are known to be Compton dominated \citep[e.g.,][]{2016ApJ...825...74P}, our correlation analysis indicates that even blazars with low mass black holes can also have a large CD ($>1$). The \fermi-LAT detected narrow line Seyfert 1 galaxies\footnote{Narrow line Seyfert 1 galaxies are known to host low-mass black holes \citep[e.g.,][]{2008ApJ...685..801Y}.} are probably a good example of this \citep[e.g.,][]{2013ApJ...768...52P}.

\subsection{Blazar Sequence}

There has been a long debate about the validity of the blazar sequence \citep[][]{1998MNRAS.299..433F}, i.e., whether such a sequence has a physical origin \citep[][]{1998MNRAS.301..451G} or is just a selection bias \citep[e.g.,][]{2012MNRAS.420.2899G}. In the physical scheme, the rate of accretion onto the central compact object can explain the observed phenomena. A luminous disk implies an efficient accretion process ($L_{\rm disk}/L_{\rm Edd}\gtrsim1\times10^{-2}$), which ionizes the BLR clouds resulting in the detection of the broad optical emission lines. Accordingly, the jet electrons interact with the dense photon field of the BLR via EC mechanism before reaching high energies and produce luminous \gm-ray emission, which makes the SEDs more Compton dominated and low-frequency peaked. On the other hand, in the low-accretion regime ($L_{\rm disk}/L_{\rm Edd}<1\times10^{-2}$), the external radiation field becomes weaker and the SED is less Compton dominated.  We test these hypotheses on our sample of CGRaBS quasars and show the results in Figure \ref{fig:blazar_sequence}.

We find a significant anti-correlation between $L_{\rm disk}$ (in Eddington units) and $\gamma_{\rm b}$ ($\rho_{\rm s}=-0.30\pm0.05$, PNC$<1\times10^{-5}$). This suggests that an increase in $L_{\rm disk}$ corresponds to a decrease in $\gamma_{\rm b}$, i.e., the shift of the SED peaks to lower frequencies. Moreover, we also find that $L_{\rm disk}$ and $R_{\rm diss}$ are strongly anti-correlated ($\rho_{\rm s}=-0.64\pm0.03$, PNC$<1\times10^{-5}$). Both anti-correlations indicate a physical connection between $L_{\rm disk}$ and the behavior of the SED in blazars. As discussed above, a stronger accretion disk emission implies a luminous BLR whose intense radiation field interacts with jet electrons making them lose energy primarily via EC-BLR process. This also hints that the stronger the disk emission, the closer the location of the dissipation region to the central black hole (with respect to the outside BLR scenario), which is observed.

In the bottom panels of Figure \ref{fig:blazar_sequence}, we show the variation of CD with respect to $L_{\rm disk}$ and $P_{\rm jet}$. In both cases, strong positive correlations are found that remain significant for CD versus $L_{\rm disk}$ and become weaker for CD versus $P_{\rm jet}$, after excluding the redshift dependence by means of a semi-partial correlation analysis (Table \ref{tab:statistics}). This implies that more Compton-dominated blazars host more powerful disks. Connecting these findings with those discussed above, we can conclude that the accretion disk plays a major role in controlling the observed properties of powerful blazars, and this supports the claim that the blazar sequence has a physical origin \citep[see also,][]{2017MNRAS.469..255G}. However, it may be worth revisiting this hypothesis by including a large sample of BL Lac objects, especially high-redshift ones \citep[see, e.g.,][]{2017ApJ...834...41K}.

\begin{figure*}[t]
\hbox{
\includegraphics[scale=0.7]{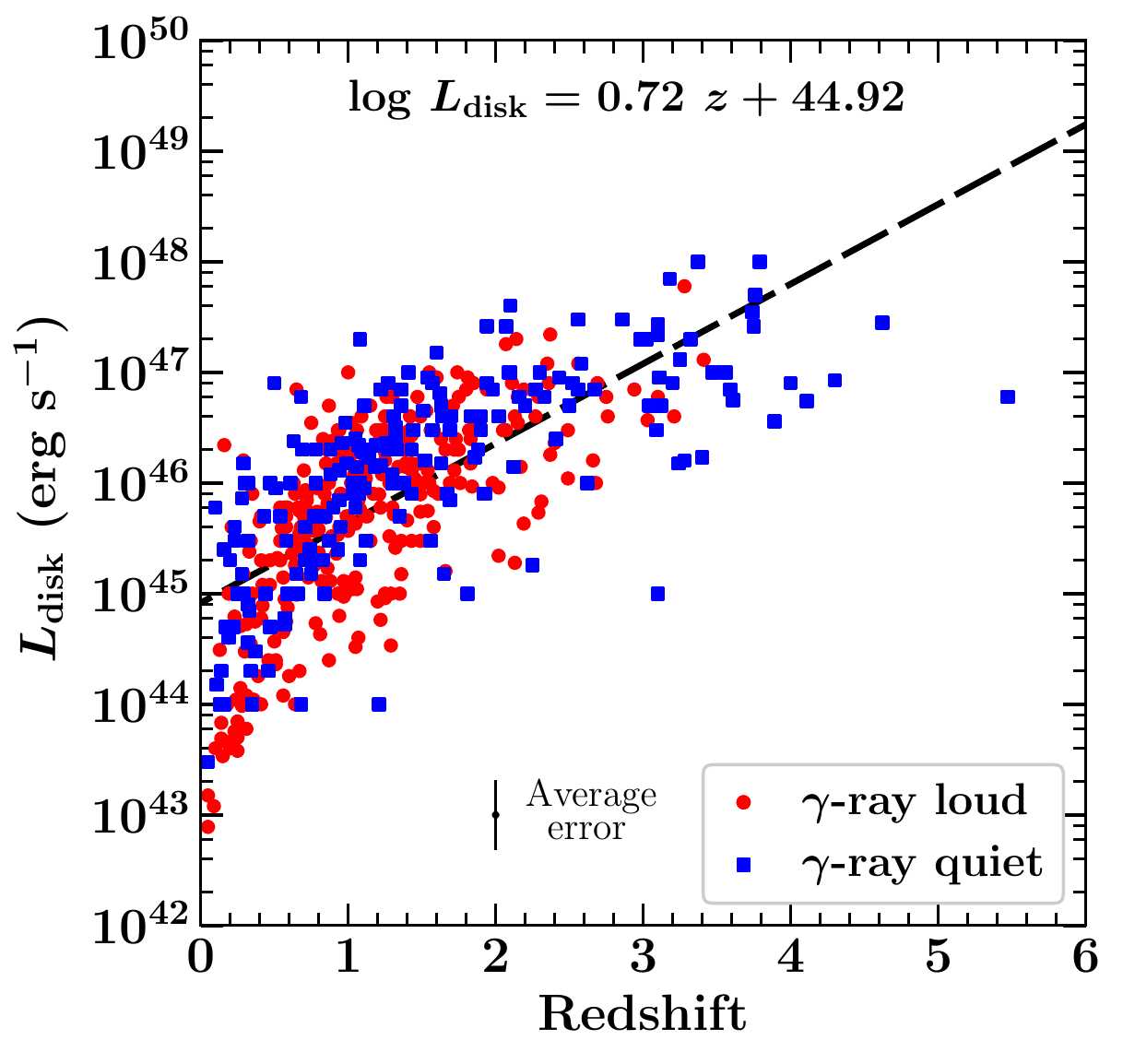}
\includegraphics[scale=0.7]{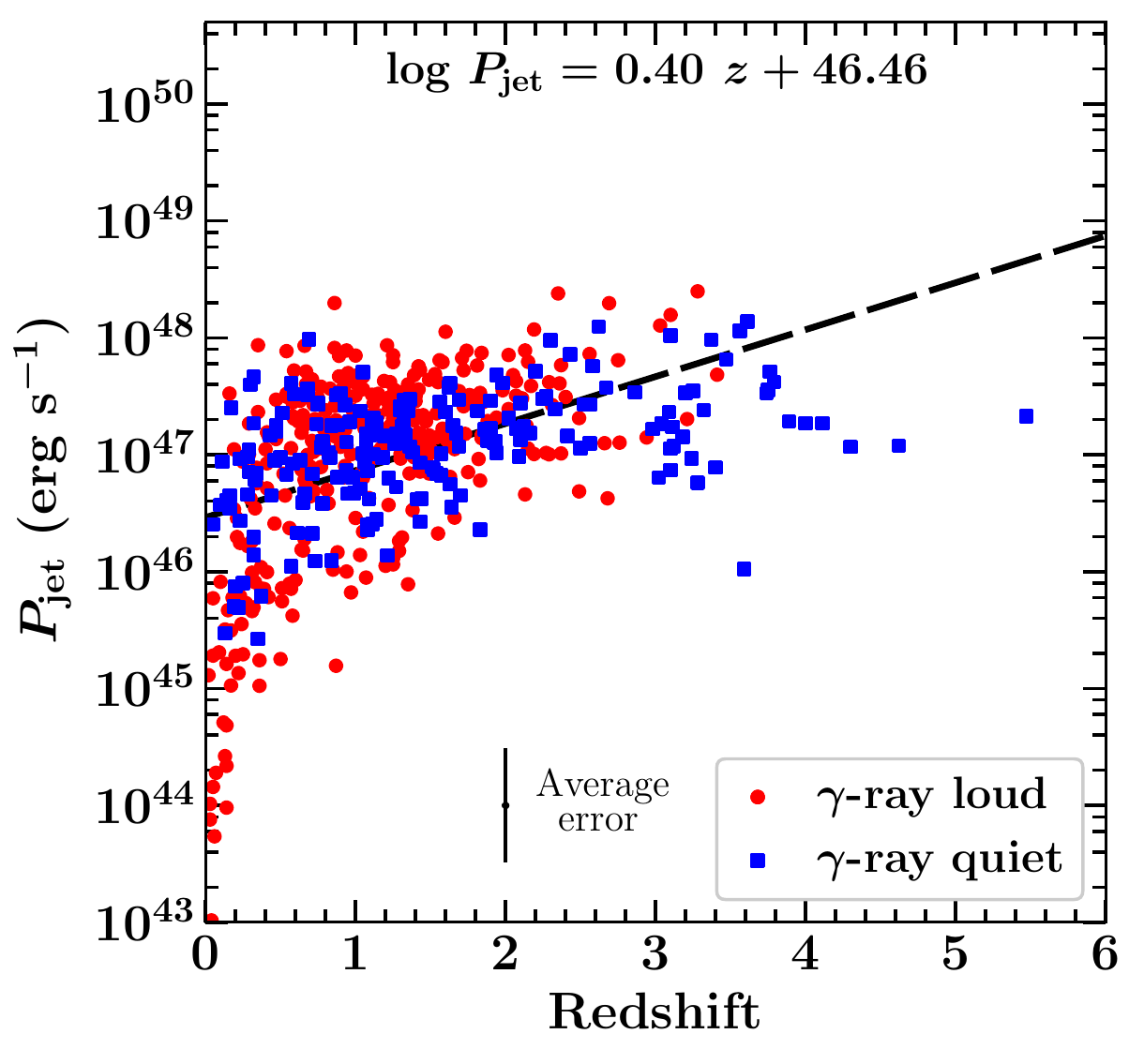}
}
\hbox{\hspace{4.2cm}
\includegraphics[scale=0.7]{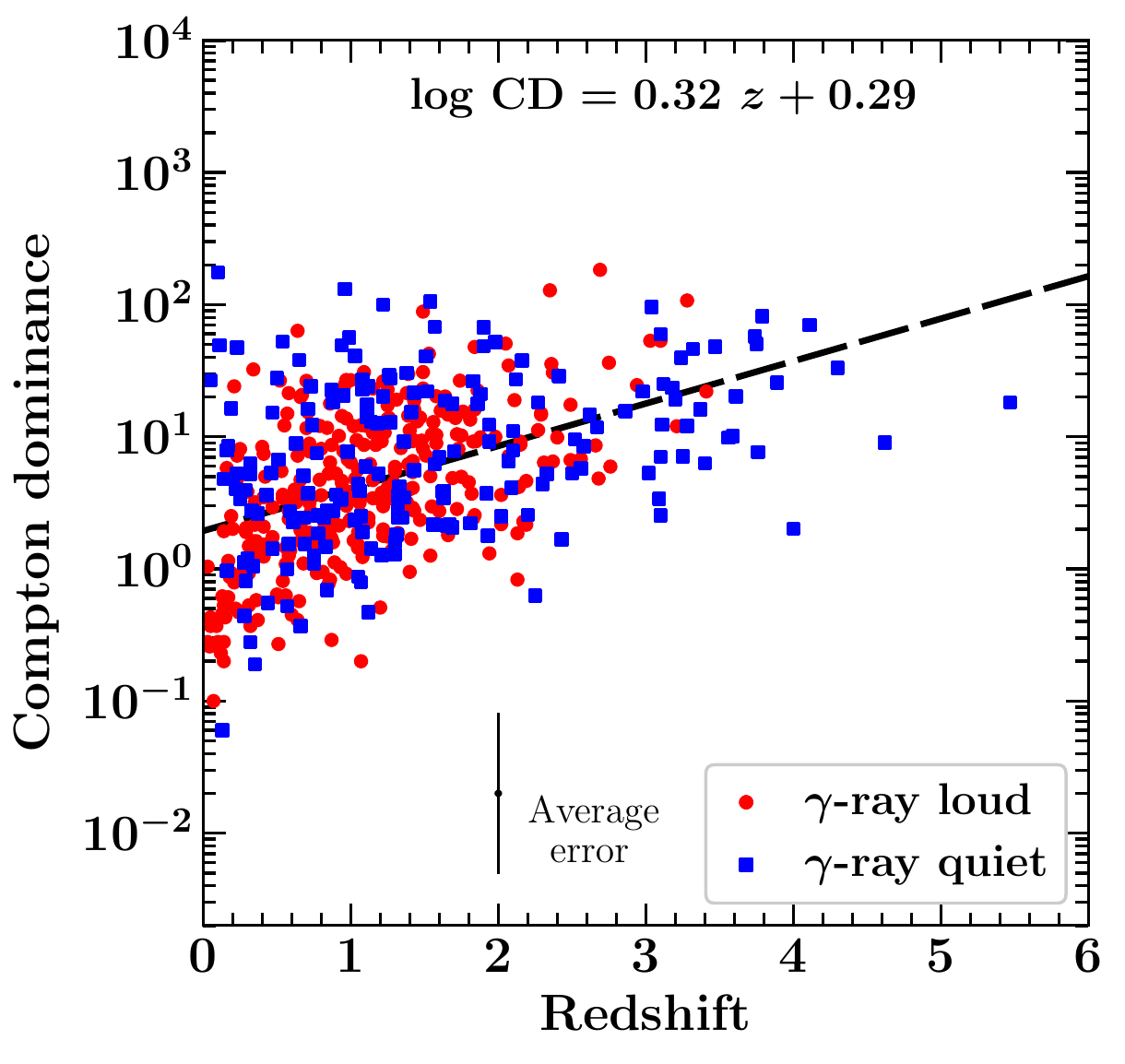}
}
\caption{Cosmic evolution of $L_{\rm disk}$, $P_{\rm jet}$ and CD as observed in \gl~and \gq~blazars.\label{fig:blazar_evolution}}
\end{figure*}

\subsection{Blazar Evolution Scenario}
We show the redshift dependence of $L_{\rm disk}$, $P_{\rm jet}$ and CD in Figure \ref{fig:blazar_evolution}, and the associated correlation parameters are reported in Table \ref{tab:statistics}. The positive correlations observed in $L_{\rm disk}$ and $P_{\rm jet}$ are most likely the Malmquist bias. However, a positive correlation of CD with redshift appears more like an evolutionary effect naturally related to redshift since CD is a redshift-independent quantity. One can argue that such a correlation could be spurious since CD exhibits a positive correlation with $L_{\rm disk}$, which itself strongly depends on redshift. However, as we have shown earlier (see Table \ref{tab:statistics} and Figure \ref{fig:blazar_sequence}), CD and $L_{\rm disk}$ are intrinsically correlated. Our results are in line with the findings of \citet[][]{2002ApJ...564...86B} and \citet[][]{2002ApJ...571..226C}, indicating the evolution of the high power, Compton-dominated sources to the less-luminous, low Compton-dominated ones. We caution, however, that a strong claim cannot be made for the following two reasons. First, there are very few objects in our sample beyond redshift 3, and without \gm-ray detection that poses difficulty in accurately measuring their IC peak luminosity and hence CD. Second, only a minor fraction of our sample are BL Lac objects, which are typically low-CD sources \citep[e.g.,][]{2010MNRAS.401.1570T}, and it is possible that such objects could be located at high redshifts \citep[see, e.g.,][]{2017ApJ...834...41K}. In our future work, we will try to address these shortcomings by considering more BL Lac sources and \gm-ray detected high-redshift blazars \citep[][]{2017ApJ...837L...5A}.

\subsection{Gamma-ray Loud versus Gamma-ray Quiet Blazars}
\begin{figure}[t]
\includegraphics[scale=0.68]{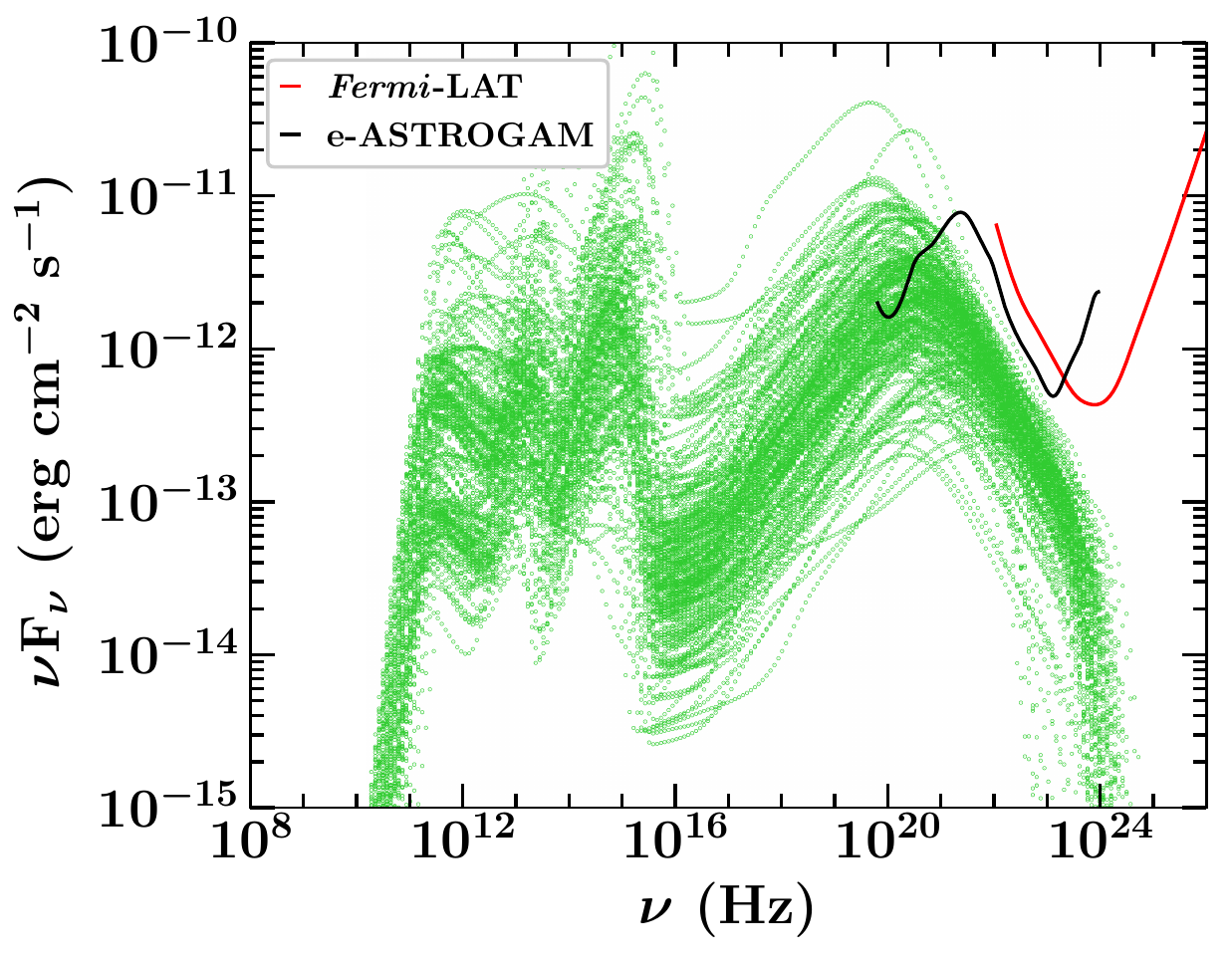}
\caption{The broadband SEDs of all of the \gq~blazars studied in this work (lime green). Overplotted are the \fermi-LAT sensitivity in 10 years of observation (red) and the 3$\sigma$ sensitivity plot for 1 year of exposure of the proposed all-sky MeV mission e-ASTROGAM (black).\label{fig:astrogam}}
\end{figure}
\fermi-LAT has detected hundreds of blazars in its first four years of observation \citep[][]{2015ApJ...810...14A}, but an even larger number are yet to be detected in \gm-rays \citep[][]{2015Ap&SS.357...75M}. It was found in several radio studies that the \gq~blazars have lower Doppler boosting factors. \citet[][]{2015ApJ...810L...9L} have reported the synchrotron peaks of the \gq~sources to be located at relatively lower frequencies ($<10^{13.4}$ Hz), implying their high-energy IC peak to lie below the energy range covered by the \fermi-LAT. These objects are therefore less likely to be \gm-ray detected. According to our analysis, the \gl~and  \gq~blazars share several similar observational features such as occupying the same region in the {\it WISE} color-color diagram (albeit with a large scatter in the \gq~objects). The $M_{\rm BH}$ and $L_{\rm disk}$ distributions of the \gl~sources are also similar with those derived from the \gq~blazars. However, the \gl~blazars are relatively brighter in the radio and X-ray bands, and they exhibit larger Doppler factors compared to their \gq~counterparts. The \gm-ray emission in powerful blazars is produced via the EC process (in the leptonic emission scenario), which is very sensitive to the Doppler boosting because of the anisotropic nature of the external radiation field in the emission region frame \citep[][]{1995ApJ...446L..63D}. Therefore, the difference in the Doppler boosting could account for the \gm-ray non-detection. Another crucial parameter that we find significantly different in the \gl~and the \gq~blazars is the break Lorentz factor, $\gamma_{\rm b}$. In our SED modeling, we have constrained $\gamma_{\rm b}$ from the location of the synchrotron peak. Since in one-zone leptonic modeling the same electron population radiates the high energy X-ray to \gm-ray emission, $\gamma_{\rm b}$ also controls the location of the IC peak according to the following relation \citep[e.g.,][]{2012MNRAS.419.1660S}

\begin{equation}
\nu_{\rm Syn}  \,  \approx {\delta \over 1+z} \gamma^2_{\rm b} \, \nu_{\rm L} \, 
\end{equation}

\begin{equation}
\nu_{\rm SSC} \, \approx {\delta \over 1+z} \gamma^4_{\rm b} \, \nu_{\rm L}
\end{equation}

\begin{equation}
\nu_{\rm EC}  \, \approx {\Gamma \delta \over 1+z} \gamma^2_{\rm b} \, \nu_{\rm seed} \,
\end{equation}

where $\nu_{\rm L}$ is the Larmor frequency and $\nu_{\rm seed}$ is the characteristic frequency of the BLR/torus photon field. As can be seen in the top right panel of Figure \ref{fig:sed_param}, \gq~blazars have smaller $\gamma_{\rm b}$ compared to \gl~sources, thus indicating their IC peak to be located at lower frequencies. In other words, they are more MeV-peaked. A shift of the IC peak to low frequencies makes their falling part of the EC spectrum  steeper and thus avoids  detection by the \fermi-LAT. The fact that the \gq~blazars are bright in the MeV band makes them suitable targets for observations from the facilities like \nustar~\citep[][]{2013ApJ...770..103H}. The ideal instrument, however, would be an all-sky survey MeV mission, e.g., e-ASTROGAM \citep[][]{2016arXiv161102232D} or AMEGO\footnote{https://pcos.gsfc.nasa.gov/physpag/probe/AMEGO\_probe.pdf}. In Figure \ref{fig:astrogam}, we plot the modeled SEDs of all of the \gq~blazars (lime green) and overplot the 3$\sigma$ flux sensitivity of e-ASTROGAM (black line) for 1 year of observation. For comparison, we also show the flux sensitivity of the \fermi-LAT (red line) for its 10 years of operation for high Galactic latitude sources. As can be seen, the chances of \fermi-LAT detection for these sources are remote unless they exhibit  high-amplitude flaring activity with a shift of the IC peak to higher frequencies \citep[see, e.g.,][for the detection of such an event that led to the detection of the FSRQ PKS 1441+25, $z=0.94$, at TeV energies]{2015ApJ...815L..22A}. On the other hand,  according to Figure \ref{fig:astrogam}, 121 \gq~blazars would be detected by e-ASTROGAM (at 500 keV) in one year of observations. Our study of \gq~blazars, therefore, presents a potential list of blazars to be detected by future MeV missions.

\section{Summary}\label{sec8}
In this work, we have performed a broadband study of a large sample of blazars included in the CGRaBS catalog \citep[][]{2008ApJS..175...97H}. Our main findings are as follows
\begin{enumerate}
\item The \gl~and the \gq~objects do not show a major difference in the {\it WISE} color-color diagram, and their X-ray spectral shapes are also similar. However, \gl~blazars are brighter in the radio and X-ray bands.
\item A comparison of $M_{\rm BH}$ and $L_{\rm disk}$ derived from the disk fitting and optical spectroscopic approaches leads to similar results, thus suggesting that disk modeling could be used as a robust diagnostic to derive $M_{\rm BH}$ and $L_{\rm disk}$ in powerful blazars.
\item Both \gl~and \gq~blazars exhibit similar distributions of  $M_{\rm BH}$ and $L_{\rm disk}$ with \gq~sources hosting slightly more massive ($\sim2.5\sigma$ significance) black holes.
\item We find a considerable difference in the Doppler factors of the \gl~and the \gq~blazars, with \gl~sources more boosted. These results confirm the earlier findings derived from  radio observations.
\item The \gl~blazars have a larger jet power in radiation compared to the \gq~objects. This implies that the jets in the \gl~blazars are more radiatively efficient.
\item We confirm that the accretion process and the jet are positively correlated, an effect that remains significant even after excluding the common redshift dependence. For a majority of the sources the jet power exceeds the accretion luminosity; however, many \gq~blazars have more powerful disks than their jets. Therefore, it is possible that we may see a deviation from the observed trend ($P_{\rm jet}>L_{\rm disk}$) when one considers a larger sample of the \gq~blazars.
\item Both $L_{\rm disk}$ and $P_{\rm jet}$ show a positive correlation with $M_{\rm BH}$, which remains strong for $L_{\rm disk}$ and becomes weaker for $P_{\rm jet}$ after excluding the common redshift dependence. On the other hand, the results of the semi-partial correlation analysis has led to the conclusion that CD does not correlate with $M_{\rm BH}$, implying that even blazars hosting low-mass black holes can be Compton dominated. The \gm-ray emitting narrow line Seyfert 1 galaxies are a good example of such objects.
\item According to our analysis, there is a physical connection between $L_{\rm disk}$ and the behavior of the blazar SEDs. In other words, we find that the blazar sequence has a physical origin, at least for the sources studied in this work.
\item We notice that the high-redshift blazars are more Compton dominated compared to their low-redshift counterparts. However, the sample size of $z>3$ blazars needs to be enlarged, and one should consider more BL Lac objects to make a strong claim about the evolution of blazars.
\end{enumerate}

\acknowledgments
We are thankful to the journal referee for constructive comments. We are also grateful to the \fermi-LAT Collaboration internal referee Rodrigo Nemmen for useful suggestions and Philippe Bruel, Elizabetta Cavazzutti, Sara Cutini, Seth Digel, Alberto Dom{\'{\i}}nguez, Marcello Giroletti, and Dave Thompson for critical reading of the manuscript. The \textit{Fermi} LAT Collaboration acknowledges generous ongoing support from a number of agencies and institutes that have supported both the development and the operation of the LAT as well as scientific data analysis. These include the National Aeronautics and Space Administration and the Department of Energy in the United States, the Commissariat \`a l'Energie Atomique and the Centre National de la Recherche Scientifique / Institut National de Physique Nucl\'eaire et de Physique des Particules in France, the Agenzia Spaziale Italiana and the Istituto Nazionale di Fisica Nucleare in Italy, the Ministry of Education, Culture, Sports, Science and Technology (MEXT), High Energy Accelerator Research Organization (KEK) and Japan Aerospace Exploration Agency (JAXA) in Japan, and the K.~A.~Wallenberg Foundation, the Swedish Research Council and the Swedish National Space Board in Sweden. Additional support for science analysis during the operations phase is gratefully acknowledged from the Istituto Nazionale di Astrofisica in Italy and the Centre National d'\'Etudes Spatiales in France. This work performed in part under DOE Contract DE-AC02-76SF00515.

This research has made use of data obtained through the High Energy Astrophysics Science Archive Research Center Online Service, provided by the NASA/Goddard Space Flight Center. Part of this work is based on archival data, software or online services provided by the ASI Data Center (ASDC). This research has made use of the XRT Data Analysis Software (XRTDAS).

This work is based on observations obtained with XMM-Newton, an ESA science mission with instruments and contributions directly funded by ESA Member States and NASA. The scientific results reported in this article are based on data obtained from the Chandra Data Archive. This research has made use of software provided by the Chandra X-ray Center (CXC) in the application packages CIAO, ChIPS, and Sherpa.

This publication makes use of data products from the Wide-field Infrared Survey Explorer, which is a joint project of the University of California, Los Angeles, and the Jet Propulsion Laboratory/California Institute of Technology, funded by the National Aeronautics and Space Administration.

\software{SAS (v15.0.0), fermiPy \citep{2017arXiv170709551W}, XSPEC \citep{1996ASPC..101...17A}, Swift-XRT data product generator \citep{2009MNRAS.397.1177E}, CIAO (v.4.9)}.

\bibliographystyle{aasjournal}
\bibliography{Master}

\clearpage
\newpage
\begin{splitdeluxetable*}{lccccccccBcccccccl}
\tablewidth{0pt} 
\tablecaption{The results of the spectral analysis of the \swift-XRT data for the \gl~blazars. \label{tab:gamma_XRT}}
\tablewidth{0pt}
\tablehead{
\colhead{Name} & \colhead{$N_{\rm H}$} & \colhead{Exp.} & \colhead{$F_{\rm X}$} & \colhead{$F_{\rm X,~low}$} & \colhead{$F_{\rm X,~high}$} & \colhead{$\Gamma_{\rm X}$}  & \colhead{$\Gamma_{\rm X,~low}$} & \colhead{$\Gamma_{\rm X,~high}$} & \colhead{$\beta_{\rm X}$} & \colhead{$\beta_{\rm X,~low}$} & \colhead{$\beta_{\rm X,~high}$} & \colhead{$\chi^2$/C-stat.} & \colhead{dof}& \colhead{$P_{\rm f-test}$}& \colhead{Model}& \colhead{fit}\\
\colhead{[1]} & \colhead{[2]} & \colhead{[3]} & \colhead{[4]} & \colhead{[5]} & \colhead{[6]} & \colhead{[7]}  & \colhead{[8]} & \colhead{[9]} & \colhead{[10]} & \colhead{[11]} & \colhead{[12]} & \colhead{[13]} & \colhead{[14]}& \colhead{[15]}& \colhead{[16]}& \colhead{[17]}
}
\startdata														     			
J0004$-$4736	& 1.34	& 5.62	   & 0.43	 & 0.31	& 0.57	  & 1.88	& 1.54	& 2.23	& ---	    & ---    & 	---    & 	26.26	& 48 & ---	  & PL	& c-stat\\
J0005+3820	    & 6.57	& 20.31	& 1.91  &	1.30	    & 2.10   & 0.03	& 0.00	& 0.36	& 1.57	& 1.12	& 1.77	& 25.33	& 30	 & $-$6.90 & LP	& chi     \\
J0011+0057	    & 2.67	& 10.53	& 0.16	 & 0.10	& 0.26	  & 1.43	& 0.86	& 2.00	& ---	    & ---	    & ---	    & 21.12	& 25	 & ---	  & PL	& c-stat \\
J0016$-$0015	& 2.72	& 3.95	   & 0.59	 & 0.41	& 1.03   & 1.29	& 0.85	& 1.72	& ---	    & ---	    & ---     & 28.15	& 32	 & ---	  & PL	& c-stat \\
J0017$-$0512	& 2.81	& 11.24  & 2.04	 & 1.88	& 2.27	  & 1.84	& 1.73	& 1.96	& ---	    & --- 	 & ---	 & 31.38	 & 26 & $-$0.22 & PL	& chi\\
\enddata
\tablecomments{The column contents are as follows. Col.[1]: name of the CGRaBS object; Col.[2]: the Galactic neutral Hydrogen column density, in 10$^{20}$ cm$^{-2}$ \citep[][]{2005AA...440..775K}; Col.[3]: net exposure, in ksec; Col.[4], [5], and [6]: observed 0.3$-$10 keV flux and its lower and upper limits, respectively, in units of 10$^{-12}$ \ergflux; Col.[7], [8], and [9]: power-law photon index (or spectral slope of the log parabola model at the pivot energy) and its lower and upper limits, respectively; Col.[10], [11], and [12]: the curvature parameter of the log parabola model and its lower and upper limits, respectively; Col.[13]: the statistics of the model fitting; Col.[14]: degrees of freedom; Col.[15]: log of the probability of null hypothesis derived from the {\tt f-test}; Col.[16]: the best fitted model, PL: power law, LP: log parabola; and Col.[17]: adopted statistics, c-stat: C-statistics \citep[][]{1979ApJ...228..939C}, and chi: $\chi^2$ fitting.\\
(This table is available in its entirety in a machine-readable form in the online journal. A portion is shown here for guidance regarding its form and content.)}
\end{splitdeluxetable*}

\begin{splitdeluxetable*}{lccccccccBcccccccl}
\tablewidth{0pt} 
\tablecaption{The results of the the \swift-XRT data analysis for \gq~blazars. \label{tab:non-gamma_XRT}}
\tablewidth{0pt}
\tablehead{
\colhead{Name} & \colhead{$N_{\rm H}$} & \colhead{Exp.} & \colhead{$F_{\rm X}$} & \colhead{$F_{\rm X,~low}$} & \colhead{$F_{\rm X,~high}$} & \colhead{$\Gamma_{\rm X}$}  & \colhead{$\Gamma_{\rm X,~low}$} & \colhead{$\Gamma_{\rm X,~high}$} & \colhead{$\beta_{\rm X}$} & \colhead{$\beta_{\rm X,~low}$} & \colhead{$\beta_{\rm X,~high}$} & \colhead{$\chi^2$/C-stat.} & \colhead{dof}& \colhead{$P_{\rm f-test}$}& \colhead{Model}& \colhead{fit}\\
\colhead{[1]} & \colhead{[2]} & \colhead{[3]} & \colhead{[4]} & \colhead{[5]} & \colhead{[6]} & \colhead{[7]}  & \colhead{[8]} & \colhead{[9]} & \colhead{[10]} & \colhead{[11]} & \colhead{[12]} & \colhead{[13]} & \colhead{[14]}& \colhead{[15]}& \colhead{[16]}& \colhead{[17]}
}
\startdata														     			
J0001+1914	    & 3.19	& 5.47	   & 0.09	& 0.00	& 0.18	& 1.37	& 0.23	& 2.74	& --- & 	---	& ---	& 7.53	    & 6	 & ---	 & PL	& c-stat\\
J0004+2019	    & 3.78	& 1.92	   & 0.49	& 0.29	& 0.91	& 1.54	& 0.85	& 2.23	& ---	&---  &---	& 15.66	& 15 & ---	 & PL	& c-stat\\
J0004+4615	    & 8.41	& 3.74	   & 0.27	& 0.12	& 0.57	& 1.12	& 0.18	& 2.04	& ---	& ---	& ---	& 15.50	& 11	 & ---	 & PL	& c-stat\\
J0006$-$0623	& 3.22	& 16.38  & 1.80	& 1.67	& 2.00	& 1.67	& 1.56	& 1.77	& ---	& ---	& ---	& 22.27	& 28	 & $-$0.06& PL	& chi\\
J0008$-$2339	& 2.31	& 4.96	   & 0.37	& 0.28	& 0.52	& 1.73	& 1.31	& 2.16	& ---	& --- & 	---	& 37.00	& 33	 & ---	 & PL	& c-stat\\
\enddata
\tablecomments{The column contents are same as in Table \ref{tab:gamma_XRT}.\\
(This table is available in its entirety in a machine-readable form in the online journal. A portion is shown here for guidance regarding its form and content.)}
\end{splitdeluxetable*}

\newpage
\begin{table*}
\caption{\chandra~and/or \xmm~data analysis.\label{tab:chandra}}
\begin{tabular}{lcccccccccccll}
\hline
Name & $N_{\rm H}$ & Exp. & $F_{\rm X}$ & $F_{\rm X,~low}$ & $F_{\rm X,~high}$ & $\Gamma_{\rm X}$ & $\Gamma_{\rm X,~low}$ & $\Gamma_{\rm X,~high}$ & $\chi^2$/C-stat. & dof & $P_{\rm f-test}$  & fit & telescope \\
\hline
 & & & & & & & \gl& & & & & & \\
 J0941$-$1335		      & 4.08	& 19.82	& 0.42	& 0.38	& 0.45	& 1.76	& 1.61	& 1.92	& 16.80	& 19 &	-0.13 & chi	& Chandra\\
J1311+5513	              & 1.68	& 3.86	    & 0.32	& 0.27	& 0.36	& 1.79	& 1.65	& 1.93	& 21.83	& 20 & 	-0.12 & chi	& XMM\\
\hline
 & & & & & & & \gq& & & & & & \\
J0148+3854		&4.79	&63.40	&0.31	&0.28	&0.33	&1.33	&1.25	&1.42	&65.33	&53	&$-$0.33	&chi	&Chandra\\
J0254+3931		&7.58	&8.79	&1.63	&1.56	&1.74	&1.70	&1.63	&1.78	&51.35	&61	&$-$0.25	&chi	&Chandra\\
J0256$-$3315	&2.43	&2.44	&0.21	&0.14	&0.32	&1.38	&1.01	&1.75	&39.71	&45	&---	&c-stat	&Chandra\\
J0324$-$2918	&1.38	&3.81	&0.16	&0.12	&0.23	&1.60	&1.28	&1.93	&50.64	&51	&---	&c-stat	&Chandra\\
J0439+0520	   &8.92	&19.87	&0.76	&0.72	&0.79	&2.20	&2.11	&2.29	&77.67	&52	&$-$0.85	&chi	&Chandra\\
J0730+4049		&6.21	&19.81	&0.18	&0.15	&0.20	&1.36	&1.18	&1.55	&109.72	&150	&---	&c-stat	&Chandra\\
J0825+6157		&3.82	&18.95	&1.50	&1.43	&1.58	&1.50	&1.44	&1.56	&88.78	&80	&$-$0.41	&chi	&Chandra\\
J0939+4141		&1.16	&7.09	&0.30	&0.26	&0.35	&1.70	&1.52	&1.90	&5.55	&10	&$-$0.36	&chi	&Chandra\\
J1058+1951		&1.69	&149.90	&2.29	&2.27	&2.33	&1.65	&1.63	&1.66	&478.85	&346	&$-$2.13	&chi	&Chandra\\
J1110+4403   	&1.30	&8.45	&0.21	&0.18	&0.25	&1.35	&1.19	&1.51	&22.30	&26	&$-$0.97	&chi	&XMM\\
J1146$-$2447	&4.61	&4.96	&0.41	&0.35	&0.48	&1.58	&1.41	&1.75	&129.61	&128	&---	&c-stat	&Chandra\\
J1430+3649		&1.04	&3.93	&0.32	&0.27	&0.38	&1.70	&1.49	&1.91	&79.56	&95	&---	&c-stat	&Chandra\\
J1431+3952		&1.09	&2.86	&1.79	&1.64	&1.99	&1.57	&1.46	&1.69	&209.18	&207	&---	&c-stat	&Chandra\\
J1727+5510		&2.72	&34.61	&0.60	&0.57	&0.65	&1.58	&1.48	&1.67	&46.50	&49	&$-$0.15	&chi	&Chandra\\
J2003$-$3251	&7.13	&13.21	&0.78	&0.75	&0.82	&1.67	&1.62	&1.72	&121.93	&141	&$-$0.17	&chi	&XMM\\
J2354$-$1513	&2.53	&16.13	&1.12	&1.08	&1.16	&1.54	&1.50	&1.57	&207.26	&211	&$-$0.16	&chi	&XMM\\
\hline
\end{tabular}
\tablecomments{The symbols have their usual meanings as given in Table \ref{tab:gamma_XRT}. The last column refers to the name of the telescope from which the X-ray measurement has been taken. It should be noted that an absorbed power-law model is preferable for all sources.}
\end{table*}

\begin{table*}
\caption{The SED parameters associated with the modeling of the broadband emission in the \gl~blazars.\label{tab:sed_param_loud}}
\begin{center}
\begin{tabular}{lccccccccccccccccc}
\hline
Name & $z$ & $\theta_{\rm v}$ & $M_{\rm BH}$ & Type & $L_{\rm disk}$ & $R_{\rm diss}$ & $R_{\rm BLR}$ & $\delta$ & $\Gamma$ & $B$ & $p$ & $q$ & $\gamma_{\rm min}$ & $\gamma_{\rm b}$ & $\gamma_{\rm max}$ & $U_{\rm e}$ & CD \\
~[1] & [2] & [3]  & [4] & [5] & [6]  & [7] & [8] & [9] & [10] & [11] & [12] & [13] & [14] & [15]& [16]&[17] &[18] \\ 
\hline
J0004$-$4736	& 0.88	& 3.0	& 8.00	& O	& 45.11	& 0.027	& 0.037	& 15.7	& 10	 & 6.4	& 1.6	& 3.9	& 3	 & 115	& 3000	& $-$1.02	& 1.6\\
J0005+3820	    & 0.23	& 3.0	& 8.70	& E	& 43.76	& 0.033	& 0.008	& 18.6	& 15	 & 0.4	& 2.2	& 3.8	& 9 	 & 543	& 5000	& $-$1.19	& 7.2\\
J0011+0057 	    & 1.49	& 3.0	& 8.78	& D	& 45.48	& 0.037	& 0.056	& 16.5	& 11	 & 1.6	& 1.8	& 3.8	& 1	 & 100	& 3000	& $-$1.10	& 30.9\\
J0016$-$0015	& 1.58	& 3.0	& 8.87	& O	& 45.93	& 0.064	& 0.094	& 17.2	& 12	 & 2.2	& 1.8	& 4.4	& 1	 & 120	& 3000	& $-$1.29	& 20.3\\
J0017$-$0512	& 0.23	& 3.0	& 7.90	& D	& 44.79	& 0.004	& 0.025	& 15.7	& 10	 & 4.0	& 1.6	& 4.2	& 10	 & 83	& 2000	& 0.51	    & 4.8\\
\hline
\end{tabular}
\end{center}
\tablecomments{The column contents are as follows: Col.[1] and [2]: name and redshift of the source; Col.[3]: viewing angle, in degrees; Col.[4]: log scale black hole mass, in solar mass units; Col.[5]: adopted method to derive $M_{\rm BH}$ and $L_{\rm disk}$, A: assumed; D: disk fitting, E: empirical relation, O: optical spectroscopy, and N: not used (see Table \ref{tab:13_syn}); Col.[6]: log scale accretion disk luminosity, in \lum; Col.[7]: dissipation distance, in parsec; Col.[8]: size of the BLR, in parsec; Col.[9] and [10]: the Doppler factor and the bulk Lorentz factor, respectively; Col.[11]: magnetic field strength, in Gauss; Col.[12] and [13]: spectral slopes of the broken power-law electron distribution before and after the break energy ($\gamma_{\rm b}$), respectively; Col.[14], [15], and [16]: the minimum, break, and the maximum Lorentz factors of the emitting electron distribution; Col.[17]: the log scale particle energy density, in erg cm$^{-3}$; and Col.[18]: Compton dominance.\\
(This table is available in its entirety in a machine-readable form in the online journal. A portion is shown here for guidance regarding its form and content.)}
\end{table*}

\begin{table*}
\caption{The SED parameters associated with the modeling of the broadband emission in the \gq~blazars.\label{tab:sed_param_quiet}}
\begin{center}
\begin{tabular}{lccccccccccccccccc}
\hline
Name & $z$ & $\theta_{\rm v}$ & $M_{\rm BH}$ & Type & $L_{\rm disk}$ & $R_{\rm diss}$ & $R_{\rm BLR}$ & $\delta$ & $\Gamma$ & $B$ & $p$ & $q$ & $\gamma_{\rm min}$ & $\gamma_{\rm b}$ & $\gamma_{\rm max}$ & $U_{\rm e}$ & CD \\
~[1] & [2] & [3]  & [4] & [5] & [6]  & [7] & [8] & [9] & [10] & [11] & [12] & [13] & [14] & [15]& [16]&[17] &[18] \\ 
\hline
J0001+1914	    & 3.10	& 3.0	& 8.70	& A	& 45.00	& 0.038	& 0.032	& 17.2	& 12	& 2.7	& 2.0	& 3.6	& 2	& 177	& 4000	& $-$0.83  & 2.5\\
J0004+2019	    & 0.68	& 3.0	& 8.70	& A	& 44.00	& 0.012	& 0.010	& 14.8	& 9	& 2.5	& 2.1	& 3.6	& 1	& 38	    & 4000	& 0.51	      & 2.4\\
J0004+4615	    & 1.81	& 3.0	& 8.70	& A	& 45.00	& 0.041	& 0.032	& 16.5	& 11	& 2.1	& 1.9	& 3.6	& 2	& 53	    & 4000	& $-$0.56  & 2.2\\
J0006$-$0623	& 0.35	& 3.0	& 8.70	& A	& 44.00	& 0.029	& 0.010	& 14.7	& 9	& 1.5	& 1.7	& 3.8	& 20	& 467	 & 3000	& $-$1.00  & 0.2\\
J0008$-$2339	& 1.41	& 3.0	& 9.30	& D	& 47.00	& 0.880	& 0.323	& 15.7	& 10	& 0.2	& 1.6	& 3.8	& 1	& 234	 & 3000	& $-$3.00  & 13.0\\
\hline
\end{tabular}
\end{center}
\tablecomments{The column contents are same as in Table \ref{tab:sed_param_loud}.\\
(This table is available in its entirety in a machine-readable form in the online journal. A portion is shown here for guidance regarding its form and content.)}
\end{table*}

\begin{table*}
\caption{The jet powers derived from the modeling of the broadband emission in the \gl~blazars.\label{tab:jet_loud}}
\begin{center}
\begin{tabular}{lccccc}
\hline
Name &  $P_{\rm ele}$ & $P_{\rm mag}$ & $P_{\rm rad}$ & $P_{\rm kin}$ & $P_{\rm jet}$\\
~[1] & [2] & [3]  & [4] & [5] & [6]  \\ 
\hline
J0004$-$4736	& 44.09	& 45.32	& 45.65	& 46.09	& 46.17\\
J0005+3820	    & 44.45	& 43.45	& 44.39	& 46.24	& 46.25\\
J0011+0057	    & 44.38	& 44.49	& 45.45	& 46.84	& 46.84\\
J0016$-$0015	& 44.73	& 45.31	& 46.12	& 47.16	& 47.17\\
J0017$-$0512	& 44.01	& 43.31	& 44.39	& 45.78	& 45.79\\
\hline
\end{tabular}
\end{center}
\tablecomments{The column contents are as follows: Col.[1] name the source; Col.[2], [3], [4], [5], and [6]: the electron, magnetic, radiative, kinetic, and total jet power, respectively. Note that $P_{\rm jet}$ = $P_{\rm ele}$ + $P_{\rm mag}$ + $P_{\rm kin}$. All jet powers are evaluated for a two-sided jet.\\
(This table is available in its entirety in a machine-readable form in the online journal. A portion is shown here for guidance regarding its form and content.)}
\end{table*}

\begin{table*}
\caption{The jet powers derived from the modeling of the broadband emission in \gq~blazars.\label{tab:jet_quiet}}
\begin{center}
\begin{tabular}{lccccc}
\hline
Name &  $P_{\rm ele}$ & $P_{\rm mag}$ & $P_{\rm rad}$ & $P_{\rm kin}$ & $P_{\rm jet}$\\
~[1] & [2] & [3]  & [4] & [5] & [6]  \\ 
\hline
J0001+1914	    & 44.75	& 45.04	& 45.80	& 47.05	& 47.05\\
J0004+2019	    & 44.84	& 43.72	& 44.48	& 47.57	& 47.57\\
J0004+4615	    & 45.00	& 44.80	& 45.39	& 47.37	& 47.37\\
J0006$-$0623	& 44.07	& 44.03	& 45.02	& 45.39	& 45.42\\
J0008$-$2339	& 44.47	& 45.34	& 45.20	& 46.67	& 46.70\\
\hline
\end{tabular}
\end{center}
\tablecomments{The column contents are same as in Table \ref{tab:jet_loud}.\\
(This table is available in its entirety in a machine-readable form in the online journal. A portion is shown here for guidance regarding its form and content.)}
\end{table*}

\begin{table*}
\caption{The list of 13 \gl~blazars which are modeled with synchrotron and SSC processes, i.e., without invoking EC mechanism.\label{tab:13_syn}}
\begin{center}
\begin{tabular}{lc}
\hline
Name & $z$ \\
\hline
J1015+4926 & 0.21 \\
J1104+3812 & 0.03 \\
J1136+7009 & 0.04 \\
J1217+3007 & 0.13 \\
J1230+2518 & 0.14 \\
J1555+1111 & 0.36 \\
J1653+3945 & 0.03 \\
J1725+1152 & 0.02 \\
J1728+5013 & 0.06 \\
J1917$-$1921 & 0.14 \\
J1959+6508 & 0.05 \\
J2009$-$4849 & 0.07 \\
J2158$-$3013 & 0.12 \\
\hline
\end{tabular}
\end{center}
\end{table*}

\clearpage

\begin{deluxetable*}{lccccc}
\tablecaption{Results of the correlation analysis, by means of a Monte Carlo simulation, performed on the sample of the \gl~and the \gq~blazars.\label{tab:statistics}}
\tablehead{
\colhead{} & \colhead{N} & \colhead{$\rho_{\rm s}$} & \colhead{PNC} & \colhead{$\rho_{\rm par}$} & \colhead{PNC}
}
\startdata
\multicolumn{6}{c}{$P_{\rm rad}$ vs. $L_{\rm disk}$}  \\
\gl              & 311  & $0.66\pm0.04$  & $<1\times10^{-5}$ & $0.22\pm0.10$  & $<1\times10^{-5}$ \\
\gq             & 191  & $0.67\pm0.05$  & $<1\times10^{-5}$ &  $0.29\pm0.20$ & $<1\times10^{-5}$ \\
All sources  & 502  & $0.63\pm 0.03$ & $<1\times10^{-5}$ & $0.19\pm 0.08$  & $<1\times10^{-5}$ \\
\hline
\multicolumn{6}{c}{$P_{\rm jet}$ vs. $\dot{M}c^2$}  \\
\gl              & 311  & $0.67\pm0.04$  & $<1\times10^{-5}$  & $0.53\pm0.08$  & $<1\times10^{-5}$ \\
\gq             & 191  & $0.21\pm0.08$  & $5.0\times10^{-2}$ & $0.02\pm0.13$  & $6.5\times10^{-1}$ \\
All sources  & 502  & $0.67\pm 0.03$ & $<1\times10^{-5}$  & $0.56\pm 0.06$  & $<1\times10^{-5}$ \\
\hline
\multicolumn{6}{c}{$M_{\rm BH}$ vs. $z$}  \\
\gl              & 228  & $0.47\pm0.06$  & $<1\times10^{-5}$  & ---  & --- \\
\gq             & 181  & $0.46\pm0.06$  & $<1\times10^{-5}$ & ---  & --- \\
All sources  & 409  & $0.50\pm 0.04$ & $<1\times10^{-5}$  & ---  & --- \\
\hline
\multicolumn{6}{c}{$L_{\rm disk}$ vs. $M_{\rm BH}$}  \\
\gl              & 228  & $0.51\pm0.07$  & $<1\times10^{-5}$  & $0.45\pm0.08$  & $<1\times10^{-5}$ \\
\gq             & 181  & $0.43\pm0.08$  & $1.8\times10^{-4}$ & $0.59\pm0.11$  & $<1\times10^{-5}$ \\
All sources  & 409  & $0.42\pm 0.05$ & $<1\times10^{-5}$  & $0.39\pm 0.06$  & $<1\times10^{-5}$ \\
\hline
\multicolumn{6}{c}{$P_{\rm jet}$ vs. $M_{\rm BH}$}  \\
\gl              & 228  & $0.29\pm0.07$  & $3.2\times10^{-3}$  & $0.22\pm0.11$  & $1\times10^{-4}$ \\
\gq             & 181  & $-0.02\pm0.08$  & $4.7\times10^{-1}$ & $-0.23\pm0.11$  & $3\times10^{-1}$ \\
All sources  & 409  & $0.26\pm 0.05$ & $4.9\times10^{-4}$  & $0.17\pm 0.09$  & $5\times10^{-4}$ \\
\hline
\multicolumn{6}{c}{CD vs. $M_{\rm BH}$}  \\
\gl              & 228  & $0.33\pm0.07$  & $5.7\times10^{-4}$  & $0.01\pm0.10$  & $8.7\times10^{-1}$ \\
\gq             & 181  & $0.45\pm0.08$  & $1.3\times10^{-4}$ & $0.18\pm0.10$  & $1.3\times10^{-2}$ \\
All sources  & 409  & $0.30\pm 0.05$ & $<1\times10^{-5}$  & $0.00\pm 0.09$  & $7.3\times10^{-1}$ \\
\hline
\multicolumn{6}{c}{$L_{\rm disk}/L_{\rm Edd}$ vs. $\gamma_{\rm b}$}  \\
\gl              & 311  & $-0.33\pm0.06$  & $4.4\times10^{-5}$  & ---  & --- \\
\gq             & 191  & $-0.28\pm0.08$  & $7.4\times10^{-3}$ & ---  & --- \\
All sources  & 502  & $-0.30\pm 0.05$ & $<1\times10^{-5}$  & ---  & --- \\
\hline
\multicolumn{6}{c}{$L_{\rm disk}/L_{\rm Edd}$ vs. $R_{\rm diss}$}  \\
\gl              & 311  & $-0.60\pm0.04$  & $<1\times10^{-5}$  & ---  & --- \\
\gq             & 191  & $-0.64\pm0.05$  & $<1\times10^{-5}$ & ---  & --- \\
All sources  & 502  & $-0.64\pm 0.03$ & $<1\times10^{-5}$  & ---  & --- \\
\hline
\multicolumn{6}{c}{CD vs. $L_{\rm disk}$}  \\
\gl              & 311  & $0.61\pm0.04$  & $<1\times10^{-5}$  & $0.36\pm0.08$  & $<1\times10^{-5}$ \\
\gq             & 191  & $0.61\pm0.05$  & $<1\times10^{-5}$ & $0.42\pm0.08$  & $<1\times10^{-5}$ \\
All sources  & 502  & $0.51\pm 0.04$ & $<1\times10^{-5}$  & $0.31\pm0.07$  & $<1\times10^{-5}$ \\
\hline
\multicolumn{6}{c}{CD vs. $P_{\rm jet}$}  \\
\gl              & 324  & $0.50\pm0.05$  & $<1\times10^{-5}$  & $0.22\pm0.07$  & $<1\times10^{-5}$ \\
\gq             & 191  & $-0.02\pm0.07$  & $4.7\times10^{-1}$& $-0.20\pm0.09$ & $3\times10^{-3}$\\
All sources  & 515  & $0.32\pm 0.05$ & $<1\times10^{-5}$  & $0.09\pm0.06$  & $6.7\times10^{-3}$ \\
\hline
\multicolumn{6}{c}{$L_{\rm disk}$ vs. $z$}  \\
\gl              & 311  & $0.70\pm0.03$  & $<1\times10^{-5}$  & ---  & --- \\
\gq             & 191  & $0.73\pm0.04$  & $<1\times10^{-5}$  & ---  & --- \\
All sources  & 502  & $0.74\pm 0.02$ & $<1\times10^{-5}$  & ---  & ---  \\
\hline
\multicolumn{6}{c}{$P_{\rm jet}$ vs. $z$}  \\
\gl              & 324  & $0.50\pm0.05$  & $<1\times10^{-5}$  & ---  & --- \\
\gq             & 191  & $0.45\pm0.06$  & $<1\times10^{-5}$  & ---  & --- \\
All sources  & 515  & $0.49\pm 0.04$ & $<1\times10^{-5}$  & ---  & ---  \\
\hline
\multicolumn{6}{c}{CD vs. $z$}  \\
\gl              & 324  & $0.53\pm0.05$  & $<1\times10^{-5}$  & ---  & --- \\
\gq             & 191  & $0.32\pm0.07$  & $1.2\times10^{-3}$  & ---  & --- \\
All sources  & 515  & $0.54\pm 0.04$ & $<1\times10^{-5}$  & ---  & ---  \\
\enddata
\tablecomments{N denotes the number of sources. Whenever applicable, we also provide the results associated with the partial correlation analysis. $\rho_{\rm s}$ and $\rho_{\rm par}$ correspond to Spearman's rank-correlation coefficients for a simple correlation and partial correlation test, respectively. The associated probabilities of no correlation are also given. We perform a semi-partial correlation test when one of the parameters is Compton dominance, a redshift-independent quantity. See the text for details.}
\end{deluxetable*}

\end{document}